\documentclass[11pt,a4paper]{article}

\usepackage{hyperref}

\usepackage{epsfig}
\usepackage[T1]{fontenc}    % Accents cods dans la fonte.
\usepackage{graphics}
\usepackage{graphicx}
\usepackage{pstricks,pst-coil,pst-fill,pst-plot}
\usepackage[fleqn]{amsmath}    % Les symboles les plus frquents.
\usepackage{amssymb,amsthm}    % Des symboles, les styles de theoremes.
\usepackage{amsfonts}   % Des fontes, eg pour \mathbb.
\usepackage{verbatim}   % Pour les codes sources en informatique.
\usepackage{mathrsfs}   % Des lettres majuscules cursives (\mathscr).
\usepackage{dsfont}
\usepackage{euscript}
\usepackage{yfonts}
\usepackage{marvosym}
\usepackage{vmargin}        % Rgler la taille de la feuille.

\setmarginsrb{2.6cm}{1.7cm}{2.6cm}{1.7cm}{1cm}{1cm}{1cm}{1.6cm}
\makeatletter
\@addtoreset{equation}{section}
\makeatother

\newcommand{\bra}[1]{\langle\,#1\,|}
\newcommand{\ket}[1]{|\,#1\, \rangle}

\newcommand{\moy}[1]{\ensuremath{\langle\, #1 \,\rangle}}

\let\ua=\uparrow
\let\da=\downarrow
\let\tend=\rightarrow

\newtheorem{prop}{Proposition}[section]

{\theoremstyle{remark}
\newtheorem{rem}{Remark}[section]}

\def\Proof{\medskip\noindent {\it Proof --- \ }}

\def\qed{\hfill\rule{2mm}{2mm}\par\medbreak}

\newcommand\beq{\begin{equation}}
\newcommand\enq{\end{equation}}
\newcommand\bem{\begin{multline}}
\newcommand\enm{\end{multline}}

\def\beqa{\begin{eqnarray}}
\def\eeqa{\end{eqnarray}}
\def\ba{\begin{array}}
\def\ea{\end{array}}

\def\det{\operatorname{det}}

\newcommand{\f}[2]{{\ensuremath{%
    \mathchoice%
    {\dfrac{#1}{#2}}
    {\dfrac{#1}{#2}}
    {\frac{#1}{#2}}
    {\frac{#1}{#2}}
}}}
\newcommand{\tf}[2]{\ensuremath{#1/#2}}
\newcommand{\pa}[1]{\ensuremath{\left(#1\right)}}
\newcommand{\paa}[1]{\ensuremath{\left\{#1\right\}}}
\newcommand{\pac}[1]{\ensuremath{\left[#1\right]}}
\newcommand{\paf}[2]{\ensuremath{\left(\f{#1}{#2}\right)}}
\newcommand{\pabb}[3]{\ensuremath{\pa{ #1 \left| \ba{c} #2 \\ #3 \ea \right .}}  }
\def\a{\alpha}

\def\be{\beta}
\def\ga{\gamma}
\def\Ga{\Gamma}
\def\de{\delta}

\def\De{\Delta}
\def\eps{\epsilon}
\def\veps{\varepsilon}
\def\la{\lambda}

\def\th{\theta}

\def\om{\omega}

\newcommand{\mc}[1]{\ensuremath{\mathcal{#1}}}
\newcommand{\mf}[1]{\ensuremath{\mathfrak{#1}}}
\newcommand{\msc}[1]{\ensuremath{\mathscr{#1}}}

\newcommand{\bs}[1]{ \ensuremath{\boldsymbol{#1}  }}

\newcommand{\ov}[1]{\ensuremath{\overline{#1}}}
\newcommand{\wt}[1]{\ensuremath{\widetilde{#1}}}
\newcommand{\wh}[1]{\ensuremath{\widehat{#1}}}

\newcommand{\Int}[2]{\ensuremath{\int\limits_{#1}^{#2}}}
\newcommand{\Oint}[2]{\ensuremath{\oint\limits_{#1}^{#2}}}

\newcommand{\sul}[2]{\ensuremath{\sum\limits_{#1}^{#2}}}
\newcommand{\pl}[2]{\ensuremath{\prod\limits_{#1}^{#2}}}

\newcommand{\R}{\ensuremath{\mathbb{R}}}

\newcommand{\Dp}[1]{\ensuremath{\partial_{#1}}}

\newcommand{\limit}[2]{\ensuremath{\underset{#1 \tend #2}{\longrightarrow} }}
\newcommand{\ex}[1]{\ensuremath{\e{e}^{#1}}}

\newcommand{\abs}[1]{\ensuremath{\left| #1 \right|}}
\newcommand{\norm}[1]{\ensuremath{\left\|#1\right\|}}

\newcommand{\dd}{\mathrm{d}}
\newcommand{\e}[1]{\ensuremath{\mathrm{#1}}}

\newcommand{\intff}[2]{\ensuremath{\left [ \, #1 \,; #2 \, \right ] }}

\newcommand{\intn}[2]{\ensuremath{[\![ \, #1 \,;\, #2 \,]\!]}}

\begin{document}

%\title{\bf Long-time and large-distance asymptotic behavior of the current-current correlators in the non-linear Schr\"{o}dinger model}
 
%\author{{\bf K.~K.~Kozlowski}\thanks{ DESY, Hamburg, Deutschland, karol.kajetan.kozlowski@desy.de}, {\bf V.~Terras}\thanks{ Laboratoire de Physique, UMR 5672 du CNRS, ENS Lyon,  France, veronique.terras@ens-lyon.fr} }  

%\date{\today} 

%\maketitle  

\begin{flushright}
LPENSL-TH-01/11

DESY 10-253
\end{flushright}

\vspace{14pt}

\begin{center}
\begin{LARGE}
{\bf  Long-time and large-distance asymptotic behavior\vspace{-1mm}\\ of the current-current correlators\vspace{2mm}\\
 in the non-linear Schr\"{o}dinger model}
\end{LARGE}
\vspace{30pt}

\begin{large}
{\bf K.~K.~Kozlowski}\footnote{ DESY, Hamburg, Deutschland,
 karol.kajetan.kozlowski@desy.de},~~
{\bf V.~Terras}\footnote{ Laboratoire de Physique, UMR 5672 du
CNRS, ENS Lyon,  France, veronique.terras@ens-lyon.fr}
\par

\end{large}

\vspace{1cm}

\today

\end{center}

\vspace{1cm}

\begin{abstract} 
We present a new method allowing us to derive the long-time and large-distance asymptotic behavior of the correlations functions of quantum integrable 
models from their exact representations.
Starting from the form factor expansion of the correlation functions in finite volume, we explain how to reduce
the complexity of the computation in the
so-called interacting integrable models to the one appearing in free fermion equivalent models.
We apply our method to the time-dependent zero-temperature current-current correlation function  in the non-linear Schr\"{o}dinger model and compute the first few terms in its asymptotic expansion.
Our result goes beyond the conformal field theory based predictions: in the time-dependent case, other types of excitations than
the ones on the Fermi surface contribute to the leading orders of the asymptotics.
\end{abstract}

%%%%%%%%%%%%%%%%%%%%%%%%%%%%%%%%%%%%%%%%%%%%%%%%%%%

\section{Introduction}
\label{sec-intro}

The successful resolution of many-body quantum integrable models through Bethe Ansatz 
\cite{Bet31,Orb58,Thi58,LieL63} naturally raised the question of the computation of their correlation functions.
In its full generality, this problem appeared first as extremely complex
due to the complicated combinatorial representation of the eigenfunctions.
However, for the particular values of the coupling constants for which the models are equivalent to free fermionic ones, the use of specific methods (such 
as the Jordan-Wigner transformation) led to numerous explicit results
\cite{LieSM61,Len64,Len66,McCW67,McC68,BarMcC71,McCW73L,WuMTB76,McCTW77a,VaiT79,JimMMS80,McCPW81,McCPS83a,KorS90,ItsIKS90,ItsIK90,ItsIKV91,ColIKT92,ItsIKV92,ColIKT93,ColIT97}.
In this context,  different types of effective representations have been obtained for the correlation functions, notably in terms of 
determinants.
It has been possible in particular to express two-point functions
in terms of Fredholm determinants of compact operators $I+V$, where the integral kernel $V$
belongs to the class of integrable integral operators \cite{Len64,Len66,KorS90,ColIKT92,CheGZ09}.
These determinants representations have been used to compute the asymptotic behavior of the correlations functions at large distances 
\cite{JimMMS80,McCPS83b,ItsIKS90,ItsIK90,ItsIKV91,ItsIKV92,ColIKT93}.

It was the understanding of the underlying algebraic structures in truly interacting integrable models which finally opened the way to the
computation of their correlation functions.
In particular, the algebraic version of the Bethe Ansatz (ABA) \cite{FadST79} enabled Izergin, Korepin and their collaborators in Leningrad to 
obtain the first exact representations for the correlation functions of the XXZ and the non-linear Schr\"odinger models away from their free 
fermion points (see \cite{BogIK93L} and references therein).
These representations, based on finite-size determinant representations for the norms of Bethe
eigenstates \cite{Gau71L,GauMcCW81,Kor82} and for more general scalar products \cite{Sla90,BogIK93L}, were however not completely explicit due 
to the introduction of dual fields to handle the combinatorial difficulties of the ABA approach.
The first explicit representations (multiple integral representations for elementary building blocks of the XXZ chain at zero temperature) were 
obtained in the 90' by Jimbo, Miki, Miwa and Nakayashiki through $q$-vertex operators and solutions of the $q$-KZ equation \cite{JimMMN92,JimM96, JimM95L}.
A few years later, these representations, together with their generalization to the case of an external magnetic field, were reproduced by the Lyon group in the ABA framework thanks
to the resolution of the so-called quantum inverse scattering problem \cite{KitMT99,KitMT00}.
The Lyon group was also able to provide finite-size determinant representations for the form factors of the finite XXZ chain \cite{KitMT99}.
Developments of the latter approach led to integral representations for the two-point function \cite{KitMST02a,KitMST05a}, to their generalizations in the temperature case \cite{GohKS04,GohKS05}, in the time-dependent case \cite{KitMST05b}, and in all these cases taken together \cite{Sak07}.
On the other hand, new interesting algebraic representations for the correlation functions were also obtained recently using reduced $q$-KZ equations \cite{BooJMST04u,BooJMST06b,BooJMST06a}, which led to the discovery of a hidden  Grassmann structure in the XXZ model
\cite{BooJMST07,BooJMST07u,BooJMST09,JimMS09,BooJMS09,BooJMS10}.

In this framework, one of the most challenging problem is the derivation, from these exact representations, of the long-distance asymptotic behavior of the correlation functions.
In general, zero-temperature correlation functions of models with a gapless spectrum are expected to decay algebraically at large distances. The exponents governing this power-law behavior, called critical exponents, are believed to be universal, {\it i.e.} not depending on the microscopic details of the model, but only on its overall symmetries. Predictions concerning these exponents were obtained via the Luttinger liquid approach \cite{LutP75,Hal80,Hal81a,Hal81b} or the connection to Conformal Field Theory \cite{Pol70,Car84,Aff85,Car86,BloCN86}, together with exact computations of finite size corrections to the spectrum \cite{DeVW85,Woy87,DesDV88,WoyET89,DesDV95}.
However, until very recently, it was not possible to confirm these (indirect) predictions via some direct computations based on exact results
for an integrable model out of its free fermion point.

To tackle this problem, one may notice that, in the ABA approach, the correlation functions can be written in terms of series of multiple 
integrals that can be though of as some multidimensional generalizations of the Fredholm determinants
which appear in the study of the free fermionic models.
One may therefore attempt to extend and adapt the large arsenal of techniques available for free 
fermionic correlators to models away from their free fermion points. The main obstacle here comes from the highly coupled nature of
the various series of multiple integrals representations for the correlation functions.
Actually, until two years ago, there were only few cases out of the free fermion point where
the analysis could have been performed to some extend \cite{KitMST02c,KitMST02d,Koz08}.
A major breakthrough was achieved in \cite{KitKMST09b}, which proposed a systematic method to perform the
asymptotic analysis of the long-distance  asymptotic behavior of the correlation functions of densities in integrable models described
by a six-vertex type $R$-matrix and solvable by the ABA.
There, the idea was to extract the asymptotic behavior of the series of multiple integrals representing the correlator
by making some
connections with the asymptotes of Fredholm determinants.
Although fully functional, the method involved several subtle steps, and in particular a quite complex summation process.
The final result itself, nevertheless, happened to be quite simply given
in terms of a properly normalized form factor \cite{KitKMST09b,KitKMST09c}.
This fact naturally suggests that the method proposed in \cite{KitKMST09b} might be simplified if one works directly at the level of the form factor expansion of the correlation functions.

It is the aim of the present article to expound such a simpler and more direct approach, based on the form factor expansion of the correlation functions in the ABA framework, 
on the finite-size determinant representation for 
the form factors in finite volume and on some of their main features in the thermodynamic limit.
The main advantage compared to \cite{KitKMST09b} is that it relies on the study of more natural physical objects: whereas in \cite{KitKMST09b} the 
correlation function was written as a series over some elementary objects expressed in terms of bare quantities (such as energy, momentum), 
the so-called cycle integrals, here we chose to pursue our study of form factors a step further, and to use some of their thermodynamic 
properties (and in particular their representation in terms of dressed physical quantities) from the very beginning.
This results into a major simplification of the summation process.
Indeed, in \cite{KitKMST09b}, the asymptotic summation of the series of cycle integrals was quite subtle. In particular, it involved a 
summation of a whole class of corrections, sub-leading to each term of the series, but nevertheless contributing to the leading term in the 
final result: they happened to be responsible, once upon summation, for the dressing of bare physical quantities.
Since in our present approach the series over form factors is already written in terms of the dressed quantities, the summation is much 
simpler: the series can be connected to a Fredholm determinant and {\em the leading asymptotic behavior of the correlation function is 
directly given by the leading asymptotic behavior of this Fredholm determinant}.
No subtle identification of sub-leading terms is needed for the result.

Let us be slightly more precise about the different steps of our method.
Its starting point is the form factor expansion of the correlation function in finite volume.
Using the determinant representation of the form factors in finite volume \cite{Sla89,KitMT99}, one can analyze their thermodynamic limit in the particle/hole picture \cite{Sla90,KitKMST09c,KitKMST10u}, and decompose their finite-size expression into some (universal) ``singular'' or ``discrete'' part, which is essentially free-fermion-like and contains the whole non-trivial scaling behavior of the form factor in the thermodynamic limit, and some (model-dependent) ``regular'' part, which admits a smooth thermodynamic limit \cite{KitKMST10u}.
The idea is to reduce the summation over form factors to a summation over their finite-size discrete parts, the (thermodynamic limit of the) regular parts being treated as some dressing functionals.
The main problem is that, for interacting models, the series we consider is highly coupled. The origin of this coupling is twofold: 
on the one hand, the regular part is a complicated function of the particle/hole Bethe roots; on the other hand,  the discrete part can
be seen as a certain functional of the shift function between the considered excited state and the ground state, and hence 
introduces non-trivial couplings between the particle and hole-type Bethe roots.
It is possible to slightly simplify the structure of the different terms by neglecting, on the level of the finite-size series, some 
contributions that should vanish in the thermodynamic limit. We thus obtain an effective (finite-size) form factor series with a simpler 
particle/hole dependence than in the original series.
This series can be linked to a decoupled one, reduced to the sum of the discrete parts of the form factors with a 
shift function that does not depend on the excited state, exactly as in a (generalized) free fermionic model.
The sum can then be computed and recast in terms of a determinant which tends to a Fredholm determinant in the thermodynamic limit.
The leading asymptotic behavior of the correlation function then follows directly from the leading asymptotic behavior of the Fredholm 
determinant \cite{Koz10ua}.

We stress that, within this method, the computation of the asymptotic expansion of the correlation functions for "truly interacting" models and 
for their free fermionic limits carry almost the same complexity, in as much as we offer a way of 
understanding the interacting case as a smooth deformation of the free fermionic one.
This joins in some sense the spirit
of the works \cite{BooJMST07,BooJMST07u,BooJMST09,JimMS09,BooJMS09,BooJMS10}
on the hidden Grassmann structure in the XXZ model with the main difference that, in our method, the free fermionic structure is only used as an intermediate step so as to carry out certain algebraic manipulations in a simpler
way and deform the range of summations to a kind of analog of the steepest descent contour.

We choose to expose our method in the framework of the quantum non-linear Schr\"odinger model, which is probably the best understood and simplest integrable interacting model.
Note however that, with minor modifications specific to the structure of the
model under investigation (more complex structure of the configuration of the pertinent Bethe roots above the ground state, the appearance of
trigonometric functions \ldots  ), this method is in principle applicable to a very wide class of integrable models
whose form factors are known and admit a finite-size determinant representation such as those obtained in 
\cite{Sla89,KitMT99,Oot04}.
We apply our method to the derivation of the first few terms in the long-distance and large-time asymptotic expansion of the correlation 
function of currents at zero temperature (see Section~\ref{sec-result}).
The case of the reduced density matrix, together with some more precise mathematical details about the method, will be considered in a further 
publication \cite{Koz10u}.
We will see that {\em the asymptotes in the time-dependent case are not only issuing from excitations in the vicinities of the Fermi 
boundaries} (which correspond to the region of the spectrum taken into account by the CFT/Luttinger liquid approach), 
{\em but also from the excitations around the saddle-point $\la_0$ of the ``plane-wave'' combination $xp(\la)-t\veps(\la)$, where $p$ and $\veps$ are respectively the dressed momentum 
and energy of the excitations}.
More precisely, the exponents governing the power-law decay of the correlation function are given in terms of the different shift functions between the ground state and excited states having one particle and one hole either on the Fermi boundaries $\pm q$ or at the saddle-point $\la_0$. The associated amplitudes are given by the corresponding, properly normalized, form factors of currents.

The paper is organized as follows.
In Section \ref{sec-NLSE}, we introduce the non-linear Schr\"odinger (NLS) model and
the different notations, relative to the description of the space of states and the thermodynamic limit of the model, that will be used 
throughout the article.
This enables us, in Section~\ref{sec-result}, to present our main result: the leading large-distance and long-time asymptotic behavior of the 
(zero-temperature) correlation function of currents.
The remaining part of the article is devoted to the derivation of this result.
We establish in Section~\ref{sec-ff-series} a (finite-size) form factor series representation for the correlation function (or more precisely 
for its generating function).
In Section~\ref{sec-fftype}, we explain how to relate our highly coupled series to a decoupled, free-fermion type series that can be summed up 
into a finite-size determinant.
Finally, in Section~\ref{sec-asympt}, we take the thermodynamic limit of the previous result, hence representing the form factor series in 
terms of a Fredholm determinant, and explain how to derive the first leading terms of the long-time/large-distance asymptotic behavior of our 
series from  those of the Fredholm determinant.
Technical details are gathered in a set of Appendices: explicit representations for form factors are given in Appendix~\ref{sec-ff}; the 
summation of the free-fermion type series is performed in Appendix~\ref{sec-free}; the notion of functional translation operator, which is used 
as a technical tool to link our series to a decoupled one, is introduced in Appendix~\ref{sec-ftrans}; in Appendix~\ref{sec-master}, we show how to 
relate the summed-up form factor series to the type of series that was obtained in \cite{KitKMST09b} by expanding the master equation; finally, 
in Appendix~\ref{sec-Natte}, we explain how to control sub-leading corrections in the asymptotic series using the so-called Natte series 
introduced in \cite{Koz10ua}.

%%%%%%%%%%%%%%%%%%%%%%%%%%%%%%%%%%%%%%%%%%%%%%%%%%%%%%%%%%%%%%%%%%%%%%%%%%%%%%%%%%%%%%%%%%%%%%%%%%%%%%
%%%%%%%%%%%%%%%%%%%%%%%%%%%%%%%%%%%%%%%%%%%%%%%%%%%%%%%%%%%%%%%%%%%%%%%%%%%%%%%%%%%%%%%%%%%%%%%%%%%%%%%%%%%%%%%%%%%%%%%%%%%%%%%%%%%%%%%%%%%%%%%%%%%%%%%%%

\section{The NLS model: notations and definitions}
\label{sec-NLSE}

The quantum non-linear Schr\"odinger model is described in terms of quantum Bose fields $\Psi(x,t)$ and $\Psi^\dagger(x,t)$ obeying to canonical equal-time  commutation relations, with Hamiltonian
\begin{equation}
H= \int\limits_0^L \left\{ \partial_x\Psi^\dagger\,\partial_x\Psi
             +c\Psi^\dagger \Psi^\dagger \Psi \Psi -h\Psi^\dagger \Psi \right\} \dd x.
\label{Ham}
\end{equation}
The model is defined on a finite interval of length $L$ (we take the thermodynamic limit $L\to\infty$ later on) 
and is subject to periodic boundary conditions.
In \eqref{Ham}, $c$ denotes the coupling constant and $h$ the chemical potential.
In this article, we focus on the repulsive regime $c>0$ in presence of a positive  chemical potential $h>0$.

Due to the conservation of the number of particles, this quantum field theory model is equivalent to a (quantum mechanical) one-dimensional many-body gas of bosons subject to point-like interactions.
Such a model was first introduced and solved by Lieb and  Liniger in 1963 \cite{LieL63,Lie63}, where it
was used as a test for Bogoliubov's theory. It was then thoroughly investigated.
In particular, in 1982,  Izergin and Korepin \cite{IzeK82} (see also \cite{BogIK93L}) proposed a lattice regularization of the
model allowing to implement the ABA scheme \cite{FadST79} in finite volume. 

In this article, we consider the zero-temperature time-dependent correlation function of currents,
\textit{i.e.} the ground state expectation value $\moy{j(x,t)\, j(0,0)}$, where $j(x,t)=\Psi^\dagger(x,t)\,\Psi(x,t)$.
Our aim is to evaluate, in the thermodynamic limit $L\to\infty$ of the model, its large-distance and long-time asymptotic behavior.
%We shall see in particular that, similarly to the time-independant case \cite{KitKMST09b,KitKMST09c}, the amplitudes of the leading terms (see Section~\ref{sec-result}) are given in terms of particular form factors of the model.
Prior to writing down our result, we introduce some necessary notations.
Namely, we describe the space of states of the model and provide definitions of several thermodynamic quantities which appear in our result.

In the algebraic Bethe ansatz framework, the eigenstates $\ket{\psi(\{\la\})}$ of the Hamiltonian \eqref{Ham} are parametrized by a set of spectral parameters $\la_1,\ldots,\la_N$ solution to the system of Bethe equations.
In their logarithmic form, these Bethe equations read
\begin{equation}\label{Bethe}
   L p_0(\la_j)+\sum_{k=1}^N\theta(\la_j-\la_k)=2\pi n_j,
   \qquad j=1,\ldots,N,
\end{equation}
in terms of a set of integer (for $N$ odd) or half-integer (for $N$ even) numbers $n_1,\ldots,n_N$.
The bare momentum $p_0(\la)$ and the bare phase $\theta(\la)$ appearing above are given by
\begin{equation}\label{bare-m-p}
 p_0(\la)=\la,
 \qquad
 \theta(\la)= i\log \Big(\frac{ic+\la}{ic-\la}\Big).
\end{equation}
The energy of the state $\ket{\psi(\{\la\})}$ is
\begin{equation}\label{energy}
 E(\{\la\})=\sum_{j=1}^N \veps_0(\la_j),\qquad\text{with}\qquad
 \veps_0(\la)=\la^2-h.
\end{equation}

The solutions to the system of Bethe equations have been studied by Yang and Yang \cite{YanY69}.
In particular, it has been proven that all solutions are sets of real numbers and that a set of solutions $\{\la_j\}$ can be uniquely parametrized by a set of (half-)integers $\{n_j\}$.

In the following, $\la_j$, $j=1,\ldots,N$, will denote the Bethe roots describing the ground state of the Hamiltonian \eqref{Ham}. They correspond to the 
solution to \eqref{Bethe} associated with the following choice of (half)-integers: $n_j=j-(N+1)/2$, where the number of particles $N$ is fixed by the value of the chemical potential $h$.
In the thermodynamic limit ($L\to\infty$, $N\to\infty$ with $N/L$ tending to some finite average density $D$), the parameters $\la_j$ 
densely fill a symmetric interval $[-q,q]$ of the real axis (the Fermi zone) with some density distribution $\rho(\la)$ solving the following linear integral equation
\begin{equation}\label{Lieb}
 \rho(\la)-\frac{1}{2\pi} \int\limits_{-q}^{q}K(\la-\mu)\,\rho(\mu)\,\dd\mu = \frac{p'_0(\la)}{2\pi},
\end{equation}
where $K$ denotes the Lieb kernel
\begin{equation}
\label{K-kern}
K(\la)=\theta'(\la)=\frac{2c}{\la^2+c^2}.
\end{equation}

Excited states of \eqref{Ham} correspond to other solutions of the Bethe equations.
For technical purpose, it is actually convenient to consider  the {\em twisted Bethe states} $\ket{\psi_\kappa(\{\mu\})}$ instead. These are 
parameterized by solutions $\mu_{\ell_1},\ldots,\mu_{\ell_{N_\kappa}}$ of the twisted Bethe equations,
\begin{equation}\label{twisted-Bethe}
 L p_0(\mu_{\ell_j})+\sum_{k=1}^{N_\kappa}\theta(\mu_{\ell_j}-\mu_{\ell_k})
 =2\pi\Big(\ell_j-\frac{N_\kappa+1}{2}\Big)+i\beta,
 \qquad j=1,\ldots,N_\kappa.
\end{equation}
Here $\beta$ is some purely imaginary parameter, $\kappa=e^\beta$, and $\ell_1<\ell_2<\cdots<\ell_{N_\kappa}$ are some integers.
Note that, the set of roots of \eqref{twisted-Bethe} being completely defined by the set of integers $\ell_j$, we have chosen to label them accordingly.
Eigenstates of the Hamiltonian \eqref{Ham} are obtained as limits when $\kappa\to 1$ of the twisted Bethe states $\ket{\psi_\kappa(\{\mu\})}$.

For purely imaginary $\beta$, the roots of \eqref{twisted-Bethe} are real.
The state parameterized by the solutions of \eqref{twisted-Bethe} with
$\ell_j=j$, $j=1,\dots,N_{\kappa}$, is called the $\kappa$-twisted
ground state in the $N_\kappa$-sector. In the thermodynamic limit,
excitations above this $\kappa$-twisted ground state correspond
to solutions such that most of the $\ell_{j}$'s coincide with their value for the
ground state: $\ell_j=j$ except for  $n$ integers (with $n$ remaining finite in the $N_\kappa\to\infty$
limit) $h_1,\dots,h_n\in\{1,\dots,{N_\kappa}\}$ for which
$\ell_{h_k}=p_k\notin \{1,\dots,{N_\kappa}\}$.
The
integers $h_k$ represent "holes" with respect to the distributions of
integers for the ground state in the $N_\kappa$-sector, whereas the
integers $p_k$ represent "particles".
For such an excited state $\ket{\psi_\kappa(\{\mu_{\ell_j}\})}$, the counting function
 \begin{equation}\label{TBE-cf}
  \wh{\xi}_\kappa(\omega)
  \equiv\wh{\xi}_\kappa(\omega \, |\, \{\mu_{\ell_j}\})
    =\frac1{2\pi}p_0(\omega)+\frac1{2\pi L}\sum_{k=1}^{{N_\kappa}}\theta(\omega-\mu_{\ell_k})
    +\frac{{N_\kappa}+1}{2L}-\frac{i\beta}{2\pi L} ,
\end{equation}
which is monotonously increasing,
defines unambiguously a set of
real "background" parameters $\mu_j$, $j\in\mathbb{Z}$, as the unique solutions to
$\wh{\xi}_{\kappa}(\mu_j)=j/L$.
Among this set of parameters, the $N_{\kappa}$ solutions of the twisted
Bethe equations with integers $\ell_j$ correspond to the solutions $\wh{\xi}_{\kappa}(\mu_{\ell_j})=\ell_j/L$, $j=1,\ldots,N_\kappa$, and the particle rapidities $\mu_{p_a}$ (respectively the hole rapidities $\mu_{h_a}$), $a=1,\ldots,n$, correspond to the solutions $\wh{\xi}_{\kappa}(\mu_{p_a})=p_a/L$ (respectively $\wh{\xi}_{\kappa}(\mu_{h_a})=h_a/L$).

Due to the particular type of correlation function we consider in this article (current-current correlation function), we will restrict our study to 
states having the same number of parameters as the ground state, {\it i.e.} $N_\kappa=N$.
To describe the position of the roots of an excited state $\ket{\psi_\kappa(\{\mu_{\ell_j}\})}$ with respect to the ground state roots $\{\la_j\}$, 
it is convenient to define the (finite-size) shift function
 \begin{equation}\label{shift-finite}
  \wh F(\omega)\equiv \wh F(\omega|\{\mu_{\ell_j}\})
        =L\bigl[\wh\xi(\omega)-\wh\xi_\kappa(\omega)\bigr]
        =\frac{1}{2\pi}\sum_{k=1}^N \big[ \theta(\omega-\la_k)-\theta(\omega-\mu_{\ell_k})\big]
        +\frac{i\beta}{2\pi},
 \end{equation}
in which $\wh\xi_\kappa$ is the excited state's counting function \eqref{TBE-cf} whereas $\wh\xi$ denotes the ground state counting function (at $\kappa=1$).
The spacing between the root $\la_j$ for the  ground state
and the background parameters $\mu_j$ defined\footnote{Since the counting function depends on the particular excited state we consider, so does the set of background parameters: these are therefore defined for \textit{each} excited state $\{\mu_{\ell_j}\}$
separately.} by $\wh{\xi}_{\kappa}(\mu_j)=j/L$ is then given by
 \begin{equation}\label{shift-TD}
   \mu_j-\lambda_j=\frac{F(\lambda_j)}{L\rho(\lambda_j)}+O\big(L^{-2}\big),
   \qquad j=1,\ldots,N,
 \end{equation}
in which $F(\lambda)$ is the thermodynamic limit of the shift function, solution of the
integral equation
 \begin{equation}\label{Int-eq-F}
    F (\lambda)-\frac{1}{2\pi}\int\limits_{-q}^qK(\lambda-\mu)\, F(\mu)\,\dd\mu
    =\frac{i\beta}{2\pi}
    -\frac1{2\pi}\sum_{k=1}^{n}\big[ \theta(\lambda- \mu_{p_k})-\theta(\lambda-\mu_{h_k}) \big].
 \end{equation}
Introducing the dressed phase $\phi(\la,\mu)$ and the dressed charge $Z(\la)$ as the respective solutions of the linear integral equations
\begin{equation}\label{d-phase}
  \phi(\la,\mu) -\frac{1}{2\pi} \int\limits_{-q}^q K(\la-\omega)\, \phi(\omega,\mu)\, \dd \omega
  = \frac{\th(\la-\mu)}{2\pi} ,
\end{equation}
\begin{equation}\label{d-charge}
  Z(\la)- \frac{1}{2\pi} \int\limits_{-q}^q K(\la-\mu)\, Z(\mu)\, \dd \mu =1,
\end{equation}  
we get
\begin{equation}\label{F-thermo}
  F (\la)= \frac{i \beta}{2\pi}Z(\la)
   -  \sum_{a=1}^{n} \big[ \phi(\la,\mu_{p_a}) - \phi(\la,\mu_{h_a}) \big] .
\end{equation}

Other important thermodynamic quantities that we need to introduce in order to formulate our result are the dressed energy $\veps(\la)$ and the dressed momentum $p(\la)$, defined as
\begin{align}\label{d-energy}
 & \veps(\la)-\frac{1}{2\pi}\int\limits_{-q}^q K(\la-\mu)\,\veps(\mu)\,\dd\mu =\veps_0(\la),
     \qquad \text{with}\quad \veps(\pm q)=0,
  \\
  \label{d-momentum}
  & p(\la)=p_0(\la)+\int\limits_{-q}^q\theta(\la-\mu)\,\rho(\mu)\,\dd\mu=2\pi\int\limits_0^\la\rho(\mu)\,\dd\mu.\end{align}
Note that expression  \eqref{d-momentum} enables us to express, in the large $L$ limit, the counting function \eqref{TBE-cf} of any excited state with a finite number of particles and holes above the $N$-particle ground state in the following form:
\begin{equation}\label{cf-p}
   \wh{\xi}_\kappa(\omega \, |\, \{\mu_{\ell_j}\})
    = \xi(\omega)+\e{O}(L^{-1}),
    \qquad\text{with}\quad
    \xi(\omega)=\frac1{2\pi}p(\omega)
    +\frac{D}{2},
\end{equation}
$D$ being the average density in the thermodynamic limit ($N/L\to D$).
In particular this means  that, in the thermodynamic limit and up to $\e{O}(1/L)$ corrections, the counting function does not depend on the particular localization of the corresponding Bethe roots $\{\mu_{\ell_j}\}$: it is the same for all particle/hole-type excited states.

%%%%%%%%%%%%%%%%%%%%%%%%%%%%%%%%%%%%%%%%%%%%%%%%%%%%%%%%%%%%%%%%%%%%%%%%%%%%%%%%%%%%%%%%%%%%%%%%%%%%%%%%%%%%%%%%%%%%%%%%%%%%%%%%%%%%%%%%%%%%%%%%%%%%%%
%%%%%%%%%%%%%%%%%%%%%%%%%%%%%%%%%%%%%%%%%%%%%%%%%%%%%%%%%%%%%%%%%%%%%%%%%%%%%%%%%%%%%%%%%%%%%%%%%%%%%%%%%%%%%%%%%%%%%%%%%%%%%%%%%%%%%%%%%%%%%%%%%%%%%%
%%%%%%%%%%%%%%%%%%%%%%%%%%%%%%%%%%%%%%%%%%%%%%%%%%

%%%%%%%%%%%%%%%%%%%%%%%%%%%%%%%%%%%%%%%%%%%%%%%%%%%%%%%%%%%%%%%%%%%%%%%%%%%%%%%%%%%%%%%%%%%%%%%%%%%%%%%%%%%%%%%%%%%%%%%%%%%%%%%%%%%%%%%%%%%%%%%%%%%%%%
%%%%%%%%%%%%%%%%%%%%%%%%%%%%%%%%%%%%%%%%%%%%%%%%%%%%%%%%%%%%%%%%%%%%%%%%%%%%%%%%%%%%%%%%%%%%%%%%%%%%%%%%%%%%%%%%%%%%%%%%%%%%%%%%%%%%%%%%%%%%%%%%%%%%%%
%%%%%%%%%%%%%%%%%%%%%%%%%%%%%%%%%%%%%%%%%%%%%%%%%%

\section{The main result: large-distance/long-time asymptotic behavior of the correlation function of currents}
\label{sec-result}

We are now in position to formulate our main result, namely the leading asymptotic behavior of the zero-temperature correlation function of currents $\moy{j(x,t)\, j(0,0) }$ in the large-distance and long-time regime, with $x/t=\text{const}$.

Let $u(\la)=p(\la)-\frac{t}{x}\veps(\la)$, where $p$ is the dressed momentum \eqref{d-momentum} and $\veps$ the dressed energy \eqref{d-energy}, the ratio $t/x$ being fixed (and non-zero).
We assume that this function has a  unique saddle-point\footnote{It is clear that the bare counterpart $u_0=p_0(\la)-\frac{t}{x}\veps_0(\la)$ of this function fulfills this property: $u_0'$ admits a unique zero $\la_{0}$ on $\mathbb{R}$, and moreover $u_0''(\la_{0})<0$. Considering the form of the integral equations \eqref{d-energy} and \eqref{d-momentum}, one easily sees that $u'$ admits at least one zero. The case of several saddle-points can in principle be treated similarly, and will give rise to several contributions.}
$\la_0$ on $\mathbb{R}$ ({\em i.e.} $u'$ admits a unique zero $\la_0$ on $\mathbb{R}$, which moreover satisfies $u''(\la_0)<0$).
One distinguishes two different regimes\footnote{We do not consider here the case $\la_0=\pm q$. Indeed, this specific case would demand some further analysis.}
according to whether $\la_0 \in ]-q,q[$ (time-like regime) or $\la_0 \notin [-q,q]$ (space-like regime).
Then, at large-distance and long-time ($x\to +\infty$, $t\to +\infty$, $x/t=\text{const}$) and in the space-like regime,
\begin{multline}\label{asympt-space}
    \moy{j(x,t)\, j(0,0)} 
    = \Big(\frac{p_F}{\pi}\Big)^2 -  \frac{ \mc{Z}^2 }{ 2\pi^2 } \frac{x^2 +t^2v_F^2 }{ \big[x^2-t^2 v^2_F\big]^2} \,
    (1+\e{o}(1))\\
    +
    \frac{ 2\cos(2x p_F) \cdot  \big| \mc{F}^{q}_{-q}   \big|^2 }{ [-i (x-v_F t ) ]^{\mc{Z}^2} \; [i (x+v_F t )]^{\mc{Z}^2} }  \,  
     (1+\e{o}(1))\\
   +  \frac{ \sqrt{2\pi}\, \ex{-i\frac{\pi}{4}} \, p'(\la_0) }{ \sqrt{t \veps''(\la_0)-x p''(\la_0)  } }
       \frac{  \ex{ix[p(\la_0)-p_F]-it\veps(\la_0)} \cdot \big| \mc{F}^{\la_0}_{q} \big|^2}
               {  [-i(x-v_F t )]^{ [F^{\la_0}_{q}(q) - 1]^2} \; [ i (x+v_F t)]^{ [F^{\la_0}_q(-q) ]^2} } \,
     (1+\e{o}(1)) \\
   +   \frac{ \sqrt{2\pi}\, \ex{-i\frac{\pi}{4}} \, p'(\la_0) }{ \sqrt{t \veps''(\la_0)-x p''(\la_0)  } }
       \frac{ \ex{ix[p(\la_0)+p_F]-it\veps(\la_0)} \cdot  \big| \mc{F}^{\la_0}_{-q} \big|^2}
              {  [-i(x-v_F t )]^{ [ F^{\la_0}_{-q}(q) ]^2} \; [ i (x + v_F t)]^{ [F^{\la_0}_{-q}(-q)+1 ]^2} }
      \, (1+\e{o}(1)) ,
\end{multline}
whereas in the time-like regime,
\begin{multline}\label{asympt-time}
    \moy{j(x,t)\, j(0,0)} 
    = \Big(\frac{p_F}{\pi}\Big)^2 -  \frac{ \mc{Z}^2 }{ 2\pi^2 } \frac{x^2 +t^2v_F^2 }{ \big[x^2-t^2 v^2_F\big]^2}
    \, (1+\e{o}(1))\\ 
    +
    \frac{ 2\cos(2x p_F)  \cdot \big| \mc{F}^q_{-q} \big|^2}{ [-i (x-v_F t ) ]^{\mc{Z}^2} \; [i (x+v_F t )]^{\mc{Z}^2} }
   \,
     (1+\e{o}(1))\\
   + \frac{ \sqrt{2\pi}\, \ex{i\frac{\pi}{4}} \, p'(\la_0) }{ \sqrt{t \veps''(\la_0)-x p''(\la_0)  } }
          \frac{  \ex{-ix[p(\la_0)-p_F]+it\veps(\la_0)} \cdot \big| \mc{F}^q_{\la_0} \big|^2}
               {  [-i(x-v_F t )]^{[F^q_{\la_0}(q) + 1]^2} \;  [ i (x+v_F t)]^{ [ F^{q}_{\la_0}(-q)]^2} } \,
     (1+\e{o}(1)) \\
   +  \frac{ \sqrt{2\pi}\, \ex{i\frac{\pi}{4}} \, p'(\la_0) }{ \sqrt{t \veps''(\la_0)-x p''(\la_0)  } }
       \frac{ \ex{-ix[p(\la_0)+p_F]+it\veps(\la_0)} \cdot \big| \mc{F}^{-q}_{\la_0} \big|^2 }
              {  [-i(x-v_F t )]^{[F^{-q}_{\la_0}(q)]^2} \; [ i (x + v_F t)]^{ [F^{-q}_{\la_0}(-q)-1]^2} } \,
        (1+\e{o}(1)) .
\end{multline}

In these expressions, $p_F= \pm p(\pm q) = \pi D$ is the Fermi momentum, $v_F=\veps^{\prime}(q)/p'(q)$ is the Fermi velocity, whereas $\mc{Z}=Z(\pm q)$ is the value of the dressed charge on the Fermi surface.
The exponents of the last two terms in \eqref{asympt-space}-\eqref{asympt-time} are given in terms of the values at the Fermi boundaries $\pm q$ of some specific shift functions \eqref{F-thermo}$F^{\mu_p}_{\mu_h}$ (at $\beta=0$)  between the ground state and an excited state with one particle at $\mu_p$ and one hole at $\mu_h$:
the shift functions
\begin{equation}
    F^{\la_0}_{\pm q}(\la)=-[\phi(\la,\la_0)-\phi(\la,\pm q)]
\end{equation}
between the ground state and an excited state with one particle at $\la_0$ and one hole at $\pm q$ in \eqref{asympt-space} (space-like regime), and the shift functions
\begin{equation}
   F^{\pm q}_{\la_0}(\la)=-[\phi(\la,\pm q)-\phi(\la,\la_0)] 
\end{equation}
between the ground state and an excited state with one particle at $\pm q$ and one hole at $\la_0$ in \eqref{asympt-time} (time-like regime).

The corresponding amplitudes are given in terms of some properly normalized form factors $\mathcal{F}^{\mu_p}_{\mu_h}$ of the current operator between the ground state and an excited state containing one particle at $\mu_p$ and one hole at $\mu_h$.
The connexion
\begin{equation}
  \big| \mc{F}^{q}_{-q} \big|^2 
   = \lim_{L\to + \infty} \Big(\frac{L}{2\pi}\Big)^{2\mathcal{Z}^2} \;
      \bigg| \frac{\bra{\psi^{q}_{-q} } \, j(0,0) \, \ket{\psi_g } }{ || \psi^{q}_{-q} || \cdot || \psi_g || } \bigg|^2,
      \label{ff0}
\end{equation}
between the amplitude $\big| \mc{F}^{q}_{- q} \big|^2$ and the square of the norm of an Umklapp form factor of the density operator between the ground state $\ket{\psi_g}$ and an excited state $\ket{\psi^{q}_{-q} }$ containing one particle and one hole at the opposite 
ends of the Fermi boundaries was already noticed in \cite{KitKMST09b,KitKMST09c}.
A similar phenomenon happens for the other terms.
More precisely, the amplitudes $\big| \mc{F}^{\la_0}_{\pm q} \big|^2$ appearing in \eqref{asympt-space} (space-like regime) correspond to the thermodynamic limit of the properly normalized norm squared of the form factor of the current operator taken between the ground state $\ket{\psi_g}$ and an eigenstate $\ket{\psi^{\la_0}_{\pm q} }$ with a particle at $\la_0$ and a hole at $\pm q$:
\begin{align}
   &\big| \mc{F}^{\la_0}_{q} \big|^2 
   = \lim_{L\to + \infty} \Big(\frac{L}{2\pi}\Big)^{1+[F^{\la_0}_{q}(q)-1]^2+[F^{\la_0}_{q}(-q)]^2} \;
      \bigg| \frac{\bra{\psi^{\la_0}_{q} } \, j(0,0) \, \ket{\psi_g } }{ || \psi^{\la_0}_{q} || \cdot || \psi_g || } \bigg|^2,
      \label{ff1}\\
    &\big| \mc{F}^{\la_0}_{-q} \big|^2 
   = \lim_{L\to + \infty} \Big(\frac{L}{2\pi}\Big)^{1+[F^{\la_0}_{-q}(q)]^2+[F^{\la_0}_{-q}(-q)+1]^2} \;
      \bigg| \frac{\bra{\psi^{\la_0}_{-q} } \, j(0,0) \, \ket{\psi_g } }{ || \psi^{\la_0}_{-q} || \cdot || \psi_g || } \bigg|^2.
      \label{ff2}     
\end{align}
Similarly, the amplitudes $\big| \mc{F}^{\pm q}_{\la_0} \big|^2$ appearing in \eqref{asympt-time} (time-like regime) correspond to the thermodynamic limit of the properly normalized  norm squared of the form factor of the current operator between the ground state $\ket{\psi_g}$ and an eigenstate $\ket{\psi^{\pm q}_{\la_0} }$ with a particle at $\pm q$ and a hole at $\la_0$:
\begin{align}
   &\big| \mc{F}_{\la_0}^{q} \big|^2 
   = \lim_{L\to + \infty} \Big(\frac{L}{2\pi}\Big)^{1+[F_{\la_0}^{q}(q)+1]^2+[F_{\la_0}^{q}(-q)]^2} \;
      \bigg| \frac{\bra{\psi_{\la_0}^{q} } \, j(0,0) \, \ket{\psi_g } }{ || \psi_{\la_0}^{q} || \cdot || \psi_g || } \bigg|^2,
      \label{ff3}\\
    &\big| \mc{F}_{\la_0}^{-q} \big|^2 
   = \lim_{L\to + \infty} \Big(\frac{L}{2\pi}\Big)^{1+[F_{\la_0}^{-q}(q)]^2+[F_{\la_0}^{-q}(-q)-1]^2} \;
      \bigg| \frac{\bra{\psi_{\la_0}^{-q} } \, j(0,0) \, \ket{\psi_g } }{ || \psi_{\la_0}^{-q} || \cdot || \psi_g || } \bigg|^2.
      \label{ff4}     
\end{align}
The explicit expressions of \eqref{ff0}-\eqref{ff4} are given in Appendix~\ref{sec-amplitudes}.

\bigskip

We now comment this result. First of all, the first two lines of each expression appear as a direct generalization, involving the relativistic combinations $x\pm v_F t$, of the large-distance asymptotic behavior of the corresponding static correlation function.  They are in agreement (up to the amplitude of the third term, computed in \cite{KitKMST09b,KitKMST09c}) with the predictions coming from the CFT/Luttinger liquid approximations. 
This is indeed not surprising, since such terms involve particle/hole excitations on the Fermi boundary and are therefore correctly taken into account by a linearization of the spectrum around this point.
However, when time is of the same order of magnitude as distance (\textit{i.e.} $\tf{x}{t} \sim \e{O}(1)$),
other types of excitations also contribute to the asymptotics, namely those involving particles (for the space-like regime) or 
holes (for the time-like regime) with rapidities located at the saddle-point $\la_0$. These contributions appear in the last two lines of \eqref{asympt-space}, \eqref{asympt-time}.

\begin{rem}
This result is also valid when $x/t \to 0$ ($t>> x$). In the $t/x \to 0$ limit  (\textit{i.e.} $t<<x$), one has $\la_0\to +\infty$, and the last two terms of \eqref{asympt-space}-\eqref{asympt-time} exhibit a very quick oscillation except around $x=0$, which reconstructs the 
$\delta(x)$-part of the equal-time correlation function.
Hence, as expected, the contribution of these terms vanishes at large distances.
\end{rem}

\begin{rem}
It is easy to see that our result coincides, at the free fermion point $c=+\infty$, with the leading asymptotic behavior obtained through a 
direct computation of the correlation function.
In that case $\la_0=x/(2t)$ and the combination of the last two terms of \eqref{asympt-space}-\eqref{asympt-time} reduces to:
\begin{equation}
   i \sqrt{\frac{\pi}{t}}\, \ex{i\alpha\frac{\pi}{4}} \ex{i\alpha(\frac{x^2}{4t}+th)}
   \bigg\{ \frac{\ex{-i\alpha[xq-t(q^2-h)]}}{x-2qt}-\frac{\ex{i\alpha[xq+t(q^2-h)]}}{x+2qt} \bigg\},
\end{equation}
where $\alpha=1$ in the space-like regime and $\alpha=-1$ in the time-like regime.
Note that this contribution is dominant with respect to the other power-law decaying terms.  
\end{rem}

%%%%%%%%%%%%%%%%%%%%%%%%%%%%%%%%%%%%%%%%%%%%%%%%%%%%%%%%%%%%%%%%%%%%%%%%%%%%%%%%%%%%%%%%%%%%%%%%%%%%%%%%%%%%%%%%%%%%%%%%%%%%%%%%%%%%%%%%%%%%%%%%%%%
%%%%%%%%%%%%%%%%%%%%%%%%%%%%%%%%%%%%%%%%%%%%%%%%%%%%%%%%%%%%%%%%%%%%%%%%%%%%%%%%%%%%%%%%%%%%%%%%%%%%%%%%%%%%%%%%%%%%%%%%%%%%%%%%%%%%%%%%%%%%%%%%%%%%%%
%%%%%%%%%%%%%%%%%%%%%%%%%%%%%%%%%%%%%%%%%%%%%%%%%%%%%%%%%%%%%%%%%%%%%%%%%%%%%%%%%%%%%%%%%%%%%%%%%%%%%%%%%%%%%%%%%%%%%%%%%%%%%%%%%%%%%%%%%%%%%%%%%%%%%%
%%%%%%%%%%%%%%%%%%%%%%%%%%%%%%%%%%%%%%%%%%%%%%%%%%

\section{Form factor expansion for the correlation function in finite volume}
\label{sec-ff-series}

%%%%%%%%%%%%%%%%%%%%%%%%%%%%%%%%%%%%%%%%%%%%%%%%%%%%%%%%%%%%%%%%%%%%%%%%%%%%%%%%%%%%%%%%%%%%%%%%%%%%%%%%%%%%%%%%%%%%%%%%%%%%%%%%%%%%%%%%%%%%%%
%%%%%%%%%%%%%%%%%%%%%%%%%%%%%%%%%%%%%%%%%%%%%%%%%%%%%%%%%%%%%%%%%%%%%%%%%%%%%%%%%%%%%%%%%%%%%%%%%%%%%%%%%%%%%%%%%%%%%%%%%%%%%%%%%%%%%%%%%%%%%%%%%%%%%%
%%%%%%%%%%%%%%%%%%%%%%%%%%%%%%%%%%%%%%%%%%%%%%%%%%%%%%%%%%%%%%%%%%%%%%%%%%%%%%%%%%%%%%%%%%%%%%%%%%%%%%%%%%%%%%%%%%%%%%%%%%%%%%%%%%%%%%%%%%%%%%%%%%%%%%
%%%%%%%%%%%%%%%%%%%%%%%%%%%%%%%%%%%%%%%%%%%%%%%%%%

The derivation of the result we have just announced is based on the form factor expansion for the correlation function.
In this section, we write down the form factor series representation for the current-current correlation function in finite volume, or more precisely for its generating function.
Each term of the series can be expressed by using the finite-size determinant representation for the form factors of the model and their relation with the overlap scalar products.
Such a series will be the starting point for our study.

Let us consider the zero-temperature correlation function of currents $\moy{j(x,t)\, j(0,0) }$, with $j(x,t)=\ex{itH} j(x,0) \, \ex{-itH}$.
Inserting the sum over the complete (see \cite{Dor93}) set of Bethe states $\ket{\psi'}$ between the two operators we obtain
\begin{align}\label{sum-ff1}
 \moy{j(x,t)\, j(0,0) }
   &= \sum_{\psi'}  \ex{-it \mc{E}_{\e{ex}}}
        \frac{\bra{\psi_g}\, j(x,0)\, \ket{\psi'}  \,\bra{\psi'}\, j(0,0) \, \ket{\psi_g}  }
                { || \psi_g ||^2 \cdot ||\psi' ||^2 },                 
\end{align}
in which $\mc{E}_{\e{ex}}$ denotes the difference of energies between the excited state $\ket{\psi'}$ and the ground state $\ket{\psi_g}$.
The matrix elements of $j(x,0)$ can be easily obtained in terms of the matrix elements of  the operator $\ex{\be Q_x}$, with 
$Q_x =\int_0^x j(y,0) \dd y$ :
\begin{equation}\label{ff-j}
  \frac{ \bra{\psi_g}\, j(x,0)\, \ket{\psi'} }{ || \psi_g || \cdot ||\psi' || }
  = \partial_x \partial_\beta \frac{ \bra{\psi(\{\la_j \})}\, \ex{\beta Q_x}\, \ket{\psi_\kappa( \{ \mu_{\ell_j} \} )} }
                                                       { \norm{\psi(\{\la_j\})}\cdot || \psi_\kappa(\{ \mu_{\ell_j} \}) || }
                                                       \Bigg|_{\beta=0}  .
\end{equation}
In their turn, the latter can be computed from the results of \cite{KitKMST07,Oot04,KorS99} in terms of the scalar products:
\begin{equation}\label{ff-Q}
   \bra{\psi(\{\la_j\})}\, \ex{\beta Q_x}\, \ket{\psi_\kappa(\{ \mu_{\ell_j} \})}
   = \ex{i x \mc{P}_{\e{ex}}^\kappa}
      \moy{ \psi(\{\la_j\}) \mid \psi_\kappa(\{ \mu_{\ell_j} \}) },
\end{equation}
with $\mc{P}_{\e{ex}}^\kappa=\sum_{j=1}^N [p_0(\mu_{\ell_j})-p_0(\la_j)]$.
In these expressions, $\{\la\}$ parametrizes the ground state whereas $\{\mu_{\ell_j} \}$ is a set of solutions to the twisted Bethe equations \eqref{twisted-Bethe} associated with the set of integers $\{\ell_j\}$, such that $\ket{\psi_\kappa(\mu_{\ell_j})}\to \ket{\psi'}$
 when $\kappa\to 1$ (we recall that $\kappa=\ex{\beta}$).
Note that it is enough to consider states with $N_\kappa=N$, as  otherwise, the corresponding matrix element is zero.

Therefore, \eqref{sum-ff1} can be rewritten as
\begin{equation}\label{sum-ff2}
 \moy{j(x,t)\, j(0,0) }
   = \sum_{\ell_1<\ldots <\ell_N}  \ex{ix\mc{P}_{\e{ex}}-it \mc{E}_{\e{ex}} }
       \bigg| \partial_y \partial_\beta \bigg\{ \ex{i y \mc{P}_{\e{ex}}^\kappa} \frac{  \moy{ \psi(\{\la_j\}) \mid \psi_\kappa(\{ \mu_{\ell_j} \}) }  }
                { \norm{\psi(\{\la_j\})}\cdot || \psi_\kappa(\{ \mu_{\ell_j} \}) ||   } \bigg\}_{\! y,\beta =0 \,} \bigg|^2,         
\end{equation}
in which we have set
$\mc{P}_{\e{ex}}=\text{lim}_{\kappa\to 1} \mc{P}_{\e{ex}}^{\kappa}$.
Using the fact that $\mc{P}_{\e{ex}}=0$ if $\ket{\psi(\{\la_j\}) }=\ket{\psi(\{ \mu_{\ell_j} \})}$ and that 
$ \moy{ \psi(\{\la_j\}) \mid \psi_\kappa(\{ \mu_{\ell_j} \}) }=0$ if $\ket{\psi(\{\la_j\}) }\not=\ket{\psi(\{ \mu_{\ell_j} \})}$ 
in virtue of the orthogonality of Bethe states \cite{Dor93}, we obtain
\begin{multline}\label{serie-corr}
   \moy{j(x,t)\, j(0,0) }
  = -\left[ \partial_\beta \mc{P}_{\e{ex}}^\kappa |_{\beta=0} \right]^2
            \\
        -\sum_{\psi'\not=\psi_g} \ex{ix\mc{P}_{\e{ex}}-it\mc{E}_{\e{ex}}}\, \mc{P}_{\e{ex}}^2
            \frac{\partial_\beta \moy{ \psi(\{\la_j\}) \mid \psi_\kappa(\{\mu_{\ell_j}\})}_{\beta=0} 
                     \cdot
                     \partial_\beta \moy{ \psi_\kappa(\{\mu_{\ell_j}\})\mid \psi(\{\la_j\})  }_{\beta=0} } 
                             { || \psi_g ||^2 \cdot ||\psi' ||^2 }
                    .
\end{multline}
On the other hand, setting $  \mc{E}_{\e{ex}}^\kappa=\sum_{a=1}^{N} [\veps_0(\mu_{\ell_a})-\veps_0(\la_a)]$ and using similar arguments,
\begin{multline}\label{fonc-gen}
  \partial_x^2 \partial_\beta^2
   \left\{ \ex{    ix \mc{P}_{\e{ex}}^\kappa-it \mc{E}_{\e{ex}}^\kappa  }
              \cdot
         \abs{    \frac{\moy{\psi(\{\la_j\})\mid \psi_\kappa(\{\mu_{\ell_j} \})}  }
                             { || \psi_\kappa(\{\mu_{\ell_j} \}) || \cdot  || \psi(\{\la_j \}) || }
                  }^2 
    \right\}_{\beta=0}    
    =    
    -2    \delta_{\psi' ,\psi_g} \left[ \partial_\beta \mc{P}_{\e{ex}}^\kappa |_{\beta=0} \right]^2
     \\
    -2(1-       \delta_{\psi' ,\psi_g}) \,
    \ex{ix\mc{P}_{\e{ex}}-it\mc{E}_{\e{ex}}}\, \mc{P}_{\e{ex}}^2
            \frac{\partial_\beta \moy{ \psi(\{\la_j\}) \mid \psi_\kappa(\{\mu_{\ell_j}\})}_{\beta=0} 
                     \cdot
                     \partial_\beta \moy{ \psi_\kappa(\{\mu_{\ell_j}\})\mid \psi(\{\la_j\})  }_{\beta=0} } 
                             {|| \psi_g ||^2 \cdot ||\psi' ||^2 }.
\end{multline}
Above we have used the fact that, for $\be \in i \R$, the solutions of the $\kappa$-twisted Bethe equations \eqref{twisted-Bethe} are real, which means that
\begin{equation}
    \ov{ \moy{\psi(\{\la\}) \mid \psi_{\kappa}(\{\mu_{\ell_j}\}) } }
    =     \moy{\psi_{\kappa}(\{\mu_{\ell_j}\}) \mid  \psi(\{\la\}) }.
\end{equation}

Let us define
\begin{equation}
    \mc{Q}_N^\kappa(x,t)
     = \sum_{ \ell_1< \dots< \ell_N } 
        \ex{ix \sum_{a=1}^N [u_0(\mu_{\ell_a})-u_0(\la_a)]}
%     \ex{ix\mc{P}_{\e{ex}}^\kappa-it\mc{E}_{\e{ex}}^\kappa}
     %\pl{a=1}{N} \paa{ \frac{ e_-^2(\la_a) }{ e_-^2(\mu_{\ell_a} ) }}
      %   \cdot
         \abs{    \frac{\moy{\psi(\{\la_j\})\mid \psi_\kappa(\{\mu_{\ell_j} \})}  }
                             { \norm{\psi_\kappa(\{\mu_{\ell_j} \})} \cdot \norm{\psi(\{\la_j \})} }
                  }^2 ,
%         S_N^{\pa{\kappa}}\pa{ \paa{\la}, \paa{\mu} }
\label{def-gen}
\end{equation}
with
$u_0(\la)=p_0(\la)-\frac{t}{x}\veps_0(\la)$.
%$ \mc{P}_{\e{ex}}^\kappa=\sum_{a=1}^{N} [p_0(\mu_{\ell_a})-p_0(\la_a)]$ and $  \mc{E}_{\e{ex}}^\kappa=\sum_{a=1}^{N} [\veps_0(\mu_{\ell_a})-\veps_0(\la_a)]$,
Comparing \eqref{serie-corr} and \eqref{fonc-gen}, we get that $\mc{Q}_N^\kappa(x,t)$ is the generating function of the time and space-dependent
zero-temperature current-current correlation function of the NLS model in finite volume:
\begin{equation}\label{gen-cur}
   \moy{j(x,t)\, j(0,0) }=\frac{1}{2} \Dp{x}^2 \Dp{\be}^2 \mc{Q}_N^\kappa(x,t) \big|_{\be=0} .
\end{equation}
We recall that, in \eqref{def-gen}, the set of parameters $\{\la_j\}$ denotes the solution of the Bethe equations \eqref{Bethe} parametrizing the ground state %$\ket{\psi(\{\la\})}$ 
of the Hamiltonian \eqref{Ham},
whereas $\{\mu_{\ell_j}\}$ stands for the unique solution of the $\kappa$-twisted logarithmic Bethe equations \eqref{twisted-Bethe} defined by the choice of integers
$\ell_1< \dots < \ell_N$.
The sum runs here through all the possible choices of ordered $N$-tuples of integers $\ell_1<\dots<\ell_N$, and therefore through all 
the excited states $\ket{\psi_{\kappa}(\{\mu_{\ell_j} \}) }$ in the $N$-particle sector \cite{Dor93}.

%%%%%%%%%%
%\subsection{Representation for the scalar products}

Each term of the form factor series \eqref{def-gen} involves a normalized scalar product between a Bethe state and a twisted Bethe state. 
For the model in finite volume, such scalar products (and form factors) admit finite-size determinant representations \cite{Sla89,KorS99}.
When the size $L$ becomes large, these scalar products (and form factors) exhibit a non-trivial behavior with respect to the size of the system \cite{Sla90,AriKMW06}. For scalar products between the ground state and an excited state with a finite number of particles and holes 
such as those described in Section~\ref{sec-NLSE}, it was shown in  \cite{KitKMST09c,KitKMST10u} how to extract the leading large $L$ asymptotic behavior 
from their finite size determinant representations.

The detailed representation for the products of form factor that appear in each term of the series \eqref{def-gen}, as well as its leading behavior in the thermodynamic limit, is recalled in Appendix~\ref{sec-ff}. In finite volume, the corresponding normalized scalar product takes the following form:
\begin{equation}\label{forme-sc}
 \abs{    \frac{\moy{\psi(\{\la_j\})\mid \psi_\kappa(\{\mu_{\ell_j} \})}  }
                             { \norm{\psi_\kappa(\{\mu_{\ell_j} \})} \cdot \norm{\psi(\{\la_j \})} }
                  }^2
   = \widehat{D}_N(\{\la_j\},\{\mu_{\ell_j}\})\cdot
       \widehat{\mathcal{G}}_N (\{\la_j\},\{\mu_{\ell_j}\}),            
\end{equation}
in which
\begin{equation}
  \widehat{D}_N(\{\la_j\},\{\mu_{\ell_j}\})
    = \prod_{j=1}^N \frac{\sin^2 [\pi \widehat{F}(\la_j) ] }
                                         {\pi^2 L^2\, \widehat\xi' (\la_j)\, \widehat\xi_\kappa' (\mu_{\ell_j})}
        \cdot \bigg[ \det_N \frac{1}{\la_j-\mu_{\ell_k}}\bigg]^2
     \label{def-DN}
\end{equation}
is the so-called discrete part of the form factor, which contains the whole non-trivial singular part of the form factor with respect to the 
system-size (and is quite universal), whereas $\widehat{\mathcal{G}}_N (\{\la_j\},\{\mu_{\ell_j}\})$ is a dressing function which, for a 
$n$-particle/$n$-hole excited state as defined in Section~\ref{sec-NLSE}, admits a smooth thermodynamic limit:
\begin{equation}
 \lim_{L,N\to\infty}\widehat{\mathcal{G}}_N (\{\la_j\},\{\mu_{\ell_j}\})
 =\mathcal{G}_n \binom{ \{\mu_{p_a}\} }{ \{\mu_{h_a}\} }.
\end{equation}
Here $\mathcal{G}_n$ is a holomorphic function  of the particle/hole rapidities $\{\mu_{p_a}\}$ and $\{\mu_{h_a}\}$.
The microscopic details of this dressing function (see Appendix~\ref{sec-ff}) depend on the model and do not really matter for our study.
We will only use the fact that $\mathcal{G}_n$ can be represented in terms of a functional
\begin{equation}
  \mathcal{G}_n \binom{ \{\mu_{p_a}\} }{ \{\mu_{h_a}\} }
  \equiv \msc{G} \bigg[ \varpi_n\bigg(\cdot \bigg| \begin{matrix} \{\mu_{p_a}\} \\ \{\mu_{h_a}\} \end{matrix} \bigg) \bigg]
\label{rep Gn comme Fnelle}
\end{equation}
of the function
\begin{equation}\label{def-varpi}
  \varpi_n\bigg(\la\,\bigg| \begin{matrix} \{\mu_{p_a}\} \\ \{\mu_{h_a}\} \end{matrix} \bigg)
  =\sum_{a=1}^n \left\{\frac{1}{\la-\mu_{p_a} }-\frac{1}{\la-\mu_{h_a} } \right\}.
\end{equation}

The behavior of the thermodynamic limit of the discrete part is more difficult to obtain (see \cite{KitKMST09c,KitKMST10u} and Appendix~\ref{sec-ff}).
Its knowledge is however unnecessary for our purpose, since our study will directly rely on the finite-size formula \eqref{def-DN}. We will simply use the fact that \eqref{def-DN} can be understood as a functional of the finite-size shift function $\wh{F}$, depending moreover on  the extra sets of particle/hole integers $\{p_a\}$ and $\{h_a\}$:
\begin{equation}
  \widehat{D}_N(\{\la_j\},\{\mu_{\ell_j}\})
  \equiv\widehat{D}_{N,n}\! \begin{pmatrix}  \{ {p_a}\} \\  \{ {h_a}\} \end{pmatrix}\! \big[\wh{F} \big].
\end{equation}
More precisely, the function $\wh{\xi}$ being fixed and the set of parameters $\la_j$ being defined as the pre-image of the set $\{j/L :  j=1,\ldots ,N\}$ by this function, it means that:
\begin{itemize}
 \item the function $\wh{\xi}_\kappa$ in \eqref{def-DN} should be understood as a functional of the shift function $\wh{F}$ through the relation $\wh{\xi}_\kappa=\wh{\xi}-L^{-1}\wh{F}$;
 \item the set of integers $\ell_j$ being fixed by the choice of $\{p_a\}$ and $\{h_a\}$ (and {\em vice-versa}), the parameters $\mu_{\ell_j}$ are obtained as their pre-image by the function $L\wh{\xi}_\kappa$.
\end{itemize}

%%%%%%%%%%%%%%%%%%%%%%%%%%%%%%%%%%%%%%%%%%%%%%%%%%
\section{Effective decoupling of the finite-size form factor series}
\label{sec-fftype}

The structure of the summations in \eqref{def-gen} is extremely intricate.
Indeed, the sets $\{\mu_{\ell_j} \}$ of Bethe parameters over which we sum up  are implicit functions of the $N$ integers $\ell_j$ labelling the 
corresponding excited states.
Although one can build on such a description so as to use the multidimensional residue theory
to recast this complex sum into a contour integral (the so-called {\em master equation} \cite{KitMST05a,KitMST05b,KitKMST07}), the later still remains difficult to 
handle.
In particular, one cannot perform the thermodynamic limit directly on the level of the master equation. 
Therefore, one has to expand the contour integral
into yet another series (the so-called multidimensional Fredholm series) that has a presumably well defined thermodynamic limit.
The asymptotic analysis of the obtained series, decomposed in terms of cycle integrals, then relies on a non-trivial summation process, resulting in particular into the dressing of bare quantities (energy, momentum) into dressed ones.
Although such an approach was successfully applied in \cite{KitKMST09b} to produce the leading asymptotic behavior of the time-independent correlation 
functions of the XXZ chain and of the NLS model (see also \cite{KozMS10u} for the temperature-dependent case), 
we wish here, so as to bypass some of the subtleties of \cite{KitKMST09b}, to propose an alternative line of though by pursuing our study directly on the form factor series.

The idea of our method is the following: the series \eqref{def-gen}  can be related to an essentially free fermion type series (see 
Appendix~\ref{sec-free}), this {\it via} some reasonable physical assumptions concerning the contribution of each of 
its terms in the thermodynamic limit, and {\it via} some formal manipulations. Indeed, it is shown in Appendix~\ref{sec-free} that if, for each excited state,
\begin{itemize}
  \item the rapidities of the particles and holes are decoupled, {\it i.e.} they are determined by a counting function that does not depend on the position of the 
  roots of the corresponding excited state,
%  \item the shift function  does not depend on the position of particle and holes,

  \item the dressing part $\wh{\mathcal{G}}_N$ of the form factors is decoupled,
\end{itemize}
then the series in finite volume can be summed up into a finite-size determinant.
%We do stress that the independence of the counting function on the considered excited state implies that the associated shift function does not depend on the excited state as well; {\it i.e.} it does  not depend on the position of the particles and holes parameterizing a given excited state.   
We explain in this section how to reduce the study of our highly coupled series to this simple case.
It means that, in our setting, the key-role in the summation is played by the universal discrete part $\wh{D}_N$ of the form factors exactly as in the free fermion case.  This fact should be put in parallel with the key-role played, 
in the master equation-based asymptotic analysis \cite{KitKMST09b}, by the Cauchy determinant part of the form factors.

%%%%%%%%%%%%
\subsection{An effective form factor series}
\label{sec-eff}

Considering that we will finally be interested in the thermodynamic limit $L\tend +\infty$ of the generating function  \eqref{def-gen}, we now slightly simplify its form factor series.
Our simplifications rely on two kinds of assumptions:
\begin{enumerate}
  \item[(i)] the contribution to the sum \eqref{def-gen} of a state having a macroscopically (with respect to $L$)  different energy and momentum from the ground state should vanish in the thermodynamic limit;
  \item[(ii)] for each term of the series, "non-discrete" quantities (\textit{i.e.}  parts of the form factor expression behaving smoothly at the thermodynamic limit) should contribute to the thermodynamic limit of the sum \eqref{def-gen} only through their leading order in $L$.
\end{enumerate}
Assumption {(i)} can be attributed in particular to the extremely quick oscillation of the phase factors for states having large excitation momenta and energies. It means notably that:
\begin{itemize}
 \item we can use the particle/hole picture to describe the large-$L$ behavior of 
 the form factors, keeping in mind that the relevant ({\em i.e.} the one not vanishing in the $L\tend +\infty$ limit)  part of the form factor expansion corresponds to a summation over states with a {\em finite}  (or at least growing much less than $L$) number of particle/hole excitations above the ground state;
 \item  we can introduce a "cut-off" (with respect to $L$) of the range of
integers on which we sum up in \eqref{def-gen} since, for very large integers, momentum and energy of the corresponding state become also very large.
\end{itemize} 
Doing this, we roughly speaking neglect correcting terms in the lattice size $L$. It is thus reasonable to assume that, on the same ground, one can also neglect some finite-size corrections  to the leading thermodynamic behavior of the form factors, as stated in (ii).
More precisely, we make the following simplifications in the series \eqref{def-gen}:
\begin{itemize}
  \item we replace the finite-size shift function $\wh{F}$, the smooth part $\wh{G}_N$ and the oscillating exponent $\sum_{a=1}^N [u_0(\mu_{\ell_a})-u_0(\la_a)]$ by their leading contribution in the thermodynamic limit;
  \item in the obtained expression, we now understand the rapidities of the particles $\mu_{p_a}$ (resp. of the holes $\mu_{h_a}$), corresponding to a particular choice of integers $\ell_1<\dots< \ell_N$, as being defined in terms of the thermodynamic limit $F$ of the shift function (and no longer in terms of its finite-size counterpart \eqref{shift-finite}) as the pre-images of $\tf{p_a}{L}$  (resp. $\tf{h_a}{L}$) by the counting function $\wh{\xi}_F=\wh{\xi}-\tf{F}{L}$;  
in other words,  the $2n$ parameters $\{ \mu_{p_a} \}$ and $\{ \mu_{h_a} \}$ are obtained as the unique solutions to the system of $2n$ equations
\beq
%
%\hspace{-1cm}%\wh{\xi}_F\pa{\mu_{p_a}}= 
%
\wh{\xi}(\mu_{p_a}) - \f{1}{L} F\!\pabb{ \mu_{p_a} }{\! \{ \mu_{p_a} \} \!\! } {\! \{ \mu_{h_a} \} \!\!}
                        = \f{p_a}{L}
\quad \; \e{and} \quad  \;
\wh{\xi}(\mu_{h_a}) - \f{1}{L} F\!\pabb{ \mu_{h_a} }{\! \{ \mu_{p_a} \} \!\! } {\! \{ \mu_{h_a} \}\!\! } = \f{h_a}{L} \;,
\label{equation part trous}
\enq
  with $a=1,\dots,n$.
  
\end{itemize}

These simplifications result into an effective series that we assume to have the {\em same value at the thermodynamic limit} as the original 
series. In other words, we conjecture that,
\begin{equation}
  \lim_{L,N\to\infty} \mc{Q}_N^\kappa(x,t)_\mathrm{eff} =  \lim_{L,N\to\infty} \mc{Q}_N^\kappa(x,t)
  \equiv  \mc{Q}^\kappa(x,t),
\end{equation}
with
\begin{multline}
   \mc{Q}_N^\kappa(x,t)_\mathrm{eff}
    = \ex{ -\frac{x \be p_F}{\pi} } \!\!
       \sum_{ \substack{\ell_1<\dots < \ell_N \\ \ell_a \in \mc{B}_L  }}
       %
 %      \prod_{a=1}^{n} \frac{e_-^{2}(\mu_{h_a}) }{ e_-^{2}(\mu_{p_a}) }
       \ex{ix\sum\limits_{a=1}^n [u(\mu_{p_a})-u(\mu_{h_a})]}
       \cdot
       \wh{D}_{N,n}\!\begin{pmatrix} \{ {p_a}\} \\ \{ {h_a}\} \end{pmatrix}\!
       \bigg[F\bigg(\! \cdot \bigg| \begin{matrix}  \{ \mu_{p_a} \} \\ \{ \mu_{h_a} \} \end{matrix} \bigg) \bigg]
              \\
       \times
       \msc{G}
       \bigg[\varpi_n\bigg(\! \cdot \bigg| \begin{matrix}  \{ \mu_{p_a} \} \\ \{ \mu_{h_a}\} \end{matrix} \bigg) \bigg].
       \label{eff-series}
\end{multline}
%
%It is the effective series \eqref{eff-series} that constitutes the starting point of our asymptotic analysis.
In \eqref{eff-series},
the sums are restricted to the set $\mc{B}_L=\big\{ \ell \in \mathbb{Z} \mid  -w_L \le \ell \le w_L \big\}$, where $w_L$ is a cut-off growing as 
$L^{1+\eps}$, for some $\eps>0$.
For a given set of integers $\{\ell_j\}$, the parameters $p_a$ and $h_a$ define the position of particles and holes as explained in Section~\ref{sec-NLSE}.
The (finite-size) representation for the form factor has been partly replaced by its thermodynamic limit.
More precisely, the dressing function and the phase factor, which both behave smoothly at the thermodynamic limit, have been replaced by their leading equivalents. 
Note that, as a consequence, it is now the dressed energy and momentum that appear in the phase factor 
(\textit{cf} Appendix~\ref{sec-ff}): $u(\la)=p(\la)-\frac{t}{x}\veps(\la)$.
In \eqref{eff-series}, we have kept  the finite-size expression of the discrete part, but we have nevertheless replaced the expression of the finite-size shift function $\wh{F}$ by its limiting value $F$ \eqref{F-thermo}.
Finally, the particles and holes' rapidities associated to a given set of integers $\ell_1<\dots<\ell_N$ are obtained (in terms of $F$) {\it via} the system of $2n$ equations \eqref{equation part trous}.

\begin{rem}
  We stress that, in \eqref{eff-series}, we are still in "finite volume" ({\em i.e.} for the moment, $L$ and $N$ are kept finite).
We have only modified the original series \eqref{def-gen} arguing that it should not affect the value of its thermodynamic limit.
\end{rem}

\begin{rem}\label{rem-finite-sums}
The introduction of the cut-off $w_L$ is convenient to avoid  problems of convergence in our further manipulations of the series.
Indeed, as far as we remain in finite volume, we now deal with finite sums only.
\end{rem}

%%%%%%%%%%%%%%%%%%%%%%%%%%%%%%%%%%%%%%%%%%%%%%%%%%%%%%%%%%%%%%%%%%%%%%%%%%%%%%%%%%%%%%%%%%%%%%%%%%%%%%
%%%%%%%%%%%%%%%%%%%%%%%%%%%%%%%%%%%%%%%%%%%%%%%%%%%%%%%%%%%%%%%%%%%%%%%%%%%%%%%%%%%%%%%%%%%%%%%%%%%%%%
%%%%%%%%%%%%%%%%%%%%%%%%%%%%%%%%%%%%%%%%%%%%%%%%%%%%%%%%%%%%%%%%%%%%%%%%%%%%%%%%%%%%%%%%%%%%%%%%%%%%%%

\subsection{Towards a free-fermion type series}
\label{sec-decoupling}

Although the effective series \eqref{eff-series} is already simpler that the original one it is still highly coupled:
\begin{itemize}
\item the thermodynamic limit of the shift function \eqref{F-thermo} still depends on the position of particles and holes, 
\item the rapidities of the particles and holes have to be computed for each excited state separately ({\it i.e.} they are excited state dependent),
\item the expression of the functional $\msc{G}$ is extremely intricate.
\end{itemize}
Our aim is now to relate $\mc{Q}_N^{\kappa}\!\pa{x,t}_{\e{eff}}$ to a decoupled series, so as to reduce its analysis to the one of the generalized free-fermion case that is carried out in Appendix~\ref{sec-free}.
This can be done by understanding the coupling between variables as the result of the action of some functional translation operators\footnote{The use of functional translation operators is simply a convenient and compact way to manipulate generalized Lagrange series (see Appendix C of \cite{KitKMST09b}). In fact, the whole reasoning that follows can instead be performed by writing explicitly the corresponding series.} (see Appendix~\ref{sec-ftrans}).

The function $\varpi_n$ \eqref{def-varpi} depends {\em linearly} on a function of the rapidities of the particles and of the holes. 
This means that one can formally express this dependence by means of the action of a functional translation operator, as explained in 
Appendix \ref{sec-ftrans}: for any sufficiently smooth functional $\mc{F}$ supported on a neighborhood $\mathcal{U}$ of the real axis,  
\begin{equation}
   \mc{F}\bigg[  \varpi_n\bigg(\! \cdot \bigg| \begin{matrix}  \{ {\mu}_{p_a}\} \\ \{ {\mu}_{h_a}\} \end{matrix} \bigg)  \bigg]
   =
   \pl{a=1}{n} \exp\Bigg\{\Int{ \R }{ } \dd \la \, \bigg[ \frac{1}{\la-\mu_{p_a}}-\frac{1}{\la-\mu_{h_a}} \bigg] \,
            \frac{ \de }{ \de\omega(\la) } \Bigg\}  \cdot 
   \mc{F} [ \omega ] \bigg|_{\omega=0} .
\label{transl ac varpi}   
\end{equation}
The thermodynamic limit of the shift function \eqref{F-thermo} also depends {\em linearly} on a function of the rapidities of the particles and holes.
However, the situation is slightly more complicated than in \eqref{transl ac varpi} since, in virtue of \eqref{equation part trous}, the parameters $\{ \mu_{p_a} \}$  and $\{ \mu_{h_a} \}$ are themselves functionals of the shift 
function. As explained in Appendix \ref{sec-ftrans}, one can still represent any smooth functional of this shift function
in terms of translation operators, provided that one imposes some operator ordering $\bs{:}\,\cdot\, \bs{:}$ and that one takes  into account the contribution of the Jacobian coming from the summation of the corresponding multi-dimensional Lagrange series (see 
formulae \eqref{L-Gamma} and \eqref{result-Lagrange}): 
\begin{equation}
   \mc{F}\bigg[  F\bigg(\! \cdot \bigg| \begin{matrix}  \{\bar{\mu}_{p_a}\} \\ \{\bar{\mu}_{h_a}\} \end{matrix} \bigg)  \bigg]
   = \, \bs{:}
   \pl{a=1}{n} \exp\Bigg\{\Int{ \R }{ } \dd \la \, [ \phi(\la,\bar{\mu}_{h_a})-\phi(\la,\bar{\mu}_{p_a}) ] \,
            \frac{ \de }{ \de\tau(\la) } \Bigg\}  \cdot 
   \mc{F} [ \nu_{\tau} ] \, \bs{ :}  \bigg|_{\tau=0} 
    J .
   % \det_{\R} \pac{ I- \f{\de\Ga\pac{\nu_{\tau}}}{\de \tau \pa{\mu}}\pa{\la} }_{\mid \tau=F} . 
\label{eq translation fction shift}
\end{equation}
%$\bs{:} \,  \cdot \, \bs{:}$.
Above the operator ordering $\bs{:}\,\cdot\, \bs{:}$  is such that, in the formal series expansion in powers of $\de / \de \tau(\la)$, 
all functional derivative operators are located on the left. 
%then computed, and finally the functions $\tau$ sent to zero.
$\nu_\tau$ is the function
\beq
\nu_{\tau}(\la)=  i\frac{\be Z(\la)}{2\pi} + \tau(\la) ,
\enq
and
$J$ is the Jacobian $J=\det_{\R}\big[ I- \tf{ \de \Ga[\nu_{\tau}](\la) }{\de \tau(\mu) } \big]_{\mid \tau=F}$ of the functional 
\begin{equation}
\Ga[\nu_{\tau}]({\la}) = \sul{a=1}{n} [ \phi(\la, \bar{\mu}_{h_a}) - \phi(\la, \bar{\mu}_{p_a}) ], \qquad 
\e{where} \quad \bar{\mu}_{\ell} \equiv \wh{\xi}^{-1}_{\nu_{\tau}}\Big(\f{ \ell }{L} \Big) 
\quad \e{with} \quad \ell \in \mathbb{Z} \, .
\label{eq expl Funct Jacobian}
\end{equation}
This Jacobian is evaluated at $\tau=F$, where $F$ coincides with the shift function occuring in the \textit{l.h.s.} of \eqref{eq translation fction shift}. 
%It is readily seen that for an $n$ particle/hole excited state with fixed $n$, it is of the type $1+\e{O}\pa{L^{-1}}$.
We stress that the bar occuring in the parameters $ \bar{\mu}_{p_a}$ (resp.  $ \bar{\mu}_{h_a}$) appearing in the \textit{r.h.s.}
of \eqref{eq translation fction shift} indicates that these are to be understood as the pre-images of $\tf{p_a}{L}$
(resp. $\tf{h_a}{L}$) by the counting function $\wh{\xi}_{\nu_{\tau}}$, as in \eqref{eq expl Funct Jacobian}.

It is easy to convince oneself that, provided that the number of particle/hole excitations is finite, $J=1+\e{O}(L^{-1})$. 
In the light of our previous arguments, only this sector of excitations is expected to contribute to the  $L\tend +\infty$
limit of the form factor expansion. As a consequence, in order to be consistent with our previous approximations, we also drop the finite-size corrections due to this Jacobian by considering from now on that our effective series (still abusively denoted by the same symbol $ \mc{Q}_N^\kappa(x,t)_\mathrm{eff}$) is in fact given by the expression \eqref{eff-series} in which each term is multiplied by the inverse of this Jacobian. 

Hence, the use of the functional derivative leads to the following representation:
\begin{multline}
      \wh{D}_{N,n}\binom{ \{ {p_a}\} }{ \{ {h_a}\} }
       \bigg[ F\bigg(\! \cdot \bigg| \begin{matrix}  \{ \mu_{p_a}\} \\ \{\mu_{h_a}\} \end{matrix} \bigg) \bigg]
       \cdot
       \msc{G}
       \bigg[\varpi_n\bigg(\! \cdot \bigg| \begin{matrix}  \{ \mu_{p_a}\} \\ \{ \mu_{h_a}\} \end{matrix} \bigg) \bigg] 
       \cdot J^{-1}
              \\
=  \;
   \bs{:} \pl{a=1}{n} \ex{\ \Int{ \R }{  } \dd \la \, [ \phi(\la,\bar{\mu}_{h_a})-\phi(\la,\bar{\mu}_{p_a}) ] \,
            \frac{ \de }{ \de\tau(\la) }  }
%   \pl{a=1}{n}\ex{ - \Int{\R}{} \dd \la [\phi(\la,\mu_{p_a}) - \phi(\la,\mu_{h_a})]  \frac{\de }{ \de \rho(\la)} }
   \pl{a=1}{n} \ex{\ \Int{ \R }{ } \dd \la \, \big[ \frac{1}{\la-\bar{\mu}_{p_a}}-\frac{1}{\la-\bar{\mu}_{h_a}} \big] \,
            \frac{ \de }{ \de\omega(\la) }  }
            \\
   \cdot
   \wh{D}_{N,n}\!\begin{pmatrix} \{ {p_a}\} \\ \{ {h_a}\} \end{pmatrix}\! [  \nu_{\tau}(\cdot)]
   \cdot
   \msc{G} [ \omega(\cdot)]  \bs{:}
    \bigg|_{\substack{\tau = 0 \\ \omega=0}} \,  .
\label{funct-shift}
\end{multline}
Here  we have insisted on the fact that $\wh{D}_{N,n}$ and $\msc{G}$ are functionals acting on the argument $\cdot$ of
$\nu_{\tau}(\cdot)$ and $\omega(\cdot)$. Also, on the \textit{l.h.s.} of \eqref{funct-shift}, we have explicitly pointed out the parametric 
dependence of $F$ and $\varpi_n$ on the parameters $\{ {\mu}_{p_a}\}$ and $\{ {\mu}_{h_a}\}$.
%However, in order to keep the formulae lighter and as long as it does not lead to confusions, we will not insist on these properties anymore.

In the \textit{r.h.s.} of \eqref{funct-shift}, the functional $\wh{D}_{N,n}$ acts on a shift function $\nu_{\tau}$ that does not depend  anymore on the 
particle/hole rapidities.
In other words, the effective shift function $\nu_{\tau}$ becomes independent of the
summation variables and hence mimics the one appearing in the generalized free fermion model studied in Appendix~\ref{sec-free}.
Moreover, the equations defining the position of the particle's/hole's rapidities also become decoupled:
the rapidity $\bar{\mu}_{\ell}$ is now the unique solution to  $\wh{\xi}_{\nu_{\tau}}(\bar{\mu}_{\ell})=\tf{\ell}{L}$, and its value does not 
depend anymore on the choice of the other integers describing the excited state, but only on the 
function $\tau$.

It is convenient to express, on a formal level, each term of the pre-factor in \eqref{funct-shift} as a ratio of two exponents
$ \ex{\,\wh{g}(\bar{\mu}_{p_a})} / \ex{\,\wh{g}(\bar{\mu}_{h_a})} $,
where $\wh{g}$ is an operator valued function (we have used the hat so as to insist on this property):
\begin{equation} 
   \wh{g}=\wh{g}_\tau+\wh{g}_\omega,
   \quad \text{with} \quad
   \wh{g}_{\tau}(\la)= - \Int{ \R }{ } \dd \mu \, \phi(\mu,\la) \, \frac{\de}{\de \tau(\mu)},
   \quad 
   \wh{g}_{\omega}(\la)=  \Int{ \R }{ } \frac{\dd \mu}{\mu-\la} \,
            \frac{ \de }{ \de\omega(\la) }  .
\end{equation}
In the following, we set
\begin{equation}
\wh{E}_-^{2}=\ex{-ix u-\wh{g}}.
\end{equation}

The operator order $\bs{:} \cdot \bs{:}$ being linear, we can exchance it with a {\em finite} summation symbol such as
the one appearing in the expression \eqref{eff-series} for $\mc{Q}_N^\kappa(x,t)_\mathrm{eff}$.
Hence, we obtain
\begin{align}
   \mc{Q}_N^\kappa(x,t)_\mathrm{eff}
   & =
   \ex{  -\frac{\be x p_F}{\pi} }
   \;   \bs{ : } \,
   \sum_{ \substack{\ell_1<\dots < \ell_N \\  \ell_j \in \mc{B}_L  } }
   \pl{a=1}{n} \frac{\wh{E}_-^{2}(\bar{\mu}_{h_a})  }{ \wh{E}_-^{2}(\bar{\mu}_{p_a}) } 
   \cdot
   \wh{D}_{N,n}\!\begin{pmatrix} \{ {p_a}\} \\ \{ {h_a}\} \end{pmatrix}\! [ \nu_{\tau}]
   \cdot
   \msc{G}[\omega]
   \,   \bs{:} 
    \bigg|_{\substack{\tau = 0 \\ \omega=0}} \,
    \nonumber\\
  &= \ex{ - \frac{\be x p_F}{\pi} } \;  \bs{: }   \,
     \,
     \pl{a=1}{N} \frac{ \wh{E}_-^2( { \bar{\mu} }_a) }{ \wh{E}_-^2(\la_a) } 
     \cdot
     X_N\big[  \nu_{\tau}, \wh{E}_-^{2} \big]
     \cdot
     \msc{G}[\omega]\,    \bs{ : }  
    \bigg|_{\substack{\tau = 0 \\ \omega=0}} \, ,
\label{relation-XN}
\end{align}
in which $X_N$ is the generalized free-fermionic functional \eqref{gen-XN} studied in Appendix~\ref{sec-free}.

%%%%%%%%%%%%%%%%%%%%%%%%%%%%%%%%%%%%%%%%%%%%%%%%
\subsection{Representation in terms of a finite-size determinant}
\label{sec-sum}

In Appendix~\ref{sec-free} we have recast the generalized free-fermionic functional $X_N[\nu,E_-^2]$  into a finite-size 
determinant \eqref{XN-det}, this without any further approximations. Such a representation is obtained through {\em purely algebraic} manipulations on the initial definition \eqref{gen-XN} for $X_N[\nu,E_-^2]$.
These two representations are still equal when applied to the (non-commutative) operator valued functions $\nu_\tau$ and $\wh{E}_-$:  it is indeed not a problem to implement the appropriate operator order at any step of the computation performed in Appendix~\ref{sec-free}.
One may therefore use the determinant representation \eqref{XN-det} to obtain new representations for our series \eqref{relation-XN}, provided that one imposes an operator order on the entries of the columns of the determinant\footnote{Such an order can for instance be implemented explicitly by expanding the determinant into a sum over permutations.} \cite{Tsi92L}.
This leads to
\begin{equation}\label{Q-det}
   \mc{Q}_N^\kappa(x,t)_\mathrm{eff}
= \ex{ - \frac{\be x p_F}{\pi} }  \bs{: } \,
     \pl{a=1}{N} \frac{ \wh{E}_-^2( \bar{\mu}_a) }{ \wh{E}_-^2(\la_a) } 
     \cdot
     \det_N \bigg[ \delta_{jk}+\frac{\wh{V}^{(L)}(\la_j,\la_k)}{L\,\wh{\xi}'(\la_k) }\bigg] \cdot
     \msc{G}[\omega]\,     \bs{ : }  
    \bigg|_{\substack{\tau = 0 \\ \omega=0}} \, ,
\end{equation}
where the (operator valued) kernel is given as
\begin{equation}
    \wh{V}^{(L)}(\la_k,\la_j)
    = 4  \frac{ \sin [\pi\nu_\tau(\la_k)]\,  \sin [\pi\nu_\tau(\la_j)] }{ 2i\pi (\la_k-\la_j)}
        \Big\{\wh{E}_+^{(L)}(\la_k) \, \wh{E}_-(\la_j)- \wh{E}_+^{(L)}(\la_j) \, \wh{E}_-(\la_k) \Big\},
\label{noyau-Lhat}
\end{equation}
with
\begin{equation}
    \wh{E}^{(L)}_+(\la)
     =i \wh{E}_-(\la) \left\{ \, \int\limits_{\bar{A}_L}^{\bar{B}_L} \hspace{-4.35mm} \diagup \hspace{1.3mm}
                                      \frac{ \dd \mu }{2\pi } \frac{ \wh{E}^{-2}_{-}(\mu) }{ \mu-\la } 
        + \frac{ \wh{E}_-^{-2}(\la) }{ 2 } \cot[ \pi \nu_\tau(\la)]   + I_L\big[ \nu_\tau,\wh{E}_-^{-2} \big] ( \la )  \right\} .
\label{E+Lhat}
\end{equation}
The expression of the functional $I_L$ can be found in \eqref{IL}.

Such an expression provides a formal re-summation of the effective form factor series \eqref{eff-series}.
In the next section, we show how to take the thermodynamic limit of \eqref{Q-det}.

%%%%%%%%%%%%%%%%%%%%%%%%%%%%%%%%%%%%%%%%%%%%%%%%%%%%%%%%%%%%%%%%%%%%%%%%%%%%%%%%%%%%%%%%%%%%%%%%%%%%%%%%%%%%%%%%%%%%%%%%%%%%%%%%%%%%%%%%%%%%%%%%%%%%%%%%
%%%%%%%%%%%%%%%%%%%%%%%%%%%%%%%%%%%%%%%%%%%%%%%%%%%%%%%%%%%%%%%%%%%%%%%%%%%%%%%%%%%%%%%%%%%%%%%%%%%%%%%%%%%%%%%%%%%%%%%%%%%%%%%%%%%%%%%%%%%%%%%%%%%%%%%%
%%%%%%%%%%%%%%%%%%%%%%%%%%%%%%%%%%%%%%%%%%%%%%%%%%%%%%%%%%%%%%%%%%%%%%%%%%%%%%%%%%%%%%%%%%%%%%%%%%%%%%%%%%%%%%%%%%%%%%%%%%%%%%%%%%%%%%%%%%%%%%%%%%%%%%%%
%%%%%%%%%%%%%%%%%%%%%%%%%%%%%%%%%%%%%%%%%%%%%%%%%%%%%%%%%%%%%%%%%%%%%%%%%%%%%%%%%%%%%%%%%%%%%%%%%%%%%%%%%%%%%%%%%%%%%%%%%%%%%%%%%%%%%%%%%%%%%%%%%%%%%%%%

\section{Thermodynamic limit and asymptotic analysis}
\label{sec-asympt}

We now build on the results of the previous section so as to derive the leading  large-distance/long-time asymptotic behavior 
of the thermodynamic limit  $\mc{Q}^{\kappa}(x,t)$ of the generating function $\mc{Q}_N^\kappa(x,t)$.

%%%%%%%%%%%%%%
\subsection{Representation in terms of a Fredholm determinant in the thermodynamic limit}
\label{sec-therm}

The next step of our study is to obtain a convenient representation for the thermodynamic limit $N,L\to +\infty$ of the generating function 
\eqref{def-gen}.
Recall at this point that the effective series \eqref{eff-series} was built so as to approach, in this limit, 
the same value $\mathcal{Q}^\kappa(x,t)$ as the original form factor series \eqref{def-gen}.
We will therefore use the representation \eqref{Q-det} to derive a suitable expression for $\mathcal{Q}^\kappa(x,t)$.

In principle, prior to taking the thermodynamic limit of a formal representation such as \eqref{Q-det}, one should first compute the effect of the translation operators: the operation of taking the thermodynamic limit is indeed \textit{a priori} only allowed on expressions defined in terms of explicit holomorphic functions ({\it i.e.} which do not contain any operator valued functions).
In Appendix~\ref{sec-master}, such a procedure is explicitly performed so as to obtain, starting from \eqref{Q-det}, a particular series representation of the thermodynamic limit $\mathcal{Q}^\kappa(x,t)$ of $\mathcal{Q}^\kappa_N(x,t)$. 
More precisely, in Appendix~\ref{sec-master}, we expand the finite-size determinant in \eqref{Q-det} into a sum over determinants of all sub-matrices of $V(\la_j,\la_k)$, factorize the Cauchy part of each of these determinants, compute the effects of the translation operators, and {\em then} take the thermodynamic limit.
This enables us to express  $\mathcal{Q}^\kappa(x,t)$ as a series of multiple integrals coinciding, in the equal-time case, with the series expansion derived from the master equation representation in  \cite{KitKMST09b}.

In fact, it is easy to convince oneself, by carrying out the computation in the reverse order, that the operation of taking the thermodynamic limit
$N,L \tend +\infty$ 
and the one of computing the effect of the translation operators
do actually commute.
Indeed, let us consider the expression  \eqref{Q-det} in which we have sent
directly $N,L\tend +\infty$, hence replacing sums by integrals  in the prefactor and the finite-size determinant representation for $X_N$ by the Fredholm determinant  \eqref{XN-Fredholm} :
\begin{equation}\label{expr-thermo}
   \ex{ \frac{-\be x p_F}{\pi} }  \bs{: } 
     \,
     \ex{ -\Int{-q}{q} \nu_{\tau}(\la) \, [ix u'(\la) +\wh{g}'(\la)] \, \dd\la }
     \det \big[ I+V[\nu_\tau, u, \wh{g} ] \big] 
     \cdot \msc{G}[\omega]
     \, 
     \bs{ : }  
    \Big|_{\substack{\tau = 0 \\ \omega=0}} \, ,
\end{equation}
where
\begin{equation}\label{kern}
  V[\nu, u, g ](\la,\mu)
  = 4 \frac{\sin[\pi\nu(\la)]\,\sin[\pi\nu(\mu)]}{2i\pi(\la-\mu)}
     \Big\{ E_+(\la)\, E_-(\mu) - E_+(\mu) \, E_-(\la) \Big\}
\end{equation}
with
\begin{align}
   &E_-(\la)= \ex{-ix\frac{u(\la)}{2} - \frac{g(\la)}{2} }, \label{E-}\\
   &E_+(\la)= i E_-(\la) \left\{ \, 
       \int\limits_\mathbb{R} \hspace{-4.3mm}\diagup \hspace{2mm}
       \frac{ \dd \mu }{2\pi } \frac{ E^{-2}_{-}(\mu) }{ \mu-\la } 
       +  \frac{ E_-^{-2}(\la) }{ 2 } \cot [\pi \nu(\la)]  \right\} . \label{E+}
\end{align}
Then, if we decompose  the Fredholm determinant in \eqref{expr-thermo} into
its Fredholm series and subsequently compute the effect of the functional translation operators, we obtain {\em the same representation \eqref{lim-Q}} as by performing all these operations in the opposite order.
This means (provided that the series \eqref{lim-Q} is convergent, which is not completely obvious but seems nevertheless a reasonable assumption) that the thermodynamic limit of the effective series \eqref{Q-det}, which is supposed to give the thermodynamic limit of the original form factor series \eqref{def-gen}, is effectively given  by the expression \eqref{expr-thermo}, {\it i.e.}
\begin{equation}\label{Q-Fred}
   \mathcal{Q}^\kappa(x,t)
   = \ex{ \frac{ -\be x p_F}{\pi} }  \bs{: } 
     \,
     \ex{ -\Int{-q}{q} \nu_{\tau}(\la) \, [ix u'(\la) +\wh{g}'(\la)] \, \dd\la }
     \det \big[ I+V[\nu_\tau, u, \wh{g} ] \big] 
     \cdot \msc{G}[\omega]\;
     \bs{ : }  
    \Big|_{\substack{\tau = 0 \\ \omega=0}} \, ,
\end{equation}
with a kernel $V$ given by \eqref{kern}-\eqref{E+}.
It also means that we can now use any other
existing representation for the Fredholm determinant so as to compute the effect of the translation operators and recover standard scalar-valued 
functions.
In fact, it is not very convenient for our purpose to expand the determinant into its Fredholm series like in Appendix~\ref{sec-master} since the latter 
does not provide any information on its large-$x$  asymptotic behavior.
This large-$x$ asymptotic behavior was studied in \cite{Koz10ua} precisely with the goal of computing the asymptotic behavior of \eqref{Q-Fred}, and a much more convenient (with respect to the 
$x\tend +\infty$ limit) representation for the Fredholm determinant was obtained there.

%%%%%%%%%%%%%%%%%%%%%%%%%%%%%%%
\subsection{Large-$x$ asymptotic behavior of the Fredholm determinant}
\label{sec-asympt-det}

The large $x$ asymptotic analysis of the Fredholm determinant with kernel \eqref{kern}-\eqref{E+} was performed in \cite{Koz10ua} using Riemann-Hilbert problem-based techniques.
There, it was proven that, under some hypothesis about the regularity and behavior of the functions $\nu$, $u$ and $g$ 
defined on some open neighborhood of the real axis (see \cite{Koz10ua} for more precisions), and provided that the function $u$ has a unique saddle-point $\la_0$ on $\mathbb{R}$ (with $\la_0\not= \pm q$),
\begin{multline}\label{as-det-leading}
   \det_{[-q,q]} \big[ I+V \big] [\nu, u, g ]
   = \exp \Bigg\{ \Int{-q}{q} \big[ ixu'(\la)+g'(\la)\big]\, \nu(\la)\, \dd\la \Bigg\}    
   \\
   \times
    \Bigg\{\, \mc{B}_x[\nu,u] \,
   +  \sum_{\eps=\pm 1 }
                  \ex{i \eps x[u(q)-u(-q)]+\eps [g(q)-g(-q)]} \,
                  \mc{B}_x[\nu+\eps,u] 
      \\
      + \frac{1}{x^{\frac32}} \sum_{\eps=\pm 1}
         \ex{i \alpha x[u(\la_0)-u(\eps q)]+\alpha [g(\la_0)-g( \eps q)]} \,
          b_1^{(\eps,\alpha)}[\nu,u] \, \mc{B}_x[\nu,u] + \mathcal{R}_x[\nu,u,g] \Bigg\}\, ,
\end{multline}
with $\alpha=+1$ in the space-like regime\footnote{The case $\la_0<-q$ was not considered in \cite{Koz10ua}.} ($\la_0>q$) 
and $\alpha=-1$ in the time-like regime ($\la_0\in ]-q,q[$).
In \eqref{as-det-leading}, the functional $\mc{B}_x$ reads
\begin{equation}
  \mc{B}_{x} [\nu,u]
  = \frac{
     \ex{i\f{\pi}{2} [\nu^2(q)-\nu^2(-q)]} \; \wt{\mc{B}}[\nu] }
    { [2qx \, (u'(q)+i0^+) ]^{\nu^2(q)} \, [2qx \, u'(-q)]^{\nu^2(-q)} } ,
 \label{BX}
\end{equation}
with $\wt{\mc{B}}[\nu]$ given by \eqref{Btilde}, and
\begin{multline}
  b_1^{(\eps,\alpha)}[\nu,u]
  =\ex{-i\alpha\frac{\pi}{4}}\,
    \frac{ [2qx\, u'(\eps q) +i0^+]^{2\eps\alpha \nu(\eps q) }  }
            { \sqrt{-2\pi u''(\la_0)}\, u'(\eps q) } \,
     \frac{(-\eps)\, \nu(\eps q)}{  (\la_0-\eps q)^2 }   \,
     \bigg( \frac{\la_0+q}{\la_0-q-i0^+} \bigg)^{\! -2\alpha\nu(\la_0)} 
     \\  
     \times   
     (\ex{-2i\pi\nu(\la_0)}-1)^{1-\alpha}  \, (\ex{-2i\pi\nu(\eps q)} -1)^\alpha \;
     \frac{ \Gamma(1-\eps\alpha\nu(\eps q) )}{\Gamma(1+\eps\alpha \nu(\eps q)) } \;  
     \ex{\alpha\wt{J}[\nu](\la_0)-\alpha\wt{J}[\nu](\eps q)},                                  
%  = -\eps\frac{\nu(\eps q) \, (\ex{-2i\pi\nu(\la_0)}-1)}{u'(\eps q) \, (\la_0-\eps q)^2  \sqrt{-2\pi u^{\prime\prime}\!(\la_0)} } \,
%       \Bigg\{ 
%       \ex{-i\frac{\pi}4}\,
%       [2qx\, u'(\eps q) +i0^+]^{2\eps \nu(\eps q) } \,
%       \frac{\ex{-2i\pi\nu(\eps q)} -1}{\ex{-2i\pi\nu(\la_0)}-1} 
%       \\
%       \quad \times
%       \bigg( \frac{\la_0+q}{\la_0-q-i0^+} \bigg)^{\! -2\nu(\la_0)} \,
%       \frac{ \Gamma(1-\eps\nu(\eps q) )}{\Gamma(1+\eps \nu(\eps q)) } \,
%       \ex{2 \Int{-q}{q} \left[ \frac{\nu(\la_0)-\nu(\mu)}{\la_0-\mu} -\frac{\nu(\eps q)-\nu(\mu)}{\eps q-\mu} \right] \dd\mu }
%       \Bigg\}^\alpha .
\end{multline}
where $\wt{J}$ is given by \eqref{Jtilde}.
$\mathcal{R}_x[\nu,u,g]$ is a remainder which is uniformly of order $\e{O}\big(\frac{\log x}{x}\big)$ in what concerns the non-oscillating corrections, of order $\e{O}\big( \mc{B}_{x}[\nu+\eps,u]\, \frac{\log x}{x}\big)$ in what concerns the oscillating corrections at 
$\ex{ i\eps x [u(q)-u(-q)] }$,
and of order $\e{O}\big( \mc{B}_{x}[\nu,u]\,   b_1^{(\eps,\alpha)}[\nu,u]\, \frac{\log x}{x^{\tf{5}{2}}}\big)$
in what concerns the oscillating corrections at $\ex{ i\a x [u(\la_0)-u( \eps q)] }$.
It was shown in \cite{Koz10ua} how to obtain a series representation for this remainder, the so-called Natte series, which possesses the property of being well ordered with respect to its large-$x$ behavior (in can be shown that its $n$-th term is (uniformly) a $\e{O}(x^{-na})$, for some $0<a<1$
that depends on $\nu$, $u$ and $g$). Hence, this series is well adapted for the study of the asymptotic behavior of the determinant (on the contrary to its Fredholm series).
The form of such a Natte series is recalled in Appendix~\ref{sec-Natte}.

\begin{rem}
Note that for $\abs{\Re\pa{\nu}} < \tf{1}{2}$ the first (non-oscillating) term  $\mathcal{B}_x[\nu,u]$ in \eqref{as-det-leading} is always leading, 
at large $x$, with respect to the other ones. 
This leading term will produce the leading asymptotic behavior of the generating function $\mathcal{Q}^\kappa(x,t)$, as we will see in the next subsection.
 However, recall that we have to differentiate twice with respect to $x$ and 
 with respect to $\beta$ at $\beta=0$ in order to obtain the correlation function $\moy{j(x,t)\, j(0,0) }$.
  By such a process, the first oscillating corrections may become leading with respect to non-oscillating terms. This is 
  the reason why we also consider these corrections in \eqref{as-det-leading}.
\end{rem}

%%%%%%%%%%%%%%%%%%%%%%%%%%%%%%%%%%
\subsection{Asymptotic behavior of the correlation function}
\label{sec-asympt-gen}

So as to obtain the asymptotic behavior of the correlation function, 
it remains to compute the effect  of the functional translation operators  on the representation \eqref{as-det-leading} of \eqref{Q-Fred}.
First, we observe that the exponential pre-factor of \eqref{Q-Fred} is canceled by the one in \eqref{as-det-leading}:
\begin{equation}\label{exponents}
\exp\Bigg\{-\Int{-q}{q} \nu_{\tau}(\mu)\, \wh{g}'(\mu)\, \dd\mu \Bigg\}
\exp\Bigg\{\Int{-q}{q} \nu_{\tau}(\mu)\, \wh{g}'(\mu)\, \dd\mu \Bigg\}
=1 .
\end{equation}
Such a simplification is justified by the fact that, when applying \eqref{exponents} on any functionals of $\nu_\tau$ or $\omega$, we observe an {\em algebraic} cancellation for
each term of the $\bs{:}\cdot \bs{:}$ ordered series expansion in $\tf{\de}{ \de \tau }$ and $\tf{\de}{ \de \om }$ for these exponents \footnote{Such an algebraic cancellation is very similar to the 
one occuring when computing $(\sqrt{I-A})^2$ through a Taylor series expansion around zero.}.
Therefore,
\begin{multline}
   \mathcal{Q}^\kappa(x,t)
   = \ex{ -\frac{\be x p_F}{\pi} }  \bs{: } 
     \,
     \Bigg\{\, \mc{B}_x[\nu_\tau,u]  + \sum_{\eps=\pm 1 }
                  \ex{i \eps x[u(q)-u(-q)]+\eps [\wh{g}(q)-\wh{g}(-q)]} \,
                  \mc{B}_x[\nu_\tau+\eps,u] 
      \\
      + \frac{1}{x^{\frac32}} \sum_{\eps=\pm 1}
         \ex{i \alpha x[u(\la_0)-u(\eps q)]+\alpha [\wh{g}(\la_0)-\wh{g}( \eps q)]} \,
          b_1^{(\eps,\alpha)}[\nu_\tau,u] \, \mc{B}_x[\nu_\tau,u] 
			\; + \; \mathcal{R}_x[\nu_\tau,u,\wh{g}]       \Bigg\}
    \cdot \msc{G}[\omega]
     \bs{ : }  \bigg|_{\substack{\tau=0 \\ \omega=0}}\, .
     \label{dot-lead-Q}
\end{multline}

It can be proved that, by computing the effect of the functional translation operators occuring in the remainder $\mathcal{R}_x[\nu_\tau,u,\wh{g}] $, one obtains corrections that are of the same order 
(\textit{i.e.} $\e{O}(\frac{\log x}{x})$  corrections to each of the terms already present in the asymptotics)
as originally in \eqref{as-det-leading}.
This is explicitly done in Appendix~\ref{sec-Natte} by using the so-called Natte series representation of the Fredholm determinant derived in \cite{Koz10ua}.
More precisely, it is shown in Appendix~\ref{sec-Natte} that the effect of the translation operators do not mix the orders in $x$ among the different terms of this (well-ordered) series.
Therefore, {\em the leading asymptotic behavior of the generating function follows directly from the above leading asymptotic behavior of the Fredholm determinant}.

Since no translation operator is applied on the first (non-oscillating) term of \eqref{dot-lead-Q}, we simply need to set $\tau=0$ and $\omega=0$ into the corresponding expressions.
The action of the translation operators $\ex{\eps [\wh{g}(q)-\wh{g}(-q)]}$ on the second term of \eqref{dot-lead-Q} results into replacing the function $\msc{G}[\omega]$ by
\begin{equation}
   \msc{G}\bigg[\varpi_1\bigg(\! \cdot \bigg| \begin{matrix}  \eps q \\ -\eps q \end{matrix} \bigg) \bigg]
   \equiv \mathcal{G}_1\binom{\eps q}{-\eps q},
\end{equation}
and the function $\nu_\tau$ by the shift function
\begin{equation}
   F\bigg(\! \cdot \bigg| \begin{matrix}  \eps q \\ -\eps q \end{matrix} \bigg) 
  % \equiv F^{\eps q}_{-\eps q}
   =\frac{i\beta}{2\pi} Z -\big[ \phi(\cdot,\eps q) -\phi(\cdot,-\eps q)\big]
   = \Big(\frac{i\beta}{2\pi}+\eps\Big) Z -\eps
\end{equation}
associated to a state with one particle and one hole located on the opposite ends of the Fermi zone.
Similarly, the action of the translation operators $\ex{\alpha [\wh{g}(\la_0)-\wh{g}(\eps q)]}$
results into replacing $\msc{G}[\omega]$ and $\nu_\tau$ respectively by
\begin{equation}
  \msc{G}\bigg[\varpi_1\bigg(\! \cdot \bigg| \begin{matrix}  \la_0 \\ \eps q \end{matrix} \bigg) \bigg]
   \equiv \mathcal{G}_1\binom{\la_0}{\eps q}
   \quad
   \text{and}
   \quad
   F\bigg(\! \cdot \bigg| \begin{matrix}  \la_0 \\ \eps q \end{matrix} \bigg) 
  % \equiv F^{\la_0}_{\eps q}
   =\frac{i\beta}{2\pi} Z -\big[ \phi(\cdot,\la_0) -\phi(\cdot,\eps q)\big]
\end{equation}
in the space-like regime $\alpha=+1$, and by
\begin{equation}
   \msc{G}\bigg[\varpi_1\bigg(\! \cdot \bigg| \begin{matrix} \eps q \\ \la_0 \end{matrix} \bigg) \bigg]
   \equiv \mathcal{G}_1\binom{\eps q}{\la_0}
   \quad
   \text{and}
   \quad
   F\bigg(\! \cdot \bigg| \begin{matrix}  \eps q\\ \la_0\end{matrix} \bigg) 
  % \equiv F^{\eps q}_{\la_0}
   =\frac{i\beta}{2\pi} Z -\big[ \phi(\cdot,\eps q) -\phi(\cdot,\la_0)\big]
\end{equation}
in the time-like regime $\alpha=-1$.
Therefore, we get
\begin{multline}
   \mathcal{Q}^\kappa(x,t)
   = \ex{ - \frac{\be x p_F}{\pi} } 
     \bigg\{ \,
       \mathcal{B}_x\!\bigg[\frac{i\beta}{2\pi} Z,u\bigg]\; \mathcal{G}_0\;
       \bigg(1+\e{O}\bigg(\frac{\log x}{x}\bigg) \bigg) 
     \\
     +\sum_{\eps=\pm 1 }
                  \ex{i \eps x[u(q)-u(-q)]} \;
                  \mc{B}_x\!\bigg[ \bigg(\frac{i\beta}{2\pi}+\eps\bigg)Z,u\bigg] \; \mathcal{G}_1\!\binom{\eps q}{-\eps q}\;
       \bigg(1+\e{O}\bigg(\frac{\log x}{x}\bigg) \bigg) 
      \\
      + \frac{1}{x^{\frac32}} \! \sum_{\eps=\pm 1} \!
         \ex{i  x[u(\la_0)-u(\eps q)]} \,
          b_1^{(\eps,+1)}\!\bigg[ F\bigg(\! \cdot \bigg| \begin{matrix}  \la_0 \\ \eps q \end{matrix} \bigg) ,u\bigg] \, 
          \mc{B}_x\!\bigg[ F\bigg(\! \cdot \bigg| \begin{matrix}  \la_0 \\ \eps q \end{matrix} \bigg) ,u\bigg]\,
          \mathcal{G}_1\!\binom{\la_0}{\eps q} \,
       \bigg(1+\e{O}\bigg(\frac{\log x}{x}\bigg) \bigg) \, \bigg\}  
     \label{lead-Q-space}
\end{multline}
in the space-like regime, and similar expressions in the time-like regime. Namely, in the time-like regime,
the exponent in the last term changes sign and $b_1^{(\eps,+1)}$, $F\big( \cdot | \genfrac{}{}{0pt}{1}{\la_0}{ \eps q} \big) $, $\mathcal{G}_1\binom{\la_0}{\eps q}$ are replaced, respectively, by $b_1^{(\eps,-1)}$, $F\big( \cdot | \genfrac{}{}{0pt}{1}{\eps q}{ \la_0} \big)$ and $\mathcal{G}_1\binom{\eps q}{\la_0}$.

In order to obtain the leading asymptotic behavior of the correlation function of currents, it remains finally to compute the 
second $x$-derivative and second $\be$-derivative at $\beta=0$ of the previous result, \textit{cf} \eqref{gen-cur}.
The derivatives of the first term (the non-oscillating one) in \eqref{lead-Q-space} produce the constant and the non-oscillating term
appearing in \eqref{asympt-space} and \eqref{asympt-time} (we have used \eqref{limG0}).
%As follows from Remark~\ref{rem-zero-beta}, the only way for obtaining a non-vanishing contribution from expressions containing $\mathcal{G}_1$ is, for generic $c$, to apply both $\beta$-derivatives on the factor $(\kappa-1)^2$.
In their turns, the derivatives of the two types of oscillating corrections produce the corresponding oscillating terms in \eqref{asympt-space}, \eqref{asympt-time}, with amplitudes given by \eqref{ff-qq}-\eqref{ff-qla0}.

%%%%%%%%%%%%%%%%%%%%%%%%%%%%%%%%%%%%%%%%%%%%%%%%%%%
%%%%%%%%%%%%%%%%%%%%%%%%%%%%%%%%%%%%%%%%%%%%%%%%%%%
\section{Conclusion}
\label{sec-concl}

In this article, we have proposed a new method to derive, starting from first principles, the leading asymptotic behavior of the two-point correlation 
functions of quantum integrable systems.
To explain the main steps of this method, we chose to focus on the example of the current-current correlation function of the quantum non-linear 
Schr\"{o}dinger model.
The case of the field/conjugated field correlator
$\moy{ \Psi^{\dagger}(x,t)\,\Psi(0,0)}$, together with a rigorous setting for carrying out all the manipulations with operator valued determinants,  will 
be given in \cite{Koz10u}.

Our result goes beyond the CFT/Luttinger liquid based predictions: the saddle-point contributions that appear in the asymptotic behavior \eqref{asympt-space}-\eqref{asympt-time} involve excitations away from the Fermi surface and cannot be neglected for $x/t$ finite.

Compared to the approach of \cite{KitKMST09b}, based on the master equation representation for the 
correlation functions, the present study relies directly on their form factor expansion. We 
would like to conclude this article by making here a few comments about the similarities and differences between the two 
methods.

Of course, since the master equation can be understood as the result of a summation over the form factors, the spirit of the two approaches is essentially the same.
In particular, in both approaches, the asymptotic analysis relies essentially on the {\em singular part} of the form factor which is explicitly extracted (in 
the form of a Cauchy determinant squared), whereas the (model-dependent) regular part is treated as a dressing part which is formally decoupled (or linked to decoupled functions) to make 
the analysis possible. This enables one to draw a link between the quantity to estimate and the Fredholm determinant of a generalized sine kernel.
The asymptotic analysis of this Fredholm determinant then leads to the asymptotic behavior of the correlation function.

However, there exist some essential differences between the two approaches.
In \cite{KitKMST09b}, the correlation function was expanded into a series whose building blocks were the so-called {\em cycle integrals}.
These cycle integrals could then be related to a Fredholm determinant, which allowed one to access to their asymptotic behavior.
The physical interpretation of these objects was however not clear, and they happened to be quite indirectly related to the correlation functions.
In fact, once the asymptotic behavior of these cycle integrals established, one had to sum up the series  so as to obtain the asymptotic 
behavior of the correlation function. In this process, the main problem was that {\em not only the leading asymptotic behavior of 
individual cycle integrals was contributing to the leading order for the correlation functions}: one had to perform a fine and non-trivial study \cite{KitKMST09a,KitKMST09b,Koz09} of the asymptotic series so as to 
gather terms {\em at all order} that finally, when summed up, were contributing to the leading order.
In fact, all these terms were rearranging themselves into some generalization of a multiple Lagrange series producing, when summed up, a {\em dressing of bare quantities} (energy and momentum) {\em into dressed ones}.
On the contrary, here, one deals from the very beginning with objects having a clear physical interpretation, the form factors.
In this context, {\em the dressed quantities appear naturally when considering the thermodynamic limit of these form factors}.
Hence, performing the summation over the form factors, we can connect the series to a Fredholm determinant that is already 
expressed in terms of these dressed quantities.
Therefore, the asymptotic study is much simpler: it happens that {\em the leading asymptotic behavior of the Fredholm determinant gives directly the leading asymptotic behavior of the correlation function}.
In other words, there is no need to resort to highly non-trivial summation as in \cite{KitKMST09b}.

In fact, we would like to stress that the whole process described in the core of this article is quite simple and direct.
Once the form factor series written down and the effective contributing part of the form factor established, the result of the summation can be 
expressed, in the thermodynamic limit, in terms of a Fredholm determinant.
The leading asymptotic behavior of the correlation function follows directly from the leading asymptotic behavior of this Fredhom determinant.
Even if, in the course of the computations, we use some functional translation operators to relate our series to a decoupled one, whenever the action of these functional translation operators has to be computed, it is quite straightforward, {\it i.e.} 
it produces a simple translation of the functional on which it acts.
On the contrary, if one wants to recover, starting from the form factor expansion, a series similar to the one studied in \cite{KitKMST09b}, the 
computations are much more involved (see Appendix~\ref{sec-master}).
In particular, the action of the translation operators on the series of Appendix~\ref{sec-master} produces  non-trivial effects:
one has to deal with summations of generalized Lagrange series, which results into an undressing of the dressed
quantities into bare ones. 
It is therefore clear that most of the mathematical complexity corresponding to this non-trivial summation has already been taken into account by the fact 
that we had some precise  description  (the particle-hole picture) of our form factors, which allowed us from the very beginning to deal with dressed quantities instead of bare ones.
Note that the simpler setting of our method enables us here to consider the time-dependent case, which is not so obvious within the approach of \cite{KitKMST09b}. 
Note also that, in principle,  there is no intrinsic obstruction preventing us from obtaining higher order terms in the asymptotic expansion for 
the correlation functions. For this it is enough to  refine the asymptotic expansion of the Fredholm determinant.  

Of course, there is a price to pay for this simplicity: the need of some clear picture to describe the form factors.
Whereas in \cite{KitKMST09b} all kinds of Bethe roots were automatically taken into account within the master equation framework (and this without any precise study of the spectrum), here we strongly rely on the fact that the spectrum of the model we consider is particularly simple: all excited states can be described in terms of particles and holes.
In order to apply our method to the case of the XXZ spin chain, for example, one has also to take into account the contribution of complex solutions.
Although it seems that, for the time-independent case, these solutions do not contribute to the leading asymptotic behavior of correlation functions (see \cite{KitKMST-ff-as}), the question remains open in the time-dependent case.
This will be the subject of a further study.

%%%%%%%%%%%%%%%%%%%%%%%%%%%%%%%%%%%%%%%%%%%%%%%%%%%%%%%%%%%%%%%%%%%%%%%%%%%%%%%%%%%%%%%%%%%%%%%%%%%%%%%%%%%%%%%%%%%%%%%%%%%%%%%%%%%%%%%%%%%%%%%%%%%%%%%%%%%%%%%%%%%%%%%%%%%%%%%%%%%%%%%%%%%%%%%%%%%%%%%%%%%%

\section*{Acknowledgements}

V. T. is supported by CNRS and by the  ANR grant ``DIADEMS''. 
K. K. K. is supported by the EU Marie-Curie Excellence Grant MEXT-CT-2006-042695.  K. K. K. would like to thank the Theoretical Physics group of the Laboratory of Physics at ENS Lyon for hospitality, which makes this
collaboration possible. V.T. would  like to thank LPTHE (Paris VI University) for hospitality.
We would like to thank N. Kitanine, J. M. Maillet and N. A. Slavnov for stimulating discussions and comments. V. T. would also like to thank M. Civelli and S. Teber for their interest in this work.

\appendix

%%%%%%%%%%%%%%%%%%%%%%%%%%%%%%%%%%%%%%%%%%%%%%%%%%%%%%%%%%%%%%%%%%%%%%%%%%%%%%%%%%%%%%%%%%%%%%%%%%%%%%

\section{The form factors and their thermodynamic limit}
\label{sec-ff}

In this appendix, we present the explicit expression and the leading thermodynamic behavior of the combinations of finite-volume form factors which appear in each term of the series~\eqref{sum-ff1}, {\it i.e.}, with the notations of Section~\ref{sec-ff-series}, of
\begin{multline}\label{comb-ff}
    \ex{-it \mc{E}_{\e{ex}}}
        \frac{\bra{\psi_g}\, j(x,0)\, \ket{\psi'}  \,\bra{\psi'}\, j(0,0) \, \ket{\psi_g}  }
                { || \psi_g ||^2 \cdot ||\psi' ||^2 }
                \\
    =    \frac{1}{2} \partial_x^2\partial_\beta^2 \bigg\{
%             \ex{ix \sum_{a=1}^N [u_0(\mu_{\ell_a})-u_0(\la_a)]}
     \ex{ix\mc{P}_{\e{ex}}^\kappa-it\mc{E}_{\e{ex}}^\kappa}
     %\pl{a=1}{N} \paa{ \frac{ e_-^2(\la_a) }{ e_-^2(\mu_{\ell_a} ) }}
      %   \cdot
         \bigg|    \frac{\moy{\psi(\{\la_j\})\mid \psi_\kappa(\{\mu_{\ell_j} \})}  }
                             { \norm{\psi_\kappa(\{\mu_{\ell_j} \})} \cdot \norm{\psi(\{\la_j \})} }
         \bigg|^2
                   \bigg\}_{\beta=0}.        
\end{multline}
These results being the complete analogues, in the case of the NLS model, of those derived  in \cite{KitKMST09c,KitKMST10u} for the XXZ chain,
we skip the details of the computations.

%%%%%%%%%%%%%%
\subsection{Thermodynamic limit of the space and time-dependent phase factor}

The thermodynamic limit of $\ex{ix\mc{P}_{\e{ex}}^\kappa-it\mc{E}_{\e{ex}}^\kappa}$ generates dressed quantities.
This limit being completely smooth, it is easy to see that
\begin{align*}
 \ex{ ix\mc{P}_{\e{ex}}^\kappa-it\mc{E}_{\e{ex}}^\kappa }
 &=
  \exp \Bigg\{ ix \int\limits_{-q}^{q} u'_0(\la)\,  F(\la)\, \dd\la 
                     +  ix \sum_{a=1}^{n} \big[ u_0(\mu_{p_a})-u_0(\mu_{h_a})\big]  \Bigg\}
(1  + \e{O}(1/L) )  \nonumber\\
  &=
  \exp\Bigg\{ -\frac{x \beta}{2\pi} \int\limits_{-q}^{q} u_0^{\prime}(\la)\, Z(\la)\,\dd\la
                     +  ix \sum_{a=1}^{n} \big[ u(\mu_{p_a})-u(\mu_{h_a})\big]  \Bigg\}
                      (1  + \e{O}(1/L) ) ,
\end{align*}
in which we have used the explicit form \eqref{F-thermo} of the shift function $F$,
and where $u$ is the following combination
of dressed quantities $u(\la)=p(\la)- \tf{t\veps(\la)}{x}$. This identification was possible due to
\begin{equation}
u(\la)= u_0(\la)-\Int{-q}{q} u_0'(\mu)\, \phi(\mu,\la)\, \dd \mu .
\end{equation}
Note that, due to the fact that $\veps_0^{\prime}$ is an odd function whereas the dressed charge $Z$ is even, one has
\begin{equation}
\Int{-q}{q} \dd \la \, Z(\la)\, u_0^{\prime}(\la) = 2 p_F .
\end{equation}
%

%%%%%%%%%
\subsection{Representation of the normalized scalar product}

Let $\{\la_j\}$ be the solution of the system of Bethe equations \eqref{Bethe} parametrizing the ground state of \eqref{Ham} and let $\{\mu_{\ell_j}\}$ be a set of Bethe roots of \eqref{twisted-Bethe} parametrizing an excited state above the $\kappa$-twisted ground state in the $N$-particle sector.
Then the normalized modulus squared of the corresponding overlap scalar product in finite volume can be represented as
\begin{equation}
 \abs{    \frac{\moy{\psi(\{\la_j\})\mid \psi_\kappa(\{\mu_{\ell_j} \})}  }
                             { \norm{\psi_\kappa(\{\mu_{\ell_j} \})} \cdot \norm{\psi(\{\la_j \})} }
                  }^2
   = \widehat{D}_N(\{\la_j\},\{\mu_{\ell_j}\})\cdot
       \widehat{\mathcal{W}}_N (\{\la_j\},\{\mu_{\ell_j}\})\cdot
       \widehat{\mathcal{A}}_N   (\{\la_j\},\{\mu_{\ell_j}\}),            
\end{equation}
where $\widehat{D}_N$, $\widehat{\mathcal{W}}_N$ and $\widehat{\mathcal{A}}_N$ are given by
\begin{align}
  &\widehat{D}_N(\{\la_j\},\{\mu_{\ell_j}\})
    = \prod_{j=1}^N \frac{\sin^2 [\pi \widehat{F}(\la_j) ] }
                                         {\pi^2 L^2\, \widehat\xi' (\la_j)\, \widehat\xi_\kappa' (\mu_{\ell_j})}
        \cdot \bigg[ \det_N \frac{1}{\la_j-\mu_{\ell_k}}\bigg]^2 ,
     \label{DN}
     \displaybreak[0]\\
  &\widehat{\mathcal{W}}_N (\{\la_j\},\{\mu_{\ell_j}\})
     =   \prod_{j,k=1}^N \frac{(\la_j-\mu_{\ell_k}-ic) \, (\mu_{\ell_j}-\la_k-ic)}
                                               {(\la_j-\la_k-ic) \, (\mu_{\ell_j}-\mu_{\ell_k}-ic)} ,
      \label{WN}
      \displaybreak[0]\\
  &\widehat{\mathcal{A}}_N     (\{\la_j\},\{\mu_{\ell_j}\})
    =       \frac{(1-\kappa)^2 }
                     {(1-\ex{-2i\pi \widehat{F}(\theta_1)}) (1-\ex{-2i\pi\widehat{F}(\theta_2)})}
             \prod_{a=1}^N \frac{(\theta_1-\la_a+ic)(\theta_2-\la_a+ic)}
                                                {(\theta_1-\mu_{\ell_a}+ic)(\theta_2-\mu_{\ell_a}+ic)}        
                       \nonumber\\
   &\hspace{3.5cm}\times
        \frac{\det_{\Ga }  \! \big[ I+\frac{1}{2i\pi} \widehat{U}_{\theta_1}^{(\la)}(\omega,\omega')\big] 
                 \cdot
                 \det_{\Ga } \! \big[ I+\frac{1}{2i\pi} \widehat{U}_{\theta_2}^{(\la)}(\omega,\omega')\big]  }
                {\det_N \! \Big[ \delta_{jk}-\frac{K(\la_j-\la_k)}{2\pi L\, \widehat\xi'(\la_k)} \Big] 
                 \cdot
                 \det_N \! \Big[ \delta_{jk}-\frac{K(\mu_{\ell_j}-\mu_{\ell_k})}{2\pi L\, \widehat\xi_\kappa'(\mu_{\ell_k})} \Big] } \,  .  
                 \label{AN}                                                                                      
\end{align}
We recall that $\widehat\xi_\kappa$ and $\widehat\xi$ denote respectively the excited state counting function \eqref{TBE-cf} and the ground state counting function at $\kappa=1$, whereas $\widehat{F}$ is the finite-size shift function \eqref{shift-finite}.
Here $\theta_1$ and $\theta_2$ are some arbitrary real parameters.
The integral operator $I+\frac{1}{2i\pi} \widehat{U}_{\theta}^{(\la)}$ acts on the closed contour $\Ga$ surrounding the ground state roots $\{\la_j\}$ and no other singularity of the kernel.
This kernel reads
\begin{align}
   &\widehat{U}_{\theta}^{(\la)}(\omega,\omega')
     = - \prod_{a=1}^N \frac{(\omega-\mu_{\ell_a}) (\omega-\la_a+ic)}
                                            {(\omega-\la_a) (\omega-\mu_{\ell_a}+ic)}
        \cdot
        \frac{K_\kappa(\omega-\omega')-K_\kappa(\theta-\omega')}{1-\ex{-2i\pi\widehat{F}(\omega)}},
    \label{Uhat-la}  
\end{align}
with
\begin{equation}
  K_\kappa(\omega)=\frac{1}{\omega+ic}-\frac{\kappa}{\omega-ic}.
  \label{Kkappa}
\end{equation}

\begin{rem}
Although each individual term, in \eqref{AN}, does depend on the set of auxiliary parameters $\theta_k$, the overall combination $\widehat{\mathcal{A}}_N$ does not.
This was proven in \cite{KitKMST09b}.
\end{rem}

The factor $\widehat{D}_N(\{\la_j\},\{\mu_{\ell_j}\})$ \eqref{DN} is the so-called discrete part of the form factor.
It contains all the non-trivial "singular part" of the form factor (see \cite{KitKMST10u}).
On the contrary, the factor
\begin{equation}\label{dressing}
   \wh{\mc{G}}_N (\{\la_j\},\{\mu_{\ell_j}\})\equiv \widehat{\mathcal{W}}_N (\{\la_j\},\{\mu_{\ell_j}\}) \cdot \widehat{\mathcal{A}}_N     (\{\la_j\},\{\mu_{\ell_j}\})
\end{equation}
admits a smooth thermodynamic limit.
It can be thought of as a dressing function. It is equal to $1$ at the free-fermion point.

In the case of an excited state $\ket{\psi_\kappa(\{\mu_{\ell_j}\})}$ with a finite number of particles and holes as described in Section~\ref{sec-NLSE}, it is easy to compute the thermodynamic limit of the dressing function $\wh{\mc{G}}_N$ (see \cite{KitKMST10u}):
\begin{equation}
  \lim_{L,N\to\infty}\widehat{\mathcal{G}}_N     (\{\la_j\},\{\mu_{\ell_j}\})
  = \mathcal{G}_n \binom{ \{\mu_{p_a}\} }{ \{\mu_{h_a}\} }
  \equiv \mathcal{W}_n \binom{ \{\mu_{p_a}\} }{ \{\mu_{h_a}\} }
    \cdot \mathcal{A}_n \! \binom{ \{\mu_{p_a}\} }{ \{\mu_{h_a}\} },
\end{equation}
with
\begin{multline}
   \mc{W}_n \binom{ \{\mu_{p_a}\} }{ \{\mu_{h_a}\} }
=
   \prod_{a,b=1}^{n} \frac{(\mu_{p_a} - \mu_{h_b}-ic) (\mu_{h_a}-\mu_{p_b}-ic) }
                                           { (\mu_{p_a} -\mu_{p_b}-ic) (\mu_{h_a}-\mu_{h_b}-ic) }
                                            \\
   \times                                         
   \ex{-2i\pi \sul{a=1}{n}\sul{\eps=\pm}{}  \{ \mc{C}[F] (\mu_{p_a}+i\eps c)  - \mc{C}[F] (\mu_{h_a}+i\eps c)\}  
   + C_0[F]}  ,
\label{def-Wn}
\end{multline}
\begin{multline}
  \mc{A}_n \binom{ \{\mu_{p_a}\} }{ \{\mu_{h_a}\} }
  =
  (1-\kappa)^2
         \frac{\ex{-2i\pi\{\mc{C}[F](\theta_1+ic)+\mc{C}[F](\theta_2+ic)\}}}
         {(1-\ex{-2i\pi F(\theta_1)})(1-\ex{-2i\pi F(\theta_2)})}  
         \prod_{j=1}^n \frac{(\theta_1-\mu_{h_j}+ic)(\theta_2-\mu_{h_j}+ic)}
                                                           {(\theta_1-\mu_{p_j}+ic)(\theta_2-\mu_{p_j}+ic)}  
         \\  
  \times
  \frac{\det  \big[ I+\frac{1}{2i\pi} U_{\theta_1}\big]   \,
           \det \big[ I+\frac{1}{2i\pi} {U}_{\theta_2}\big]   }
          {\det^2\big[ I-\frac{1}{2\pi} K\big]  }    .
\label{def-An}
\end{multline}
In these expressions, $F$ denotes  the shift function $F$ \eqref{Int-eq-F}, $\mc{C}[\nu]$ is the rational Cauchy transform of the function $\nu$ on $[-q,q]$,
\begin{equation}
   \mc{C}[\nu ](\la)=\Int{-q}{q} \frac{ \dd \mu }{2i\pi} \frac{\nu(\mu)}{\mu-\la},
\label{def-Cauchy}
\end{equation}
whereas $C_0$ is the following functional:
\begin{equation}
   C_0[\nu]= -\Int{-q}{q}\dd \la\, \dd \mu\, \frac{ \nu(\la)\, \nu(\mu) }{ (\la-\mu-ic)^2} .
\label{def-C0}
\end{equation}
The integral kernel $U_{\theta}$ takes the form
\begin{equation}
U_{\theta}(\om,\om')
= - \prod_{a=1}^{n} \frac{  (\om-\mu_{p_a}) (\om-\mu_{h_a}+ic) }{ (\om-\mu_{h_a}) (\om-\mu_{p_a}+ic)  }
      \frac{\ex{2i\pi \mc{C}[F](\om)} \{ K_{\kappa}(\om-\om')- K_{\kappa}(\theta-\om') \} }
       {\ex{2i\pi \mc{C}[F](\om+ic)} (1-\ex{-2i\pi F(\om)})}.
\label{def-Ulambda}
\end{equation}

\begin{rem}\label{rem-dependence}
The dependence on the particle/hole rapidities of $\mc{G}_n$ is twofold: explicitly in the above expressions, and also in the shift function $F$. The latter is a holomorphic function of $\{ \mu_{p_a} \}$ and $ \{ \mu_{h_a} \}$, \textit{cf} \eqref{F-thermo}.
 Therefore, $\mathcal{G}_n$ is itself a holomorphic function of  $\{ \mu_{p_a} \} $ and $ \{ \mu_{h_a} \}$.
  This function can, in fact, be understood as a functional $\msc{G}[\varpi_n]$ of the function $\varpi_n$ \eqref{def-varpi} (note that the dependence in $n$ is exclusively contained in $\varpi_n$).
  For this it is enough to observe that
\begin{equation}
    F(\la) = \frac{i\beta}{2\pi}Z(\la)-\Int{\Gamma(\mathbb{R})}{} \frac{\dd\mu}{2i\pi} \phi(\la,\mu)\,
    \varpi_n\bigg(\mu\bigg| \begin{matrix} \{\mu_{p_a}\} \\ \{\mu_{h_a}\} \end{matrix} \bigg),
\end{equation} 
where the contour $\Ga(\R)$ surrounds the real axis counterclockwise, and that
\begin{equation}
\pl{a=1}{n} \ex{f(\mu_{p_a}) - f(\mu_{h_a}) } = \exp\Bigg\{ \Int{\Ga(\R)}{} \f{ \dd \la }{2i\pi} f(\la)\, 
 \varpi_n\bigg(\la\bigg| \begin{matrix} \{\mu_{p_a}\} \\ \{\mu_{h_a}\} \end{matrix} \bigg) \Bigg\}
\end{equation}
for any holomorphic functions $f$ with a sufficiently mild growth at infinity on $\Ga(\R)$.
In particular, we have
\begin{multline}
   \prod_{a,b=1}^{n} \frac{(\mu_{p_a} - \mu_{h_b}-ic) (\mu_{h_a}-\mu_{p_b}-ic) }
                                           { (\mu_{p_a} -\mu_{p_b}-ic) (\mu_{h_a}-\mu_{h_b}-ic) }
                                           \\
   = \exp\Bigg\{-\Int{\Gamma(\mathbb{R})}{} \frac{\dd\la\,\dd\mu}{(2i\pi)^2} 
         \log(\la-\mu-ic)\, 
         \varpi_n\bigg(\la\bigg| \begin{matrix} \{\mu_{p_a}\} \\ \{\mu_{h_a}\} \end{matrix} \bigg)\,
         \varpi_n\bigg(\mu\bigg| \begin{matrix} \{\mu_{p_a}\} \\ \{\mu_{h_a}\} \end{matrix} \bigg)
         \Bigg\}.                                 
\end{multline}
\end{rem}

\begin{rem}\label{rem-zero-beta}
The explicit factor $(1-\kappa)^2$ in \eqref{def-An} stresses that, for generic configurations 
of the parameters $\{\mu_{p_a}\}$ and $\{\mu_{h_a}\}$, 
$\mathcal{G}_n$, considered as a function of $\beta$, has in general a zero of order $2$ at $\beta=0$.
Note however that there exist some exceptions, in particular at the free fermion point $c=+\infty$, when $\mathcal G$ is identically $1$ or, for general $c$, when the excited state coincides with the ground state in the limit $\beta\to 0$ ({\it i.e.} for $n=0$). In the latter case, the limit can be taken as in \cite{KitKMST09b} and we obtain
\begin{equation}\label{limG0}
   \lim_{\beta\to 0} \mathcal{G}_0=1.
\end{equation}   
\end{rem}

The study of the thermodynamic behavior of the factor $\widehat{D}_N$ \eqref{DN} is slightly more technical.
Proceeding as in \cite{KitKMST10u}, one obtains that, for a $n$ particle/hole excited state,
\begin{multline}
  \widehat{D}_N (\{\la_j\} ,\{\mu_{\ell_j}\} )
  = \bigg(\frac{L}{2\pi}\bigg)^{-F^2(q)-F^2(-q)-2n}\; 
  [2q\,p'(q)]^{-F^2(q)-F^2(-q)} \; \wt{\mc{B}}[F]     
  \\
  \times
      \Gamma^2\!\begin{pmatrix}
            \{p_j\} \, , \{p_j-N+F(\mu_{p_j} )\} \, , \{ h_j+F(\mu_{h_j})\} \, , \{N+1-h_j-F(\mu_{h_j})\} \\
            \{p_j-N\} \, , \{p_j+F(\mu_{p_j})\} \, , \{ h_j\} \, , \{N+1-h_j \} 
                          \end{pmatrix}
 \\                         
  \times
    \det_n^2\bigg[ \frac{1}{\mu_{p_j}-\mu_{h_k} } \bigg] \;
    \prod_{j=1}^n \frac{\sin^2[\pi F(\mu_{h_j})]\,
                             \ex{J[F](\mu_{p_j})-J[F](\mu_{h_j}) }}
                             {\pi^2\, p'(\mu_{p_j}) \, p'(\mu_{h_j})}  \;
       \bigg(1+\e{O}\bigg(\frac{\log L}{L}\bigg)\bigg) .  
       \label{dis-ff}                                              
\end{multline}
The expression \eqref{dis-ff} is given in terms of the below functional of the shift function \eqref{F-thermo}:
\begin{equation}\label{Btilde}
   \wt{\mc{B}}[\nu]
    = \ex{ C_1[\nu ]   } \frac{  \, G^2(1+\nu (q)) \, G^2(1-\nu (-q))  }
           { (2\pi)^{\nu (q) - \nu(-q)}  } 
\end{equation}
where $G$ is the Barnes $G$ function and
\begin{multline}
   C_1[\nu]
   = \frac{1}{2} \Int{-q}{q} \dd \la \, \dd \mu \,  
                                  \frac{\nu^{\prime}(\la) \nu(\mu)-\nu^{\prime}(\mu) \nu(\la) }{\la-\mu}
                                  \\
      +\nu(q) \Int{-q}{q} \frac{\nu(q) - \nu(\la)}{q-\la} \dd \la
      +\nu(-q)\Int{-q}{q}\frac{\nu(-q) - \nu(\la)}{q+\la}  \dd \la  .
\label{def-C1}      
\end{multline}
We have also defined
\begin{equation}\label{def-J}
 J[F](\la)= 2F(\la)\,\log \bigg[ \frac{\la-q}{\la+q}\frac{p(\la)-p(-q)}{p(\la)-p(q)}\bigg]
 +\wt{J}[F](\la),
\end{equation}
with
\begin{equation}\label{Jtilde}
  \wt{J}[\nu](\la)=2\Int{-q}{q} \frac{\nu(\mu)-\nu(\la)}{\mu-\la}\,\dd\mu.
\end{equation}
Finally,
\begin{equation}
   \Gamma \begin{pmatrix} \{ a_k\} \\ \{ b_k \} \end{pmatrix}
    =\prod_{k=1}^n \frac{\Gamma(a_k)}{\Gamma(b_k)},
\end{equation}
with the prescription that, should some of the particles in \eqref{dis-ff} have their rapidities to the left of the Fermi zone, the arguments of the $\Gamma$-functions have to be understood as limits
\begin{equation}
   \frac{\Gamma(p_k)}{\Gamma(p_k-N)}=\lim_{\eps\to 0} \frac{\Gamma(p_k+\eps)}{\Gamma(p_k-N+\eps)}.
\end{equation}

Note that in \eqref{dis-ff} the thermodynamic limit has only been  taken partly. Indeed,  since the complete thermodynamic behavior of \eqref{dis-ff} depends on whether the corresponding particles and holes remain or not at a finite distance from the Fermi boundaries.
In the next subsection, we particularize these expressions to the special form factors with one particle and one hole which appear in the results \eqref{asympt-space}, \eqref{asympt-time}.

%%%%%%%%%%%%%%%%%%%%%%%%%%%
\subsection{Explicit value of the amplitudes}
\label{sec-amplitudes}

We collect here the explicit values of the non-universal amplitudes appearing in \eqref{asympt-space}-\eqref{asympt-time}, which correspond to the normalized one particle/hole form factors \eqref{ff0}-\eqref{ff4}.
They can be easily obtained from the expressions \eqref{def-Wn}-\eqref{dis-ff} and read
\begin{align}
\big|  \mc{F}^{q}_{-q}  \big|^2
&=\big|  \mc{F}^{-q}_{q}  \big|^2
    \nonumber\\
 &= -\frac{ 2 p_F^2}{\pi^2} \, \frac{ \wt{\mc{B}}\big[ F^{q}_{-q}\big]\;
       \ex{\wt{J} [ F^{q}_{-q} ](q)- \wt{J} [ F^{q}_{-q}](-q)} }
               {[2 q \, p'(q)]^{   [ F^{q}_{-q}(q)+1]^2+[ F^{q}_{-q}(-q) +1]^2 } }\,
       \Gamma^2\big(1+ F^{q}_{-q}(q)\big)\,    \Gamma^2\big(1+ F^{q}_{-q}(-q)\big) 
       \nonumber\\
 &\hspace{5cm}\times      
       \partial_\beta^2 \sin^2 \Big[ i\frac{\beta}{2}\mc{Z}+\pi F^{q}_{-q}(-q)   \Big]\,    \mathcal{G}_1\binom{q}{-q} \bigg|_{\beta=0}          
   \nonumber \\  
 &= -\frac{ 2 p_F^2}{\pi^2} \, \frac{ \wt{\mc{B}}\big[ F^{-q}_{q}\big]\;
       \ex{\wt{J} [ F^{-q}_{q} ](-q)- \wt{J} [ F^{-q}_{q}](q)} }
               {[2 q \, p'(q)]^{   [ F^{q}_{-q}(q)-1]^2+[ F^{q}_{-q}(-q) -1]^2 } }\,
       \Gamma^2\big(1- F^{-q}_{q}(q)\big)\,    \Gamma^2\big(1- F^{-q}_{q}(-q)\big) 
       \nonumber\\
 &\hspace{5cm}\times      
       \partial_\beta^2 \sin^2 \Big[ i\frac{\beta}{2}\mc{Z}+\pi F^{-q}_{q}(q)   \Big]\,    \mathcal{G}_1\binom{-q}{q} \bigg|_{\beta=0} , 
   \label{ff-qq}
\end{align}
in terms of the shift functions
$
   F^{\pm q}_{\mp q}(\la)=\phi(\la,\mp q)-\phi(\la,\pm q)
       = \pm Z(\la)\mp 1;
$
%
%\begin{multline}
%   \label{ffla0q}
%\big|  \mc{F}^{\la_0}_q  \big|^2
%= -\bigg(\frac{p(\la_0)-p_F}{\la_0-q}\bigg)^{\! 2} \bigg(\frac{\la_0-q}{\la_0+q}\bigg)^{\! 2 F^{\la_0}_q(\la_0)}
%   \frac{[2q\, p'(q)]^{-[F^{\la_0}_q(q)]^2- [F^{\la_0}_q(-q)]^2 +2 F^{\la_0}_q(q) } }{2 \pi^2\, p'(\la_0)\, p'(q) }\; 
%      \wt{\mc{B}}\big[ F^{\la_0}_{q}\big]
%        \\
% \times
%       \ex{\wt{J} [ F^{\la_0}_{q} ](\la_0)- \wt{J} [ F^{\la_0}_{q}](q) }  \;
%        \Gamma^2\big(1- F^{\la_0}_{q}(q)\big)\;     
%       \partial_\beta^2 \sin^2 \Big[ i\frac{\beta}{2}\mc{Z}+\pi F^{\la_0}_{q}(q)   \Big]\,    \mathcal{G}_1\binom{\la_0}{q} \bigg|_{\beta=0} , 
%\end{multline}
%
\begin{multline}
   \label{ffla0-q}
\big|  \mc{F}^{\la_0}_{\eps q}  \big|^2
= -\bigg(\frac{p(\la_0)-\eps p_F}{\la_0-\eps q}\bigg)^{\! 2} \bigg(\frac{\la_0-q}{\la_0+q}\bigg)^{\! 2 F^{\la_0}_{\eps q}(\la_0)}
   \frac{[2q\, p'(q)]^{-[F^{\la_0}_{\eps q}(q)]^2- [F^{\la_0}_{\eps q}(-q)]^2 +2 \eps F^{\la_0}_{\eps q}(\eps q) } }{2 \pi^2\, p'(\la_0)\, p'(q) }
   \\  
   \hspace{5cm}
   \times
      \wt{\mc{B}}\big[ F^{\la_0}_{\eps q}\big] \;
      \ex{\wt{J} [ F^{\la_0}_{\eps q} ](\la_0)- \wt{J} [ F^{\la_0}_{\eps q}](\eps q) }  \;
      \Gamma^2\big(1-\eps  F^{\la_0}_{\eps q}(\eps q)\big)
   \\
   \times        
       \partial_\beta^2 \sin^2 \Big[ i\frac{\beta}{2}\mc{Z}+\pi F^{\la_0}_{\eps q}(\eps q)   \Big]\,    \mathcal{G}_1\binom{\la_0}{\eps q} \bigg|_{\beta=0} , 
\end{multline}
in terms of the shift functions $F^{\la_0 }_{\eps q}=\phi(\la,\eps q)-\phi(\la,\la_0)$, with $\eps=\pm 1$;
\begin{multline}
   \label{ff-qla0}
\big|  \mc{F}_{\la_0}^{\eps q}  \big|^2
= -\bigg(\frac{p(\la_0)-\eps p_F}{\la_0-\eps q}\bigg)^{\! 2} \bigg(\frac{q-\la_0}{q+\la_0}\bigg)^{\! -2 F_{\la_0}^{\eps q}(\la_0)}
   \frac{[2q\, p'(q)]^{-[F_{\la_0}^{\eps q}(q)]^2- [F_{\la_0}^{\eps q}(-q)]^2 -2 \eps F_{\la_0}^{\eps q}(\eps q) } }{2 \pi^2\, p'(\la_0)\, p'(q) }
   \\  
   \hspace{5cm}
   \times
      \wt{\mc{B}}\big[ F_{\la_0}^{\eps q}\big] \;
      \ex{\wt{J} [ F_{\la_0}^{\eps q} ](\eps q)- \wt{J} [ F_{\la_0}^{\eps q}](\la_0) }  \;
      \Gamma^2\big(1+\eps  F_{\la_0}^{\eps q}(\eps q)\big)
   \\
   \times        
       \partial_\beta^2 \sin^2 \Big[ i\frac{\beta}{2}{Z}(\la_0)+\pi F_{\la_0}^{\eps q}(\la_0)   \Big]\,    \mathcal{G}_1\binom{\eps q}{\la_0} \bigg|_{\beta=0} ,
\end{multline}
in terms of the shift functions $F_{\la_0 }^{\eps q}=\phi(\la,\la_0)-\phi(\la,\eps q)$, with $\eps=\pm1$.
We recall that the functionals $\wt{\mc{B}}$ and $\wt{J}$ are respectively given by \eqref{Btilde} and \eqref{Jtilde}, and that the expressions of the dressing functions $\mathcal{G}_1$ are obtained through \eqref{def-Wn}-\eqref{def-An}.

\begin{rem}
For generic parameters, the $\beta$-derivatives will apply directly to the factor $(1-\kappa)^2$ of $\mathcal{G}_1$ (see remark~\ref{rem-zero-beta}). However, in the free fermion point $c=+\infty$, they will apply on the sinus squared, since in that case the corresponding shift functions vanish (and $\mathcal{G}_1\equiv 1$).
There may exist other particular values of the parameters for which the shift function becomes an integer but the overall expressions are anyway smooth (we may then have to apply the $\beta$-derivatives to other factors such as the Barnes functions).
\end{rem}

%\begin{rem}
%These expressions are given for generic values of the parameters, and in particular of $q$. There may happen that, for very specific values of the parameters, the shift function take integers values, which may cause some apparent singularities, but the overall expression is anyway smooth in these limits.
%\end{rem}

%%%%%%%%%%%%%%%%%%%%%%%%%%%%%%%%%%%%%%%%%%%%%%%%%%%%%%%%%%%%%%%%%%%%%%%%%%%%%%%%%%%%%%%%%%%%%%%%%%%%%%

\section{Study of a generalized free fermionic generating function}
\label{sec-free}

In this appendix, we consider the case of a generalized free-fermionic model for which the effective generating function \eqref{eff-series} can be represented as a finite-size determinant.
Namely, we suppose here that the shift function does not depend on the position of the roots parametrizing the corresponding excited state and that the dressing function $\mc{G}_n$ is separated (which happens in particular at the free fermion point $c=+\infty$ of the NLS model).

More precisely, for a given counting function $\wh{\xi}$ (for example the ground state counting function of the NLS model), which defines a set of real parameters $\la_j$, $j=1,\ldots,N$ by $\wh{\xi}(\la_j)=j/L$, we consider the functional
\begin{equation}
   X_N\big[\nu, E_-^{2}\big] 
   =      
      \sum_{ \substack{\ell_1<\dots <\ell_N  \\ \ell_j \in \mc{B}_L } } 
       \prod_{j=1}^N \frac{E_-^2(\la_j)}{E_-^2(\mu_{\ell_j})}
       \cdot
       \prod_{j=1}^N \frac{\sin^2 [\pi \nu(\la_j) ] }
                                         {\pi^2 L^2\, \widehat\xi' (\la_j)\, \widehat\xi_\nu' (\mu_{\ell_j})}
        \cdot \bigg[ \det_N \frac{1}{\la_j-\mu_{\ell_k}}\bigg]^2 ,
\label{gen-XN}
\end{equation}
where $E_-^{-1}$ and $\nu$ are holomorphic functions on some open neighborhood of $\mathbb{R}$.
The above expression should be understood as follows:
\begin{itemize}
 \item the function $\wh{\xi}_\nu$ is defined by $\wh{\xi}_\nu=\wh{\xi}-L^{-1}\nu$;
 \item the multiple sum runs through all the possible choices of $N$-tuples of integers $\ell_1<\dots <\ell_N$ belonging to the set  $\mc{B}_L=\{ j\in\mathbb{Z} \mid -w_L < j < w_L\}$, where $w_L \sim L^{1+\eps}$, $\eps>0$,  is some cut-off which goes to $+\infty$ with $L$. 
 \item for a given set of integers $\ell_j$, the parameters $\mu_{\ell_j}$ are obtained as the pre-image of $\ell_j/L$ by the function $\wh{\xi}_\nu$. This means, in particular, that they depend on $\nu$.
 \end{itemize}
Moreover, we restrict our study those functions  $\nu$ that  make $\wh{\xi}_\nu$ a "good" counting function ({\em i.e.} ensuring a one-to-one correspondence between the rapidities and the integers).

\begin{prop}
\label{prop-XN-det}
The functional $X_N\big[ \nu, E_-^2\big]$ can be recast as the following finite size determinant:
\begin{equation}
   X_N\big[ \nu, E_-^2\big]  = \det_N \bigg[ \de_{jk} +   \frac{ V^{(L)}(\la_j,\la_k) }{L \, \wh{\xi}'(\la_k)  }  \bigg].
\label{XN-det}
\end{equation}
The corresponding finite-size kernel is given as
\begin{equation}
    V^{(L)}(\la,\mu)
    = 4  \frac{ \sin [\pi\nu(\la)]\,  \sin [\pi\nu(\mu)] }{ 2i\pi (\la-\mu)}
        \Big\{E_+^{(L)}(\la) \, E_-(\mu)- E_+^{(L)}(\mu) \, E_-(\la) \Big\},
\label{noyau-L}
\end{equation}
where
\begin{equation}
    E^{(L)}_+(\la)
     =i E_-(\la) \left\{ \, \int\limits_{A_L}^{B_L} \hspace{-4.35mm} \diagup \hspace{1.3mm}
                                      \frac{ \dd \mu }{2\pi } \frac{ E^{-2}_{-}(\mu) }{ \mu-\la } 
        + \frac{ E_-^{-2}(\la) }{ 2 } \cot[ \pi \nu(\la)]   + I_L\big[ \nu, E_-^{-2} \big] ( \la )  \right\} ,
\label{E+L}
\end{equation}
In this expression, the last term $I_L\big[ \nu, E_-^{-2} \big] ( \la )$ corresponds to the integral
\begin{equation}
    I_L\big[ \nu,E_-^{-2} \big] ( \la )
     = \int\limits_{ \msc{C}_{\ua} } \frac{ \dd z }{ 2\pi }  \frac{ E_-^{-2}(z) }{ z -\la }
                                                        \frac{  1  }{ 1-\ex{-2i\pi L \wh{\xi}_{\nu}(z)  } } 
        +
        \int\limits_{ \msc{C}_{\da} } \frac{ \dd z }{ 2\pi }  \frac{ E_-^{-2}(z)  }{ z -\la }
                                                        \frac{  1  }{  \ex{ 2i\pi L \wh{\xi}_{\nu}(z) } -1  } .
\label{IL}
\end{equation}
Finally, the integration endpoints $A_L$ and $B_L$ are such that $L\wh{\xi}_\nu(A_L)=-w_L-\tf{1}{2}$ and $L\wh{\xi}_\nu(B_L)=w_L+\tf{1}{2}$, and $\msc{C}_{\ua/\da}$ are some oriented contours, included in the joint domain of holomorphy of $\nu$ and $E_-^{-1}$. 
Theses contours join the points $A_L$ and $B_L$ through the upper/lower half plane respectively, it has been  depicted on Fig.~\ref{contour-IL}.
%
%
%%%%%%%%%%%%%
\begin{figure}[h]
\begin{center}

\begin{pspicture}(6,4)

\psline[linestyle=dashed, dash=3pt 2pt](1.5,2)(4.5,2)

%Label des point q et la 0.

\psdots(1.5,2)(4.5,2)

\rput(1.1,1.9){$A_L$}
\rput(4.9,1.9){$B_L$}
\rput(4.9,2.6){$\msc{C}_{\ua}$}
\rput(1.1 ,1.1){$\msc{C}_{\da}$}

\pscurve(1.5,2)(1.5,1.5)(1.5,1)(3,1)(4.5,1)(4.5,1.5)(4.5,2)(4.5,2.5)(4.5,3)(3,3)(1.5,3)(1.5,2.5)(1.5,2)

\psline[linewidth=2pt]{->}(3,1)(3.1,1)
\psline[linewidth=2pt]{->}(3.1,3)(3,3)

\end{pspicture}

\caption{Contour $\msc{C}_{\ua/\da}$. \label{contour-IL}}
\end{center}
\end{figure}
%%%%%%%%%%%%%
%
%
\end{prop}

\Proof
The summand in \eqref{gen-XN} is a symmetric function of the $N$ summation variables $\mu_{\ell_j}$ which vanishes on the diagonals 
$\ell_j=\ell_k$, $j\not= k$.
Therefore, we can replace the summation over the fundamental simplex $\ell_1<\dots<\ell_N$ in the $N^{\e{th}}$ power cartesian product 
$\mc{B}_L^N$ by a summation over the whole space $\mc{B}_L^N$, provided that we divide the result by $N!$.
The summation domain being now symmetric, we can invoke the antisymmetry of the determinants so as to replace one of the Cauchy determinants by 
$N!$ times the product of its diagonal entries. This last operation produces a separation of variables, which enables us to recast the result 
into a single determinant by introducing the sum over $\ell_j$ into the $j^{\text{th}}$ line of the determinant:
\begin{equation}
   X_N\big[ \nu , E_-^{2}\big]
   = \prod_{j=1}^N \frac{ 4 \sin^2 [\pi \nu (\la_j)] }{ \wh{\xi}^{\prime}(\la_j)}
      \cdot  \det_{N} M,
\end{equation}
with
\begin{multline}
   M_{j k}
   =  \de_{j,k } \sum_{\ell \in \mc{B}_L }
       \frac{ E_-^{2}(\la_j ) \cdot E_-^{-2}(\mu_\ell) }
               { 4\pi^2 L^2 \, \wh{\xi}^{\prime}_\nu (\mu_\ell)\, (\mu_\ell - \la_j )^2 }
     \\
    + (1-\de_{j,k})
        \sum_{ \ell \in \mc{B}_L  }
        \frac{ E_-^{2}(\la_j) \cdot E_-^{-2}(\mu_\ell)  }
                {4\pi^2 L^2\, \wh{\xi}^{\prime}_\nu (\mu_\ell)\,  (\la_j-\la_k) }
       \bigg[   \frac{ 1 }{ \mu_\ell - \la_j   }  - \frac{ 1}{  \mu_\ell - \la_k  }  \bigg] .
       \label{el-M}
\end{multline}
The above discrete sums can be expressed in terms of Hilbert transforms (plus corrections that vanish in the $L\to +\infty$ limit).
Indeed,
\begin{align}
  \sum_{\ell\in\mc{B}_L}  \frac{ E_-^{-2}(\mu_\ell) }{2\pi L\,\wh{\xi}_\nu'(\mu_\ell) (\mu_\ell-\la)}
  &=  \frac{-i E_-^{-2}(\la)}{ \ex{2i\pi L\,\wh{\xi}_\nu(\la)}-1}
        +\int\limits_{\msc{C}_{\ua}\cup\msc{C}_{\da} } \hspace{-2mm}
          \frac{E_-^{-2}(z)\,\dd z}{2\pi (z-\la)(\ex{2i\pi L\,\wh\xi_\nu(z)}-1)}
      \nonumber\\
  &=  \frac{-i E_-^{-2}(\la)}{ \ex{2i\pi L\,\wh{\xi}_\nu(\la)}-1} 
        -\int\limits_{\msc{C}_{\ua}}   \frac{ E_-^{-2}(z)\, \dd z}{2\pi (z-\la) }
        +  I_L\big[\nu, E_-^{-2}\big](\la) 
      \nonumber\\
  &=  %V.P. \!\!\! \int\limits_{-U_L}^{U_L}  \!\!   
          \int\limits_{A_L}^{B_L} \hspace{-4.35mm} \diagup \hspace{1.3mm}
          \frac{ E_-^{-2}(z)\, \dd z}{2\pi (z-\la )}
        -\frac{E_-^{-2}(\la)}{2} \cot \! \big[\pi L \wh\xi_\nu(\la)\big]
        + I_L \big[ \nu, E_-^{-2}\big](\la)    .  
        \label{id-1}
\end{align}
Differentiating  the above expression with respect to $\la$ leads to
\begin{multline}
  \sum_{\ell\in\mc{B}_L}  \frac{ E_-^{-2}(\mu_\ell) }{2\pi L\,\wh{\xi}_\nu'(\mu_\ell) (\mu_\ell-\la)^2}
  = \frac{\partial}{\partial \la}   \int\limits_{A_L}^{B_L} \hspace{-4.35mm} \diagup \hspace{1.3mm}
          \frac{ E_-^{-2}(z)\, \dd z}{2\pi (z-\la )}
    -\frac{\partial_\la E_-^{-2}(\la)}{2} \cot \! \big[\pi L \wh\xi_\nu(\la)\big]
    \\
    +\frac{E_-^{-2}(\la)\, \pi L \wh{\xi}'_\nu(\la)}{2\sin^2 [\pi L \wh\xi_\nu(\la)] }
    +\partial_\la I_L \big[\nu, E_-^{-2}\big](\la)  .
    \label{id-2}
\end{multline}
By applying \eqref{id-1}, \eqref{id-2} in \eqref{el-M}, and using the fact that $L \wh{\xi}_{\nu}(\la_j)=L\wh{\xi}(\la_j)- \nu(\la_j)=j- \nu(\la_j)$,
we obtain
\begin{align}
   M_{j k} 
    &=    \de_{j,k} \, \frac{ \wh{\xi}'(\la_j)}{4 \sin^2[ \pi \nu(\la_j)]  } 
    + \frac{E_-(\la_j)}{E_-(\la_k)}
       \frac{ E_+^{(L)}(\la_j)\, E_-(\la_k)- E_+^{(L)}(\la_k)\, E_-(\la_j) }{ 2i\pi L (\la_j -\la_k) } 
       \notag\\
    &= \frac{E_-(\la_j)}{E_-(\la_k)} \cdot
         \frac{ \wh{\xi}'(\la_k)}{4 \sin[ \pi \nu(\la_j)] \, \sin[ \pi \nu(\la_k)]  } 
         \left\{ \delta_{j,k}+ \frac{V^{(L)}(\la_j,\la_k)}{L \, \wh\xi'(\la_k)} \right\},
\label{calcul-M}
\end{align}
in which $E^{(L)}_+$ is given by \eqref{E+L}. Note that we recovered the derivative of the $\la$-type counting function $\wh{\xi}$ thanks to
the identity $\wh{\xi}_{\nu}=\wh{\xi}-L^{-1}\nu$.
This ends the proof of the proposition.
\qed

Let us now suppose that, in the thermodynamic limit $L,N\to+\infty$ with $N/L\to D$ ($D$  finite), the roots $\la_j$ condensate on some symmetric interval $[-q,q]$ of the real axis with the density $\rho(\la)=\lim_{L,N\to\infty}\wh{\xi}'(\la)$ of the NLS model.
Let us moreover suppose that, in this limit, the counting function $\wh\xi_\nu$ takes the form \eqref{cf-p} in terms of the dressed momentum $p(\la)$ of the NLS model.
We demand in addition that 
 $\log E_-^{-2}$ has at most a polynomial growth in $\la$ when $\Re(\la)\to\pm\infty$.
Then the thermodynamic limit of $X_N\big[\nu,E_-^{2}\big]$ is well defined and given by the following Fredholm determinant:
\begin{equation}\label{XN-Fredholm}
 \lim_{M,N\to\infty}X_N\big[\nu,E_-^{2}\big]
 =\det_{[-q,q]} [ I + V]
 \end{equation}
with kernel
\begin{equation}
   V(\la,\mu) = 4  \frac{ \sin [\pi\nu(\la)] \,  \sin[\pi\nu(\mu)]}{ 2i\pi (\la-\mu)} 
                         \big\{E_+(\la) \, E_-(\mu)-E_+(\mu) \, E_-(\la) \big\}  .
\label{kern-V}
\end{equation}
Here, the function $E_+(\la)$ reads
\begin{equation}
     E_+(\la)
      =i E_-(\la) \left\{ \, 
       \int\limits_\mathbb{R} \hspace{-4.3mm}\diagup \hspace{2mm}
       \frac{ \dd \mu }{2\pi } \frac{ E^{-2}_{-}(\mu) }{ \mu-\la } 
       +  \frac{ E_-^{-2}(\la) }{ 2 } \cot [\pi \nu(\la)]  \right\} .
\label{def-E+}
\end{equation}

Indeed, it follows from the form \eqref{cf-p} of the counting function $\wh{\xi}_\nu$ that $A_L$ and $B_L$ tend respectively to $-\infty$ and $+\infty$ in the thermodynamic limit.
Moreover, it can easily be shown that $I_L[\nu, E_-^{2}]= \e{O}(L^{-1})$.
The thermodynamic limit of $X_N[\nu,E_-^2]$ is therefore a direct consequence of the 
fact that $\det_{N}[ \de_{k\ell}+ \e{o}(L^{-1}) ] \tend 1$, whenever the $\e{o}(L^{-1})$ symbol is uniform in the entries.

%%%%%%%%%%%%%%%%%%%%%%%%%%%%%%%%%%%%%%%%%%%%%%%%%%%
\section{Functional Translation operator}
\label{sec-ftrans}

In this appendix, we provide a heuristic approach to the notion of functional translation operator that we use in Section~\ref{sec-decoupling} to express our highly coupled form factor series in terms of a decoupled one.
As we will see, such an object is in fact a convenient tool to manipulate generalized multi-dimensional Lagrange series (see Appendix C of \cite{KitKMST09b}). 
We refer to~\cite{Koz10u} for a more explicit and rigorous construction.

The notion of one-dimensional translation operator, which acts on some holomorphic function $g$ as
\begin{equation}
    \ex{\a \Dp{\omega}} \cdot g(\omega) \big|_{\omega=0}
   =   \sum_{n \geq 0} \frac{\alpha^n}{ n! }\, g^{(n)}(0)=g(\a),
\end{equation}
can easily be extended to the multidimensional case:
\begin{equation}
 \pl{p=1}{s} \ex{\a_p  \Dp{\omega_p}} \cdot g_s(\omega_1,\dots, \omega_s) \big|_{\omega_p=0}
 = g_s(\a_1,\dots, \a_s).
\end{equation}
By analogy, we can thus define a translation operator acting on functionals as
\begin{equation}\label{def-ftrans}
T_{\ga} \cdot U[\tau] \big|_{\tau=0}
           =  \exp\paa{ \Int{\mathcal{C}}{}\dd\omega\, \ga(\omega)\, \frac{ \de }{ \de \tau(\omega)}  } U[\tau] \Big|_{\tau=0} =U[\ga].
\end{equation}
Here, the integral is taken on a contour $\mathcal{C}$ (typically, an interval of the real axis), and $U$ is a functional acting on holomorphic functions defined on a neighborhood $\mathcal{U}$ of $\mathcal{C}$.

Formula \eqref{def-ftrans} can be understood as the result of the action of finite dimensional translation operators on a finite dimensional approximation of $U$.
More precisely, let us consider a discretization $t_1,\dots, t_s$ of $\mathcal{C}$ and a collection of holomorphic 
functions $U_s(\{z_i\}_{i=1}^{s})$ such that, for any holomorphic function $\gamma$,
$U_s ( \{ \ga(t_i)\}_{i=1}^{s} ) \limit{s}{+\infty} U[\ga]$.
Then, we can represent the translation operator
in terms of limits of the finite variable  case:
\begin{equation}
T_{\ga} \cdot U[\tau] \big|_{\tau=0} 
=  \lim_{s\to +\infty} \pl{p=1}{s}\exp\paa{ \ga(t_p)\, \f{ \Dp{} }{ \Dp{}\tau(t_p)}  } \; U_s\big(\{\tau(t_k)\}_{1}^{s} \big) \Big|_{\tau=0}  .
\end{equation}

Such discretizations allow us to compute the action of more complicated translation operators. Let us define, as in Section~\ref{sec-decoupling}, the operator ordered version $\bs{:} \mc{O} \bs{:}$ of some expression $\mathcal{O}$ containing functional derivatives of the type $\tf{\de }{ \de \tau(\la)}$ (such as in \eqref{def-ftrans}) as being the expression where all functional derivative operators are placed on the left (in each term of the Taylor series expansion of $\mathcal{O}$).
Then, let us consider, for some functionals $\Ga$ and $U$, the following generalization of \eqref{def-ftrans}: 
\begin{align}
 \mathcal{L}
  &= \; \bs{:}\,  \exp\Bigg\{ \Int{ \mathcal{C} }{} \Ga[\tau](\mu) \, \frac{\de }{ \de \tau(\mu)} \, \dd \mu \Bigg\} \,
                                                U[\tau] \, \bs{:}  \, \Big|_{\tau=0} 
                                                \label{L-Gamma}\\
  &\equiv \sul{n=0}{+\infty} \f{1}{n!} \Int{\mathcal{C} }{} \dd^n \mu  \, \pl{i=1}{n}\f{ \de }{ \de \tau(\mu_i)}
                 \cdot
                 \Bigg\{ \pl{p=1}{n} \Ga[\tau](\mu_p)  \; U[\tau] \Bigg\} \Bigg|_{\tau=0}  .
\end{align}
After discretization one gets
\begin{equation}
 \mathcal{L} = \lim_{s\to +\infty} \mathcal{L}_s
\end{equation} 
with
\begin{align}
  \mathcal{L}_s
  &=  {\bs{:}} \pl{p=1}{s} \exp\paa{ \Ga_s( \{\tau_k\}_1^s) (\tau_p)\, \Dp{\tau_p}} U_s( \{\tau_k\}_1^s) 
        \Big|_{\tau_k=0} \!\! { \bs{:} }
    \\
  &\equiv
\sul{ \substack{ n_1,\ldots,n_s \\ =0  } }{+\infty } \pl{\ell=1}{s} \f{ \Dp{\tau_{\ell} }^{n_{\ell}} }{n_{\ell}!}
  \cdot
\Bigg\{ \pl{p=1}{s} \Ga_s^{n_p}( \{\tau_k\})(\tau_p) \; U_s(\{\tau_k\}) \Bigg\} \Bigg|_{\tau_k=0} ,
\end{align}
in which we have set $\tau_k \equiv \tau(t_k)$.
The last series is a multi-dimensional Lagrange series which can be computed as
\begin{equation}
  \mathcal{L}_s = \f{ U_s( \{\nu(t_k)\} ) }{ \det_s \big[ \de_{jk} -\Dp{\tau_k}\Ga_s(\{\tau_{\ell}\})(t_j) \big]_{\tau_k=\nu(t_k)} } \, ,
\end{equation}
where $\nu(t_k)$, $k=1,\ldots,s$, are obtained as the solutions of the system $\nu(t_k) = \Ga_s(\{\nu(t_{\ell})\})(t_k)$, $k=1,\ldots,s$.
The $s\tend +\infty$ limit can be taken, at least formally. It yields
\begin{equation}\label{result-Lagrange}
  \mathcal{L}= \f{ U[\nu] }{ \det_{ \mathcal{C} } \Big[I-\f{\de }{\de \tau(\mu)} \Ga[\tau](\la)  \Big]_{\tau=\nu} } \, ,
\end{equation}
in which $\nu$ is the solution to the equation
\begin{equation}
    \nu(\mu) = \Ga[\nu](\mu) .
\end{equation}
%

%%%%%%%%%%%%%%%%%%%%%%%%%%%%%%%%%%%%%%%%%%%%%%%%%%%
%%%%%%%%%%%%%%%%%%%%%%%%%%%%%%%%%%%%%%%%%%%%%%%%%%%
\section{The Master equation issued-like series representation}
\label{sec-master}

In this appendix we obtain, starting from \eqref{Q-det}-\eqref{E+Lhat}, an alternative series representation for  $\mc{Q}_N^\kappa(x,t)_\mathrm{eff}$.
This representation is in the spirit of the series obtained in \cite{KitKMST09b} for the equal-time correlation functions.

\subsection{The finite-size series}

Expanding the determinant in \eqref{Q-det} into its finite
Fredholm series (this expansion is an immediate consequence of the Laplace expansion for determinants), we obtain:
\begin{equation}
   \mc{Q}_N^\kappa(x,t)_\mathrm{eff}
    = \ex{- \frac{\be p_F}{\pi} x}
       \bs{:}   \prod_{j=1}^{N} \frac{ \wh{E}_-^{2}(\bar{\mu}_j) }{ \wh{E}_-^{2}(\la_j) }
       \sum_{n=0}^{N} \frac{1}{n!} 
       \sul{ i_1,\dots, i_n=1   }{N}  \pl{s=1}{n} \frac{1}{L \wh{\xi}'(\la_{i_s})  } \,
    \det_{n} \! \big[ \widehat{V}^{(L)}(\la_{i_j},\la_{i_k}) \big]  \,  \msc{G}\pac{\om}
    \bs{:} 
    \Big|_{\substack{\tau = 0 \\ \omega=0}} ,
\label{QnKappa Mult Fred}    
\end{equation}
in which $\hat{V}^{(L)}(\la,\mu)$ is given by \eqref{noyau-Lhat}.
It is convenient to use the analytic properties of the operator valued mappings $\wh{E}_{+}^{(L)}$ and $\wh{E}_-$ so as to re-cast this finite-size kernel
$V^{(L)}$ into a more compact contour integral representation:
\begin{equation}
\wh{V}^{(L)}(\la,\mu) = 4 \sin [ \pi \nu_{\tau}(\la) ] \sin [ \pi \nu_{\tau}(\mu) ]\, \wh{E}_-(\la) \, \wh{E}_-(\mu)
\Oint{ \msc{C}_q }{}  \f{\dd z }{ (2i\pi)^2 } 
 \f{ \wh{E}_+^{(L)}(z)\,  \wh{E}_-^{-1}(z)  }{(z-\la) (z-\mu)} ,
\label{V-fonct}
\end{equation}
where $\msc{C}_q$ surrounds $[-q,q]$ (see Fig.~\ref{contours}), $\la,\mu \in \intff{-q}{q}$ and
\beq
 \wh{E}_+^{(L)}(z)\,  \wh{E}_-^{-1}(z) 
   = i\Int{\msc{C}^{(L)}}{} \frac{\dd y}{2\pi} \frac{ f^{(L)}(y,\nu_{\tau}(y)) }{y-z} \, \wh{E}_-^{-2}(y).
 \label{def-O}
\enq
In this last expression the contour $\msc{C}^{(L)}=\msc{C}^{\pa{L}}_E \cup \wt{\msc{C}}_q \cup \msc{C}_{\ua} \cup \msc{C}_{\da}$
consists of a union of four contours. Two of them, $\msc{C}_{\ua/\da}$, are as depicted in Fig.~\ref{contour-IL}, and 
the remaining two, $\msc{C}_E^{\pa{L}}\cup \wt{\msc{C}}_q$, appear in Fig.~\ref{contours}. In particular, the contour $\wt{\msc{C}}_q$
encircles the contour $\msc{C}_q$ of \eqref{V-fonct}, which enables us to recast the integral kernel $\wh{V}^{\pa{L}}(\la,\mu)$
in a more compact form. The integrand in \eqref{def-O} involves the function 
\beq
f^{(L)}(y,\nu_{\tau}(y)) = \bs{1}_{\msc{C}^{\pa{L}}_E}(y) - \f{ \bs{1}_{\wt{\msc{C}}_q}(y)  }{ \ex{-2i\pi \nu_{\tau}(y)}-1 }
 +  \f{ \bs{1}_{\msc{C}_{\da}}(y)  }{ \ex{2i\pi L \wh{\xi}_{\nu_{\tau}}(y)}-1 }
 +  \f{ \bs{1}_{\msc{C}_{\ua}}(y)  }{ 1-\ex{-2i\pi L \wh{\xi}_{\nu_{\tau}}(y)}}\;.
\enq
\begin{figure}[h]
\begin{center}

\begin{pspicture}(6,2.5)

\psline(0.5,0.5)(1,0.5)
\psline(1,0.5)(1,2.5)
\psline(1,2.5)(5,2.5)
\psline(5,2.5)(5,0.5)
\psline(5,0.5)(5.5,0.5)

%Label des point q

\psdots(0.5,0.5)(2,0.5)(4,0.5)(5.5,0.5)

\rput(2.2,0.7){$-q$}
\rput(3.7,0.7){$q$}
\rput(0.5,0.2){$\wh{A}_L$}
\rput(5.5,0.2){$\wh{B}_L$}

\rput(4.1,1.7){$\wt{\msc{C}}_q$}
\rput(3,0.8){$\msc{C}_q$}
\rput(0.6,1.7){$\msc{C}_E^{(L)}$}

\psline[linewidth=2pt]{->}(3.1,1.1)(3,1.1)

\pscurve(1.6,0.5)(1.7,0.9)(2.5,1.1)(3,1.1)(4,1)(4.5,0.5)(4.5,0.3)(4,0.2)(2.5,0.2)(1.7,0.2)(1.6,0.5)

\psline[linewidth=2pt]{->}(3.1,1.5)(3,1.5)

\psline[linewidth=2pt]{->}(3.1,2.5)(3.2,2.5)

\pscurve[linestyle=dashed](1.3,0.5)(1.4,1.1)(2.3,1.4)(3,1.5)(4,1.4)(4.5,1) (4.8,0.5)(4.8,0.2)(4,0)(2.5,0)(1.7,0)(1.3,0.5)
\end{pspicture}

\caption{Contours $\msc{C}_E^{(L)}$, $\msc{C}_q$ and $\wt{\msc{C}}_q$.
The endpoints $\wh{A}_L$ and $\wh{B}_L$ are such that $L\wh{\xi}_{\nu_{\tau}} (\wh{A}_L)=-w_L-\tf{1}{2}$ and $L\wh{\xi}_{\nu_{\tau}}(\wh{B}_L)=w_L+\tf{1}{2}$.\label{contours}}
\end{center}
\end{figure}

Factorizing the integrals over $z$ \eqref{V-fonct} out of the determinant in \eqref{QnKappa Mult Fred}  and 
then using the symmetry of the summand in order to reconstruct a second Cauchy determinant, we get
\begin{multline}
  \mc{Q}_N^\kappa(x,t)_\mathrm{eff}
   = \ex{-\f{\be p_F}{\pi} x}  
  \bs{:}    \pl{\ell=1}{N}\f{ \wh{E}_-^{2}(\bar{\mu}_{\ell}) }{ \wh{E}_-^{2}(\la_{\ell}) } \,
  \sul{n=0}{N} \f{1}{ \pa{n!}^2} \sul{  i_1,\dots, i_n  =1   }{N}
  \pl{s=1}{n}  \left\{ \f{4\sin^2[\pi  \nu_{\tau}(\la_{i_s})]  }{L \wh{\xi}'(\la_{i_s})} \, \wh{E}_-^{2}(\la_{i_s}) \right\} \\
\times
      \Oint{ \msc{C}_q }{} \hspace{-1mm}\f{\dd^n z }{\pa{2i\pi}^{2n}} \, \Int{\msc{C}^{\pa{L}}}{} \hspace{-1mm} \f{ \dd^n y}{ \pa{-2i\pi}^n }
 \pl{s=1}{n} \bigg\{ \f{ f^{(L)}(y_s,\nu_{\tau}(y_s)) }{ y_s-z_s } \wh{E}_-^{-2}(y_s) \bigg\}     
           \cdot
            \det^{2}_n\pac{ \f{1}{z_j-\la_{i_k}} } \,   \msc{G}[\om] 
\bs{:} 
    \bigg|_{\substack{\tau = 0 \\ \omega=0}} \,.
\label{eq-Cauchy}
\end{multline}

Since
\begin{equation}
   \wh{E}_-^2= \ex{-ixu-\wh{g}_\tau-\wh{g}_{\om}},
\end{equation}
it remains to compute the effect of the two types of functional translation operators occuring in \eqref{eq-Cauchy}. 
We start by taking  the $\om$-translation into account, namely 
\begin{align}
\pl{a=1}{n} \f{ \ex{\wh{g}_{\om}(y_a)} }{ \ex{\wh{g}_{\om}(\la_{i_a} ) } } 
\pl{a=1}{N} \f{ \ex{\wh{g}_{\om}(\la_a)} }{ \ex{\wh{g}_{\om}(\bar{\mu}_a ) } } \cdot \msc{G}[\om] \Big|_{\om=0}
&=
\msc{G}\bigg[ \varpi_{N+n}\! \pabb{ \cdot }{ \{\la_j\}_{1}^{N} \cup \paa{y_s}_{s=1}^{n} }{ \{ \bar{\mu}_j\}_{1}^{N} \cup \{\la_{i_s}\}_{s=1}^n } \bigg] \nonumber\\
 &=     \mc{G}_{N}  \binom{ \{\la_j\}_{1}^{N} \setminus \{\la_{i_s}\}_{s=1}^n\cup \paa{y_s}_{s=1}^{n} }{ \{ \bar{\mu}_j\}_{1}^{N}  } \;.
\end{align}
Above, we have first computed the action of the translation operators according to \eqref{transl ac varpi}, 
then used the equality \eqref{rep Gn comme Fnelle}. 
We should now compute the action of the second set of translation operators involving $\wh{g}_{\tau}$. 
Since the parameters $\bar{\mu}_\ell\equiv \bar{\mu}_{\ell}[\nu_{\tau}]= \wh{\xi}_{\nu_{\tau}}^{-1}(\tf{\ell}{L}) $, $\ell=1,\dots,N$, at which the $\wh{g}_{\tau}$ are evaluated, are themselves functionals of 
$\tau$, we should take care of the operator ordering.
The expression we have to compute is of the type:
\begin{multline}
\bs{:}
   \sul{n=0}{ N }  \f{1}{ \pa{n!}^2}
   \sul{   i_1,\dots, i_n =1   }{N}   \;    \Oint{ \msc{C}_q }{} \hspace{-1mm}\f{\dd^n z }{(2i\pi)^{2n}}  \hspace{-1mm}
   \Int{ \msc{C}^{\pa{L}} }{} \hspace{-2mm} \f{ \dd^n y }{ (-2i\pi)^n }
\exp\bigg\{ \Int{\R}{} \Ga_{\! \{ i_a \} }[\tau](\om) \f{ \de }{ \de \tau(\om) }  \dd \om  \bigg\}  
\\   
\cdot \mc{R}_{\{i_a\}}^{(n)}\binom{ \{y_s\}_1^n }{ \{z_s \}_1^n  } [\nu_{\tau} ]  \bs{:} 
    \bigg|_{ \tau = 0 } \,, 
\end{multline}
in which $\mc{R}_{\{i_a\}}^{\pa{n}}$ is a smooth functional of $\nu_{\tau}$ that, moreover, depends on the integration variables $\{ y_s \}$
and $\{ z_s \}$, and the functional $\Ga_{\!\{ i_a \}}[\tau](\la)$ driving the functional translation reads 
\beq
\Ga_{\!\{ i_a \}}[\tau](\la) =  \sul{\ell=1}{N} \pac{ \phi(\la, \bar{\mu}_{\ell}[\nu_{\tau}] ) - \phi(\la, \la_{\ell} ) }
\; + \; \sul{s=1}{n} \pac{ \phi(\la, \la_{i_s} ) - \phi(\la, y_s )} .
%\label{def-fnelle Gamma ia}
%
\enq
According to Appendix~\ref{sec-ftrans} (see \eqref{L-Gamma}), computing the action of a functional translation operator 
with weigths $\Ga_{\!\{ i_a \}}[\tau](\la)$ as above amounts to summing up a multi-dimensional Lagrange series.
In our case, the result is obtained by replacing the functional argument $\nu_{\tau}$ of $\mc{R}_{\{i_a\}}^{(n)}$
by the function $\nu_{\wh{\mf{r}}}$, in which $\wh{\mf{r}}$  is the solution to the non-linear functional equation 
$\wh{\mf{r}}(t)=\Ga_{\!\{ i_a \}}[\, \wh{\mf{r}}\, ](t)$,
and then by dividing the obtained expression by the corresponding functional Jacobian $J_{\!\{ i_a \}}[\, \wh{\mf{r}}\, ]=\det_{\R}\big[I- \tf{ \de \Ga_{\!\{ i_a \}}[\tau](\la) }{ \de \tau(\la)}  \big] \big|_{ 
\tau=\wh{\mf{r}}}$ (see \eqref{result-Lagrange}).

Finally, $\mc{Q}_N^\kappa(x,t)_\mathrm{eff}$ admits the following representation: 
\begin{multline}
 \mc{Q}_N^\kappa(x,t)_\mathrm{eff}
 = \ex{-\frac{\beta p_F}{\pi} x}
   \sul{n=0}{ N }  \f{1}{ \pa{n!}^2}
   \sul{   i_1,\dots, i_n =1  }{N}   \;    \Oint{ \msc{C}_q }{} \f{\dd^n z }{(2i\pi)^{2n}}  \hspace{-1mm}
   \Int{ \msc{C}^{\pa{L}} }{} \hspace{-2mm} \f{ \dd^n y }{ \pa{-2i\pi}^n }
   \frac{\ex{ ix\,  \mc{U}_{\! \{ i_a \} }^{(L)} ( \{ \la_a\} , \{ \bar{\mu}_a \}, \{ y_s \} ) }   }{J_{\!\{ i_a \}}[\, \wh{\mf{r}}\, ]}  
   \\
%
%\exp\bigg\{ \Int{\R}{} \Ga_{ \{ i_a \} }\!\pac{\tau}\!\pa{\om} \f{ \de }{ \de \tau\pa{\om} }  \dd \om  \bigg\}     
%
\times \pl{s=1}{n} \paa{ \f{4 \sin^2[\pi\nu_{\wh{\mf{r}}}(\la_{i_s})]}
 				{L \wh{\xi}'(\la_{i_s}) \pa{y_s-z_s}} f^{(L)}(y_s,\nu_{\wh{\mf{r}}}(y_s)) }  
 \\
 \times
 \det^{2}_n\pac{ \f{1}{z_j-\la_{i_k}} }  \, 
 \mc{G}_{N}  \binom{ \{\la_j\}_{1}^{N} \setminus \{\la_{i_s}\}_{s=1}^n\cup \paa{y_s}_{s=1}^{n} }{ \{ \bar{\mu}_j [ \nu_{\wh{\mf{r}}} ] \}_{1}^{N}  } .
\label{Q-formel}
\end{multline}
In \eqref{Q-formel}, we have set 
\beq
\mc{U}_{ \{ i_a \} }^{\pa{L}}\! \pa{ \{ \la_a\} , \{ \bar{\mu}_a [ \nu_{\wh{\mf{r}}} ] \}, \{ y_s \} } =
\sul{ \ell=1  }{N} \pac{ u(\la_{\ell}) -  u (\bar{\mu}_{\ell}[ \nu_{\wh{\mf{r}}} ]) }
\; + \; \sul{s=1}{n} \pac{ u(y_s) - u(\la_{i_s}) }
%
%\label{def-fnelle Gamma ia}
%
\enq
and explicitly insisted  on the fact that $\bar{\mu}_\ell \equiv \bar{\mu}_\ell[\nu_{\wh{\mf{r}}}]$, $\ell=1,\dots,N$, are now functionals of 
the solution $\wh{\mf{r}}$ to $\wh{\mf{r}}(t)=\Ga_{\!\{ i_a \}}\!\pac{ \, \wh{\mf{r}} \, }\!\pa{t}$.

\subsection{Taking the thermodynamic limit}

For finite $N$, \eqref{Q-formel} gives a rather implicit representation. We are however interested in computing the thermodynamic limit $L,N\tend +\infty$
of this expression (in fact it is only in this limit that the effective series \eqref{Q-formel} is supposed to coincide with the original form factor series \eqref{def-gen}).  In this limit of interest \eqref{Q-formel} can be considerably simplified.

In particular, the non-linear functional equation $\wh{\mf{r}}(t)=\Ga_{\{ i_a \}}\!\pac{\,\wh{\mf{r}}\,}\!\pa{t}$ for $\wh{\mf{r}}$
turns into a linear integral equation for $\mf{r}(t)$, with $\wh{\mf{r}}(t) \limit{N,L}{\infty} \mf{r}(t)$:
\begin{equation}\label{lim-r_goth}
  \mf{r}(t)=\nu_{\mf{r}}(t)-i\frac{\beta Z(t)}{2\pi}
               =\Int{-q}{q} \partial_2 \phi(t,\la)\, \nu_{\mf{r}}(\la)\,\dd\la + \sul{s=1}{n} [ \phi(t,\la_{i_s}) - \phi(t,y_s) ].
\end{equation}
Indeed, in the thermodynamic limit,  the Bethe roots $\la_j$ for the ground state condensate on $[-q,q]$ with the density $\rho(\la)$, whereas
the parameters $\bar{\mu}_j[\nu_{\wh{\mf{r}}}]$ (defined in terms of the counting function $\wh{\xi}_{\nu_{\wh{\mf{r}}}}$), 
are shifted with respect to the $\la_j$'s according to:
\begin{equation}\label{shift-mu}
   \bar{\mu}_j[\nu_{\wh{\mf{r}}}]-\la_j=\frac{\nu_{\mf{r}} ( \la_j ) }{L\rho(\la_j)} + \e{O}(L^{-2}) \; , 
   \qquad j=1,\ldots,N.
\end{equation}
Above, we have denoted the thermodynamic limit of the function $\nu_{\wh{\mf{r}}}$ by $\nu_{\mf{r}}$.
The linear integral equation \eqref{lim-r_goth} is readily solved as soon as one observes that  the derivative  $\partial_2 \phi$ of the dressed phase  \eqref{d-phase} with respect to its second variable is actually related to the resolvent $R$ of the Lieb kernel ($\partial_2 \phi(t,\la)=-R(t,\la)$, with $(I-K/2\pi)(I+R)=I$):
\begin{equation}\label{nu-r_thermo}
  \nu_{\mf{r}}(t)=i\frac{\beta}{2\pi} + \frac{1}{2\pi}\sum_{s=1}^n [\theta(t-\la_{i_s})-\theta(t-y_s)].
\end{equation}
It also follows from \eqref{lim-r_goth} that the Jacobian of the transformation (see \eqref{result-Lagrange}) is simply given by
\begin{equation}\label{jacobien}
   J=\det_{[-q,q]} \big[ I-\partial_2\phi\big]
     =\det_{[-q,q]}^{-1} \Big[ I -\f{K}{2\pi} \Big].
\end{equation}

The expression \eqref{nu-r_thermo} for the thermodynamic limit of $\nu_{\wh{\mf{r}}}(t)$ along with the estimations for the 
shift \eqref{shift-mu} of the parameters $\bar{\mu}_j\pac{ \nu_{\wh{\mf{r}}}}$ with respect to the $\la_j$'s allows one to compute the thermodynamic limit of 
$\mc{U}_{ \{ i_a \} }^{(L)}( \{ \la_a\} , \{ \bar{\mu}_a \}, \{ y_s \} ) $. Namely,
\bem
\mc{U}_{ \{ i_a \} }^{(L)}\! ( \{ \la_a\} , \{ \bar{\mu}_a \}, \{ y_s \} )  \limit{N,L}{\infty}   
- \! \Int{-q}{q} \! u'(\la)\,\nu_{\mf{r}}(\la)\,\dd\la
+\sul{a=1}{n} u(y_a) - u(\la_{i_a})
\\ = -i\frac{\beta p_F}{\pi}+
\sul{a=1}{n} [ u_0(y_a)-u_0(\la_{i_a})]  . \!
\end{multline}
To obtain this limit, we have used the definitions \eqref{d-energy}-\eqref{d-momentum} of $\veps$ and $p$
so as to re-cast the expressions involving integrals of $u=p- t \veps/x$ in terms of $u_0=p_0- t \veps_0/x$.

The thermodynamic limit of $\mc{G}_N$ can be computed along the lines of \cite{KitKMST10u}.
Indeed, one has
\beq
F \pabb{\la} { \{\la_j\}_{1}^{N} \cup \paa{y_s}_{s=1}^{n} }{ \{ \bar{\mu}_j\}_{1}^{N} \cup \{\la_{i_s}\}_{s=1}^n } \limit{N,L}{\infty}  \nu_{\mf{r}}\pa{\la} \;.
\enq
The latter implies that 
\beq
\pl{j=1}{N}  \f{ \om - \bar{\mu}_j+i\eps c }{ \om - \la_j+i\eps c }\cdot \ex{-2i\pi \mc{C}\pac{F}\pa{\om+i\eps c}} \limit{N,L}{\infty}  1\;, \qquad 
\e{for} \quad \eps \in \paa{\pm 1, 0} \;.
\label{ident simpl prod thermo lim}
\enq
By applying identity \eqref{ident simpl prod thermo lim} to the various products entering in the definition of $\mc{G}_N$, one can convince oneself that 
\bem
\det_{[-q,q]} \big[ I -\tf{K}{2\pi} \big] \; \pl{s=1}{N}4 \sin^2\! \pac{\pi \nu_{\mf{r}}\!\pa{\la_{i_s}}}\;
 \mc{G}_{N}  \binom{ \{\la_j\}_{1}^{N} \setminus \{\la_{i_s}\}_{s=1}^n\cup \paa{y_s}_{s=1}^{n} }{ \{ \bar{\mu}_j\}_{1}^{N}  } \\
\limit{N,L}{\infty}   \; 
\pl{s=1}{n} \pac{ 1-\kappa \f{V_-}{V_+}\!\pabb{\la_{i_s} }{\! \paa{\la_{i_a}}_1^n\!\!  } {\! \paa{y_a}_1^n\!\! }   }
\cdot
\mc{F}_n\binom{  \{\la_{i_a} \}_1^n  }{ \{y_s\}_1^n } \;.
\end{multline}
Above, we have set 
\begin{equation}
   V_{\pm}\!\pabb{\mu}{\! \paa{\la_j}_{j=1}^N \!\! }{ \! \paa{z_j}_{j=1}^N \!\! } 
   = \pl{a=1}{N} \f{ ic \mp \pa{\mu-\la_a}  }{ ic \mp \pa{\mu-z_a}  } ,
\end{equation}
and, agreeing that the auxiliary arguments of $V_{\pm}$  are undercurrent by those of $\mc{F}_n$, 
\bem
\mc{F}_n\binom{  \{\la_s\}_1^n  }{ \{y_s\}_1^n } = 
\f{\pa{1-\kappa}^2}{ \det[ I-\tf{K}{2\pi}] } 
\pl{s=1}{n}  \pac{ 1- \f{V_+(\la_{s} ) }{\kappa V_-(\la_{s} ) }  } 
\pl{j=1}{2} \bigg\{ \f{ V_-(\th_j) \, \det_n \big[ \de_{jk} +  \wh{V}^{ (\th_j)}_{jk} \big]  } { 1- \kappa \tf{V_- (\th_j) }{ V_+ (\th_j) } } \bigg\} \\
\times \pl{a,b=1}{n}  \f{ \pa{y_a-\la_{b} -ic}  \pa{y_b-\la_{a} -ic} }{  \pa{y_a-y_b -ic}  \pa{\la_{b}-\la_{a} -ic} } ,
\label{def-Fcal}
\end{multline}
with
\beq
\wh{V}^{\pa{\th}}_{k\ell} = - \, \f{ \pl{s=1}{n} \pa{\la_{k}-y_s}  }{  \pl{ \substack{ s=1 \\ \not= k}  }{n } \pa{ \la_{k}-\la_s }  }
\f{ K_{\kappa}(\la_{k}-\la_{{\ell}}) -  K_{\kappa}(\th-\la_{{\ell}})   }{ V_-^{-1} (\la_{k}) - \kappa V_+^{-1} (\la_{k}) } \;.
\enq
%
%
%
%with the auxiliary arguments being, again, undercurrent by those of $\mc{F}_n$. 

%

It only remains to deal with the $y$-integrations over $\msc{C}^{\pa{L}}$. In this limit, the integrals over $\msc{C}_{\ua/\da}$ give
vanishing contributions and hence the $y$-type integration contour boils down to the contour $\msc{C}= \msc{C}_E^{(\infty)}\cup \wt{\msc{C}}_q$
with a weight function that now reads 
\beq
f\!\pa{y,\nu_{\mf{r}}\pa{y}} = \bs{1}_{ \msc{C}_E^{(\infty)} } \pa{y} 
- \f{ \bs{1}_{\wt{\msc{C}}_q}\!\pa{y}  }{ \ex{-2i\pi \nu_{\mf{r}}\pa{y}}-1 } \;.
\enq
The contour $\msc{C}_E^{(\infty)}$ corresponds to the $L\tend +\infty$ limit of the contour $\msc{C}_E^{\pa{L}}$
depicted on Fig.~\ref{contours}. 
%The parameter $\de>0$ occuring in the definition of $\msc{C}$ is such that $\R +i \de$ is always above $\wt{\msc{C}}_q$. 

Therefore, in the thermodynamic limit,
$\mc{Q}_N^\kappa(x,t)_\mathrm{eff}  \limit{N,L}{\infty} \mc{Q}^\kappa(x,t) $, with
\begin{multline}\label{lim-Q}
     \mc{Q}^\kappa(x,t) =
     \sul{n=0}{+\infty} \f{ \pa{-1}^n  }{ \pa{n!}^2} \Int{-q}{q}  \f{ \dd^n \la}{ (2i\pi)^n} 
     \Oint{ \msc{C}_q }{} \f{\dd^n z }{(2i\pi)^n} 
     \Int{ \msc{C} }{}  \f{ \dd^n y }{ \pa{2i\pi}^n } \pl{a=1}{n} \ex{ix \pac{ u_0(y_a)-u_0(\la_a)}}  
     \\
 \times 
  \pl{s=1}{n} \bigg\{ \f{ f(y_s,\nu_{\mf{r}}(y_s)) }{ y_s-z_s}\,
  \pac{ 1-\kappa \f{V_+}{V_-}\!\pabb{\la_{s} }{\! \{\la_{a}\} \!\! } { \!\{y_a\} \!\! }  }
  \bigg\}
  \det^{2}_n\bigg[ \f{1}{z_j-\la_{k}} \bigg] \,
  \mathcal{F}_n\binom{ \{ \la_a \} }{\{y_a\} } \, .
\end{multline}
Here, we remind that $\nu_{\mf{r}}$ is given by \eqref{nu-r_thermo}
and is a function of $\{y_s\}$ and $\{\la_j\}$.

%%%%%%%%%%%%%%%%%%%%%%%%%%%%%%%%%%%%%%%%%%%%%%%%%%%
\subsection{The equal-time case}

We now show that in the equal-time case, one recovers the series obtained from the master equation in \cite{KitKMST09b}. 
This stems from the fact that, at $t=0$, the $y$-integrals over $\msc{C}^{\pa{\infty}}_E$ do not contribute. 
Indeed, when $t=0$, we can deform the contour $\msc{C}^{\pa{\infty}}_E$ into $\mathbb{R}+i\de$, with $\de>0$
and small but such that the line $\R+i\de$ is above $\wt{\msc{C}}_q$. 

In order to prove this assertion, we first build on the symmetries of the integrand in \eqref{lim-Q} so as to split
the $y$-integrations into those along $\R+i\de$ ($y_a$, $a=1,\dots,k$) and those along $\wt{\msc{C}}_q$ ($y_a$, $a=k+1,\dots,n$), with $k=1,\dots, n$.
The integrals over $\wt{\msc{C}}_q$ can then be computed by residues. One eventually obtains
\begin{multline}
\mc{Q}^\kappa(x,0) =
     \sul{n=0}{+\infty} \sul{k=0}{n}  \f{ \pa{-1}^{k}  }{ n!\, k! (n-k)!} \Int{-q}{q} \! \f{ \dd^n \la}{ (2i\pi)^n} 
     \Oint{ \msc{C}_q }{} \! \f{\dd^n z }{(2i\pi)^n} 
     \Int{ \R + i \de }{} \!\!\! \f{ \dd^{k} y }{ (2i\pi)^{k} }
     \pl{a=k+1}{n} \hspace{-2mm} \ex{ix [ z_a-\la_a]} 
     \pl{a=1}{k} \f{ \ex{ix [y_a-\la_a]} }{ y_a-z_a }  \\
 \times  \det^{2}_n\bigg[ \f{1}{z_j-\la_{k}} \bigg]   \,
\f{   \pl{s=1}{n} \pac{ 1-\kappa \f{V_-}{V_+}\!\pabb{\la_{s} }{\! \paa{\la_{a}}_1^n\!\!  } {\! \{y_a\}_1^k \cup \{z_a\}_{k+1}^{n}  \!\! }   }  }
{   \pl{a=k+1}{n} \pac{\kappa \f{V_-}{V_+}\!\pabb{z_{a} }{ \!\paa{\la_{a}}_1^n\!\!  } {\!\! \{y_a\}_1^k \cup \{z_a\}_{k+1}^{n}  \!\! }   -1 }  }\;
 \mathcal{F}_n\binom{ \{ \la_a \}_1^n }{ \{y_a\}_1^k \cup \{z_a\}_{k+1}^{n} }   \;.
\label{QThermoDecInty}
\end{multline}

When considered as a function of $y_1,\dots, y_k$, the integrand in \eqref{QThermoDecInty} has no poles (or even other singularities)
in the half-planes $\Im\pa{y_k} \geq \de$. Indeed, no poles can arize from the $\th_j$ dependent terms in $\mc{F}_n$ in as much as this function
does not depend on $\th_j$, \textit{cf} \cite{KitKMST09b}. Also the potential singularities of the determinants at the zeroes of 
$1 - \kappa  (\tf{V_-}{   V_+})(\la_a)$ are only apparent due to the presence of the  pre-factors 
$\prod_s [1 - \kappa^{-1}(\tf{V_+}{  V_-}) (\la_s)]  [1 -  \kappa (\tf{V_-}{   V_+}) (\la_s)]$
distributed in between \eqref{def-Fcal} and \eqref{QThermoDecInty}.
The potential poles at $y_a=\la_b \pm ic$ introduced by these pre-factors are cancelled by the zeroes of the double product in 
the last line of  \eqref{def-Fcal}. In its turn, the poles in the upper-half plane at $y_a=z_s + ic $, with $a=1,\dots, k$ and $s=k+1,\dots,n$, that are 
introduced by the double product in \eqref{def-Fcal}, are compensated\footnote{The poles at $y_a=z_s-ic$ lying in the lower 
half-plane are not explicitly compensated but this is irrelevant for our purposes} by the same poles present in   
$\prod_{s=k+1}^{n}[ 1 - \kappa (\tf{V_-}{ V_+})(z_s)]$. 
Therefore the only singularities in the $y$ variables  correspond to the zeroes of 
\beq
  \kappa \f{V_+}{V_-}\!\pabb{z_{s} }{\! \paa{\la_{a}}_1^n\!\!  } { \!\{y_a\}_1^k \cup \{z_a\}_{k+1}^{n}\!\! }  -1 \;.
\enq
However, one can always squeeze the $z$-integration contours in \eqref{QThermoDecInty} so that $\Im(z_a)= 0^{\pm}$, $a=1,\dots,n$. In such a 
 situation, it is readily seen that for $y_a \in \R+i\de$, $\de>0$, one has 
\beq
\abs{  \kappa \f{V_+}{V_-}\!\pabb{\la_{s} }{\! \paa{\la_{a}}_1^n \!\! } {\! \{y_a\}_1^k \cup \{z_a\}_{k+1}^{n}\!\! } } >1 \;.
\enq
In other words, the aforementioned equation has no solutions in the upper half-plane. 
This lack of singularities allows one to simultaneously deform the $y$-integration contours  from $\R+i\de$
to $\R +i \De$, with $\De>0$ and as large as desired. Due to the presence of the oscillatory factors $\ex{ix y_a}$, 
these integrals will contribute as $\ex{-\De x}$. Therefore,  by sending $\De \to +\infty$, we see that these contributions vanish.

We have thus proven that, in the $t=0$ case, the series of multiple integral representation for 
$\mc{Q}^\kappa(x,0)$ boils down to 
\begin{multline}
    \mc{Q}^\kappa(x,0) 
    =
    \sul{n=0}{N} \f{(-1)^n }{ (n!)^2} \Int{-q}{q}\! \f{ \dd^n \la}{ (2i\pi)^n}
    \Oint{ \msc{C}_q }{}\! \f{\dd^n z }{(2i\pi)^n}   \pl{s=1}{n}  \ex{ix(z_s-\la_s ) } \;
 \det^{2}_n\pac{ \f{1}{z_j-\la_{k}} } \;
 \mc{F}_n \binom{  \{z_a\} }{ \{\la_{a} \}  }    
   \\
\times    
 \pl{s=1}{n} \f{   1- \kappa \f{V_-}{ V_+}\pabb{\la_s }{\! \{\la_{a}\}\!\!  } {\! \{z_a\} \!\! }    }
 {  1- \kappa \f{V_-}{ V_+}\pabb{z_s }{ \!\{\la_{a}\}\!\!  } {\! \{z_a\} \!\! }    }  .
\label{equation pour fonction generatrice avant simplification finale}
\end{multline}
In order to prove our assertion, it remains to show that the last line in  \eqref{equation pour
fonction generatrice avant simplification finale} does not contribute once that the $z$-integrals are taken.
In virtue of the symmetry of the integrand, one can replace one of the Cauchy determinants by $n!$ times the products of its 
diagonal entries. By expanding the remaining Cauchy determinant into a sum over permutations, one sees that 
one has to compute double poles at $z_{p_k}=\la_{p_k}$ with $p_1<\dots < p_{\ell}$ with $p_i \in \{1,\dots n \}$ and $1 < \ell <n$. 
All other poles will be simple, leading to the equality between the two sets $\{z_a \}_{a \not= p_k} = \{ \la_a \}_{a\not=p_k}$. 
These double poles will produce first order derivatives with respect to $z_{p_k}$ at $z_{p_k}=\la_{p_k}$. 
Therefore, the effect of these double poles can be taken into account by setting\footnote{The integrand is symmetric with respect to 
the integration variables $z_a$ and $\la_a$.} $z_a=\la_a$ for $a \not= p_k$, $k=1,\dots,\ell$,
and $z_{p_k}=\la_{p_k}+\eps_{p_k}$, and then taking the first order $\eps_{p_k}$-derivatives of the obtained expression at $\eps_{p_k}=0$. 
As a consequence, only the linear in each $\eps_{p_k}$ order of the integrand will contribute. However, under such a substitution, 
one can readily convince oneself that 
\begin{equation*}
 \pl{s=1}{n} \f{   1- \kappa \f{V_-}{ V_+}\!\pabb{\! \la_s \! }{\!\! \{\la_{a}\} \!\!\! } {\!\! \{z_a\} \!\!\! }    }
 {  1- \kappa \f{V_-}{ V_+}\!\pabb{\! z_s \! }{\!\! \{\la_{a}\}  \!\!\! } {\!\! \{z_a\} \!\!\! }    }  
=
\pl{s=1}{\ell} \! \bigg\{1+ \f{i \kappa}{\kappa\! -\! 1} \sul{k=1}{\ell} \eps_{p_s}\eps_{p_k} K^{\prime}(\la_{p_s}-\la_{p_k}) + \e{O}(\eps_{p_k}^2) \bigg\}
=
1+\e{O}(\eps_{p_k}^2).
\end{equation*}
The linear order in the $\eps_{p_k}$'s vanishes. Therefore, we are led to 
\beq
    \mc{Q}^\kappa(x,0) 
    =
    \sul{n=0}{N} \f{(-1)^n }{ (n!)^2} \Int{-q}{q} \! \f{ \dd^n \la}{ (2i\pi)^n}
    \Oint{ \msc{C}_q }{} \! \f{\dd^n z }{(2i\pi)^n}   \pl{s=1}{n}  \ex{ix(z_s-\la_s ) } 
 \det^{2}_n\pac{ \f{1}{z_j-\la_{k}} } \,
 \mc{F}_n \binom{  \{z_a\} }{ \{\la_{a} \}  }    .
\enq
Once upon taking the complex conjugate we recover, word for word, the series obtained in \cite{KitKMST09b}. 

\begin{rem}
Note that our conventions correspond to $x\mapsto -x$  with respect to the work \cite{KitKMST09b}.
\end{rem}

%%%%%%%%%%%%%%%%%%%%%%%%%%%%%%%%%%%%%%%%%%%%%%%%%%%
%%%%%%%%%%%%%%%%%%%%%%%%%%%%%%%%%%%%%%%%%%%%%%%%%%%
\section{Controlling sub-leading corrections: the Natte series representation}
\label{sec-Natte}

The Natte series representation for the Fredholm determinant $\det [I+V]$ is obtained \cite{Koz10ua} from a specific representation of the solution to
the Riemann--Hilbert problem associated with the integrable integral operator $I+V$.
As this particular representation for the solution of the Riemann--Hilbert problem is obtained by a series of 
contour deformations on the so-called initial Riemann--Hilbert problem associated with the integral operator $I+V$, 
it can be seen that the Natte series stems from a certain number of \textit{algebraic transformations}\footnote{Such as contour deformations or
algebraic summations.} carried out on the initial Fredholm series for the determinant.
This fact allows one, at least on the formal ground, to use this Natte series representation in \eqref{Q-Fred}. The rigorous justification of the 
possibility to use the Natte series is given in \cite{Koz10u}.

\subsection{The formula for the remainder}

In fact, the Natte series corresponds to a representation of the remainder $\mc{R}_x[\nu,u,g]$. The latter is expressed as a series of multiple 
integrals. The integrands appearing in this series have good properties with respect to the large-$x$ limit.

More precisely, 
the Natte series for the sine kernel \eqref{kern} takes the following form:
\beq
\mc{R}_x[\nu,u,g]    \; = \; 
\sul{n \geq 1}{} \sul{ \mc{K}_n }{} \sul{ \mc{E}_n(\mathbf{k} ) }{} \;
\Int{ \msc{C}_{\{ \eps_{\mf{t}}}\} }{} \!\! \f{ \dd^{n} z_{\mf{t}} }{ (2i\pi)^n}\,
 H_{n;x}( \{\eps_{\mf{t}}\}, \{z_{\mf{t}}\} ) [\nu]  \pl{\mf{t}\in J(\mathbf{k})}{} \ex{ \eps_{\mf{t}} g(z_{\mf{t}}) } \; .
\label{serie-Natte}
\enq
The second sum appearing  above runs through all the $n$-tuples $\mathbf{k}$ belonging to 
\beq
\mc{K}_n  =  \bigg\{ \mathbf{k}=(k_1,\dots, k_{n}) \; : \; k_a \in \mathbb{N} \; , a=1,\dots, n \quad \e{and} \;
 \e{such} \; \e{that} \quad  \sul{s=1}{n} s k_s   =n \bigg\} \;.
\enq
For each element $\mathbf{k}$ of $\mc{K}_n$, one defines the associated set of triplets $J(\mathbf{k})$ (with cardinal $n$),
\beq
J(\mathbf{k})=\bigg\{ \mf{t}=\pa{t_1,t_2,t_3} \, , \, t_1\in \intn{1}{n} \, , \, t_2 \in \intn{1}{k_{t_1}} \, , \, t_3 \in \intn{1}{t_1} \bigg\} \; ,
\enq
which provides a convenient way of labelling sets of $n$ variables. The third sum runs through all the possible 
choices of elements belonging to the set 
\beq
\mc{E}_n(\mathbf{k}) = \bigg\{  \{\eps_{\mf{t}}\}_{\mf{t} \in J(\mathbf{k})} \: : \; 
\eps_{\mf{t}} \in \{\pm 1, 0\} 
 \quad \e{and} \quad \sul{t_3=1}{t_1}\eps_{\mf{t}} =0 \quad \forall (t_1,t_2)\in \intn{1}{n} \times \intn{1}{k_{t_1}}  \bigg\} \;.
\nonumber
\enq
In other words, $\mc{E}_n(\mathbf{k})$ consists of sets of $n$ parameters $\eps_{\bf{t}}\in \{\pm 1, 0\} $, indexed by triplets
$\mf{t}=(t_1,t_2,t_3)\in J(\mathbf{k})$ and subject to summation constraints.
Finally, there is an $n$-fold integral appearing in the $n^{\e{th}}$ summand of \eqref{serie-Natte}.
The integration variables $z_{\bs{t}}$ are, again, indexed by the triplets $\mf{t}=(t_1,t_2,t_3)$ of $J(\mathbf{k})$.
The contours of integration $\msc{C}_{\{\eps_{\mf{t}}\} }$ depend on the set $\{\eps_{\mf{t}}\} \in \mc{E}_n(\mathbf{k})$. 
They are realized as $n$-fold Cartesian products of one-dimensional compact curves corresponding to various deformations of 
$\R$.

The integrand  $H_{n;x}(\paa{\eps_{\bs{t}}} ; \paa{z_{\bs{t}}} )[ \nu ]$ is a smooth functional of $\nu$ which
is also a function of the integration variables $z_{\mf{t}}$. This functional depends on the  choice of the parameters $\{\eps_{\mf{t}}\}$ from 
$\mc{E}_n(\mathbf{k})$ and on $x$. It is a holomorphic function of the integration
variables $\paa{z_{\mf{t}}}$ belonging to some open neighborhood of the integration contour $\msc{C}_{\{ \eps_{\mf{t}} \} }$.

The Natte series converges for $x$ large enough in as much as, for $n$ large, 
\begin{equation}
\Bigg|\,
\Int{ \msc{C}_{ \{ \eps_{\bs{t}}  \} } }{} \!\! \f{ \dd^{n} z_{\mf{t}} }{ (2i\pi)^n} \,
 H_{n;x} ( \paa{\eps_{\mf{t}}} , \paa{z_{\mf{t}}} )[ \nu ]  \pl{\mf{t}\in J(\mathbf{k}) }{} \ex{ \eps_{\mf{t}} g(z_{\mf{t}}) }   \Bigg|     \leq
 c_2\paf{c_1}{x}^{ nc_3 } ,
\end{equation}
where $c_1$ and $c_2$ are some constants depending on the values taken by $u$ and $g$ in some small neighborhood of $\R$ and on the values taken  by 
$\nu$ on a small neighborhood of $[-q,q]$. The constant $c_3>0$ only depends on $\nu$.

Finally, one has $H_{1;x}=\e{O}(x^{-\infty})$ and, for $n\geq 2$,
\begin{multline}
\hspace{-1mm}  H_{n;x}( \{\eps_{\mf{t}}\} , \{z_{\mf{t}}\} ) [ \nu ]
 = \e{O}(x^{-\infty})  +  
\sul{b=0}{ \pac{\tf{n}{2}} } \sul{p=0}{b} \sul{ m=b-\pac{\f{n}{2}} }{ \pac{\tf{n}{2}}-b} 
%
%\paf{ \mf{e}\pa{q} }{ \mf{e}\pa{-q} }^{m-\bs{\eta} b} \paf{ \mf{e}\pa{\la_0} }{ \mf{e}\pa{-q} }^{\bs{\eta} b}
%
 \paf{ \ex{ix [u(\la_0)-u(-q)] } }{ x^{2 [\nu(-q)] } }^{\!\! \a b} 
\paf{ \ex{ix [u(q)-u(-q)] } }{ x^{2[ \nu(q)+\nu(-q)] }  }^{ \!\! m-\a p} \\
\times
  \f{1}{x^{n-\f{b}{2}}} \cdot  \; H_{n;x}^{m,p,b}( \paa{\eps_{\mf{t}}} , \paa{z_{\mf{t}}} ) [ \nu ]   \, ,
\label{ecriture forme detaille fonction Hn}
\end{multline}
where, for all admissible values of $m, b ,p$, 
\beq
H_{n;x}^{m,p,b}( \paa{\veps_{\mf{t}}} , \paa{z_{\mf{t}}} )[ \nu ] = \e{O}( x^{ n \wt{w} - \de_{n,2}} )
\label{estimation Hnx}
\enq
uniformly on the integration contour. Above, we have set 
\beq
\wt{w} =2 \sup\bigg\{ \abs{ \Re\pac{ \nu(z) - \nu(\tau q)} }^{}_{} \; : \; \abs{z-\tau q} \leq \de \;, \;  \tau=\pm   \bigg\}  \;,
\enq
where $\de>0$ can  be taken as small as desired. 
These functions $H_{n;x}^{m,p,b}\pa{ \paa{\veps_{\mf{t}}} , \paa{z_{\mf{t}}} } \pac{ \nu }$ are such that their asymptotic expansion into
inverse powers of $x$ possesses poles that are encircled by the part of the contour $\msc{C}_{\{ \eps_{\bs{t}} \} }$  producing algebraic (in $x$ large 
enough) contribution to the integral. As a consequence, the $n$-fold integration occurring in the $n^{\e{th}}$ term of the series can produce $z$ 
derivatives. It was shown in \cite{Koz10ua} that the total order of these derivative is at most $n$. Once that these derivatives are taken, 
the estimates in \eqref{estimation Hnx} change from $\e{O}(x^{n\wt{w}-\de_{n,2}})$ to $\e{O}(x^{-\de_{n,2}} \log^n x)$. 

\subsection{The Natte series for the determinant}

We now use the Natte series representation \eqref{serie-Natte}  for the remainder 
$\mc{R}_x[\nu,u, \wh{g} ]$ to prove that the subleading corrections appearing in the large-$x$ asymptotic behavior of the (non operator-dependent)
Fredholm determinant and included in $\mc{R}_x[\nu,u, g ]$ still remain corrections with respect to the leading terms once that 
one computes the action of the functional translation operators. 
By expanding the Fredholm determinant in \eqref{Q-Fred} into its Natte series, we obtain 
\bem
   \mathcal{Q}^\kappa(x,t)
   = \ex{ -\frac{\be x p_F}{\pi} }  \bs{: } 
     \,
     \Bigg\{\, \mc{B}_x[\nu_\tau,u]  + \sum_{\eps=\pm 1 }
                  \ex{i \eps x[u(q)-u(-q)]+\eps [\wh{g}(q)-\wh{g}(-q)]} \;
                  \mc{B}_x[\nu_\tau+\eps,u] 
      \\
      + \frac{1}{x^{\frac32}} \sum_{\eps=\pm 1}
         \ex{i \alpha x[u(\la_0)-u(\eps q)]+\alpha [\wh{g}(\la_0)-\wh{g}( \eps q)]} \;
          b_1^{(\eps,\alpha)}[\nu_\tau,u] \; \mc{B}_x[\nu_\tau,u]  \\
      + \sul{n \geq 1}{} \sul{ \mc{K}_n }{} \sul{ \mc{E}_n(\mathbf{k}) }{}\,
\Int{ \msc{C}_{\{ \eps_{\bs{t}} \} } }{} \!\! \f{ \dd^{n} z_{\mf{t}} }{ (2i\pi)^n} \pl{\mf{t}\in J(\mathbf{k})}{} \hspace{-2mm} 
\big[ \ex{ \eps_{\mf{t}} \wh{g}(z_{\mf{t}}) } \big] \;  
 H_{n;x}( \{\eps_{\mf{t}}\}, \{z_{\mf{t}}\} ) [\nu_{\tau}] \; \mc{B}_x[\nu_\tau,u]          
 \Bigg\}    \cdot \msc{G}[\omega]
     \bs{ : }  \Bigg|_{\substack{\omega=0 \\ \tau=0}}\,  .
\label{natte1}
\end{multline}
Note that in \eqref{natte1}, we have already simplified the two exponents. Also, we remind that $\a=1$ in the space-like regime and
$\a=-1$ in the time-like regime. The action of the translation operators occuring in the first two lines has already been computed in 
Section~\ref{sec-asympt-gen}. In order to compute this action in the last line of 
\eqref{natte1}, it is convenient to introduce the notations 
\beq
a_{ \{\eps_{\mf{t}} \} }= \#\{ \mf{t} \; : \; \eps_{\mf{t}} =1 \} \quad \e{and} \quad 
\{ z_{\mf{t}}^{\bs{\pm} } \} = \paa{ z_{\mf{t}} \; : \mf{t} \in J(\mathbf{k}) \;\; \e{such} \; \e{that} \;  \pm \eps_{\mf{t}} >0} \;.
\enq
It then readily follows that 
\beq
\pl{\mf{t}\in J(\mathbf{k})}{} \hspace{-2mm}  \big[ \ex{ \eps_{\mf{t}} \wh{g}_{\om}(z_{\mf{t}}) } \big]  \cdot \msc{G}[\om] \Big|_{\omega=0} = 
\mc{G}_{ a_{ \{\eps_{\mf{t}} \} } }  \binom{ \{z_{\mf{t}}^{ \bs{+} } \}  }{  \{z_{\mf{t}}^{ \bs{-} } \}  } \;, 
\label{eqn reconst G indice a eps t}
\enq
and
\beq
\bs{:} \!\!
\pl{\mf{t}\in J(\mathbf{k}) }{} \hspace{-2mm}  \big[ \ex{ \eps_{\mf{t}} \wh{g}_{\tau}(z_{\mf{t}}) } \big]  \cdot 
H_{n;x}( \{\eps_{\mf{t}}\}, \{z_{\mf{t}}\} ) [\nu_{\tau}] \; \mc{B}_x[\nu_\tau,u]
\bs{:} \Big|_{\tau=0} \!\! =
H_{n;x}( \{\eps_{\mf{t}}\}, \{z_{\mf{t}}\} ) [ F_{\! \{\eps_{\mf{t}} \} } ] \; \mc{B}_x[ F_{\! \{ \eps_{\mf{t}} \} } ,u] .
\enq
Note that, in order to obtain \eqref{eqn reconst G indice a eps t},  we have used the fact that $\eps_{\mf{t}}\in \{\pm 1, 0 \}$  for all $\mf{t} \in 
J(\mathbf{k})$ and that these parameters are subject to the condition $\sum_{\mf{t}\in J(\mathbf{k})} \eps_{\mf{t}}=0$. As a consequence, 
$\#  \{  z_{\mf{t}}^{\bs{+}} \}=\#  \{  z_{\mf{t}}^{\bs{-}} \}$. Also we have set
\beq
F_{\! \{\eps_{\mf{t}} \} }  (\la;\{ z_{\mf{t}} \} )
\equiv  F_{\! \{\eps_{\mf{t}} \} } (\la)
 =\nu_{\eps_{\mf{t}}} (\la,\{z_t\})
=\f{ i\be Z(\la)}{2\pi} - \sul{ \mf{t} \in J(\mathbf{k}) }{} \eps_{\mf{t}} \, \phi(\la,z_{\mf{t}})   .
\enq
This leads to the following representation for the generating function:
\begin{multline}
\mc{Q}^\kappa(x,t)
 = \mc{Q}^{\kappa;\a}_{\e{asym}} (x,t) 
			\; + \; \sul{n \geq 1}{} \sul{ \mc{K}_n }{} \sul{ \mc{E}_n(\mathbf{k}) }{} \,
\Int{ \msc{C}_{\{ \eps_{\bs{t}} \} } }{} \f{ \dd^{n} z_{\mf{t}} }{ (2i\pi)^n} \\
\times
    H_{n;x}( \{\eps_{\mf{t}}\}, \{z_{\mf{t}}\} ) [ F_{\! \{\eps_{\mf{t}} \} }  ] \; \mc{B}_x[ F_{\! \{\eps_{\mf{t}} \} }  ,u] \; 
    \mc{G}_{ a_{ \{\eps_{\mf{t}} \} } }  \binom{ \{z_{\mf{t}}^{ \bs{+} } \}  }{  \{z_{\mf{t}}^{ \bs{-} } \}  } \; .
\end{multline}
Here, $\mc{Q}^{\kappa;\a}_{\e{asym}}(x,t)$ denotes the leading asymptotic part that is given in \eqref{lead-Q-space} or by its similar 
expression in the time-like regime (without the $\e{O}\big( \tf{(\ln x)}{x} \big)$  remainders since these are now explicitly 
taken into account in the rest of the above equation). 

Note that the $n$-fold integration occuring in the $n^{\e{th}}$ term of the series can produce $z$ derivatives whose total order is at most $n$. 
It thus follows from the representation for the functional $H_{n;x}$ \eqref{ecriture forme detaille fonction Hn}
and from the form of the estimates \eqref{estimation Hnx} that, indeed, the remainder produces corrections of the form 
written in \eqref{lead-Q-space}.

\end{document}